\documentclass[11pt]{article}
\usepackage{import}

\usepackage[dvipsnames]{xcolor}

\pdfoutput=1
\usepackage{jheppub} 
\usepackage{amsmath,amssymb,epsfig,amsfonts}
\usepackage{epsf}
\usepackage{makeidx}
 \usepackage{slashed}
    \usepackage{verbatim} 
    \usepackage{float}
    \restylefloat{table}
    \usepackage{pdflscape}
    \usepackage{tikz}
    \usetikzlibrary{shadows.blur}
    \usepackage{pifont}
\usepackage{array}
    \usepackage{youngtab}
    \usepackage{multirow}
    \usepackage{ulem}
    \usepackage{cleveref}
    \usepackage{tensor}
    \usepackage{todonotes}
    \usepackage{mathrsfs}
    \usepackage{changepage}
    \usepackage{anyfontsize}
   \usepackage{caption}
\usepackage{orcidlink}
\usepackage{academicons}
\usepackage{fontawesome}
\usepackage{ytableau}
\usepackage{subcaption}
\usepackage{tikz-3dplot}

\newcommand{\fleft}{\left.}
\newcommand{\fright}{\, \right|_F\ }

\usetikzlibrary{scopes,fadings,shadings}
\tikzset{
    canvas/.style={draw,left color=blue!35!black,right color=white},
    sign/.style={align=center,fill=white,fill opacity=0.2,text opacity=1,text=white},
    axis label/.style={midway,below,sloped}
}

\usetikzlibrary{shadows}
\usetikzlibrary{decorations.pathreplacing,calligraphy}

\usepackage{pgfplots}
\pgfplotsset{compat=1.17}

\usepgfplotslibrary{fillbetween}

\makeatletter

\newcommand{\be}{\begin{equation}}
\newcommand{\ee}{\end{equation}}
\newcommand{\dd}{\mathrm{d}}
\newcommand{\D}{\mathcal{D}}

\newcommand{\me}{\mathrm{e}}
\newcommand{\ii}{\mathrm{i}}
\newcommand{\vol}{\mathrm{vol}}

\newcommand{\del}{\partial}

\newcommand{\lp}{\ell_{\mathrm{p}}}
\newcommand{\nn}{\nonumber}

\newcommand{\hook}{\mathbin{\rule[.2ex]{.4em}{.03em}\rule[.2ex]{.03em}{.9ex}}}

\newcommand{\Rmnum}[1]{\expandafter\@slowromancap\romannumeral #1@}
\makeatother

\def\bea{\begin{eqnarray}}
\def\eea{\end{eqnarray}}

\newcommand{\ex}{\mathrm{e}}
\newcommand{\diff}{\mathrm{d}}
\newcommand{\R}{\mathbb{R}}
\newcommand{\Z}{\mathbb{Z}}
\newcommand{\C}{\mathbb{C}}

\newcommand{\KI}{K^{I}}
\newcommand{\wI}{w^{I}}
\newcommand{\pI}{{x}^{I}}
\newcommand{\wvarphiI}{\psi^{I}}
\newcommand{\alphaI}{\alpha^{I}}
\newcommand{\mI}{m^I}

\newcommand{\wvarphi}{\psi}

\newcommand{\Ftwo}{F^{(2)}}
\newcommand{\Dtwo}{D^{(2)}}
\newcommand{\Fzero}{F^{(0)}}
\newcommand{\Dzero}{D^{(0)}}
\newcommand{\Sigmazero}{\Sigma^{(0)}}

\newcommand{\piorb}{p_i^{\mathrm{orb}}}
\newcommand{\porb}{p^{\mathrm{orb}}}

\newcommand{\dN}{d^N}
\newcommand{\dS}{d^S}

\newcommand{\pnut}{\mathfrak{n}}

\newcommand{\Sigmabulk}{\Sigma_{\mathrm{bulk}}}
\newcommand{\K}{K}
\newcommand{\abulk}{a^{\mathrm{bulk}}}

\newcommand{\sG}{}

 	\definecolor{hotmagenta}{rgb}{1.0, 0.11, 0.81}

\makeatother

\begin{document}


\baselineskip=18pt  
\numberwithin{equation}{section}  
\allowdisplaybreaks  

\thispagestyle{empty}

\vspace*{1cm} 
\begin{center}
{\fontsize{20pt}{23pt}\selectfont\textbf{Localizing punctures in M-theory}\vspace{10mm}}
 \renewcommand{\thefootnote}{}
\begin{center}
\fontsize{12pt}{20pt}\selectfont{Christopher Couzens\textsuperscript{\orcidlink{0000-0001-9659-8550}}\footnotetext{\href{mailito:christopher.couzens@maths.ox.ac.uk}{christopher.couzens@maths.ox.ac.uk}}, Alice L\"uscher\textsuperscript{\orcidlink{0009-0001-8231-3080}}\footnotetext{\href{mailito:alice.luscher@maths.ox.ac.uk}{alice.luscher@maths.ox.ac.uk}} and James Sparks\textsuperscript{\orcidlink{0000-0003-3699-5225}} \footnotetext{\href{mailito:james.sparks@maths.ox.ac.uk}{james.sparks@maths.ox.ac.uk} }}

\end{center}
\vskip .2cm

 \vspace*{.5cm} 
 \fontsize{12pt}{14pt}\selectfont{{ Mathematical Institute, University of Oxford,\\ Andrew Wiles Building, Radcliffe Observatory Quarter,\\ Woodstock Road, Oxford, OX2 6GG, U.K.}\\}
  
 {\tt {}}

\vspace*{0.8cm}
\end{center}

\renewcommand{\thefootnote}{\arabic{footnote}}
 
\begin{center} {\bf Abstract } 
\end{center}
\noindent 
We use equivariant localization and holography to study four-dimensional $\mathcal{N}=1$ superconformal field theories arising from M5-branes wrapped on a punctured Riemann surface. We explain how, given a Riemann surface with marked points, one can glue in a ``puncture geometry'' locally around each point.
Using equivariant localization we show that the central charge consists of a bulk contribution plus localized puncture contributions.
We recover and generalize the known results for locally $\mathcal{N}=2$ preserving punctures, and derive new results for genuinely locally $\mathcal{N}=1$ preserving punctures. 

\newpage

\tableofcontents
\printindex

\section{Introduction and overview}

Rich classes of interacting superconformal field theories (SCFTs) may be engineered by embedding them into string theory or M-theory. Many such constructions also have  holographic duals in supergravity, 
describing a strong coupling limit of the SCFT. The resulting interplay between field theory, string theory and holography has led to many interesting developments,
including the construction of new infinite
classes of SCFTs, new exact
methods in supersymmetric field theories, and
new constructions in supergravity and holography.

In this paper we continue this development, focusing
on four-dimensional (4d) SCFTs 
arising from M5-branes 
wrapped on a Riemann surface $\Sigma_g$ of genus $g$. 
For a \emph{smooth} Riemann surface,  general
$\mathcal{N}=1$ holographic duals of this type were 
constructed in 
\cite{Bah:2012dg}, generalizing the seminal work of \cite{Maldacena:2000mw}. 
On the other hand, 
the particular $\mathcal{N}=2$ solution 
of \cite{Maldacena:2000mw}
admits a vast family of 
$\mathcal{N}=2$ generalizations for 
\emph{punctured}
Riemann surfaces \cite{Gaiotto:2009we, Gaiotto:2009gz}. 
Here the ``puncture'' is a set of $\mathcal{N}=2$ preserving boundary conditions (specified by a Young diagram of $\mathrm{SU}(N)$ \cite{Gaiotto:2009we}) 
around a given marked point on 
$\Sigma_g$. The enhanced superconformal symmetry gives considerable control in this case, both in field theory and in the holographic duals. In particular the explicit form of 
$\mathcal{N}=2$ AdS$_5$ solutions in 
\cite{Gaiotto:2009gz} allows one to construct 
essentially the most general such supergravity solution in closed form. 

One of the motivations of the present work was to 
develop a more systematic understanding of punctures in holography, and in particular to extend the well-understood $\mathcal{N}=2$ punctures to $\mathcal{N}=1$.  The geometric structure of general $\mathcal{N}=1$ AdS$_5 \times M$ solutions of M-theory was derived in \cite{Gauntlett:2004zh}, 
where $M$ is the 6d internal space. 
For example, for the 
smooth Riemann surface solutions in \cite{Bah:2012dg},
$M$ is the total space of an $S^4$ bundle (the normal sphere to the M5-branes) over $\Sigma_g$. 
Although many explicit solutions 
are known, 
the general $\mathcal{N}=1$ equations, and in particular adding ``puncture data'' to \cite{Bah:2012dg}, are far too 
difficult to hope to solve in closed form.  

Equivariant localization in supergravity \cite{BenettiGenolini:2023kxp} is a newly discovered structure that leads to a method for computing supergravity observables, 
crucially without solving 
the supergravity equations explicitly. 
In particular, 
 the central charge and  scaling dimensions of certain BPS operators 
 for AdS$_5\times M$ solutions have been computed this way in \cite{BenettiGenolini:2023ndb}. It has also been applied to various other setups in \cite{BenettiGenolini:2023kxp,BenettiGenolini:2023yfe,BenettiGenolini:2023ndb,BenettiGenolini:2024kyy,BenettiGenolini:2024xeo,Cassani:2024kjn,Colombo:2023fhu,Couzens:2024vbn,Martelli:2023oqk,Suh:2024asy,Hristov:2024cgj,BenettiGenolini:2024hyd,BenettiGenolini:2024lbj,Couzens:2025ghx,Colombo:2025ihp,Cassia:2025jkr}. Here recall that all 4d $\mathcal{N}=1$ SCFTs possess a conserved U$(1)_R$ symmetry, and this is reflected in the dual supergravity solutions by the existence of a canonically defined Killing vector $\xi$ on $M$. The central charge and other observables may then be computed using 
localization for the action of $\xi$ on $M$, using 
 the Berline--Vergne--Atiyah--Bott fixed point formula \cite{BV:1982, Atiyah:1984px}. 
 We will show how equivariant localization allows one to efficiently recover results for $\mathcal{N}=2$ punctures, but in a way which does not depend on the enhanced $\mathcal{N}=2$ supersymmetry or by solving any supergravity equations. This will allow us to easily extend to
$\mathcal{N}=1$ punctures.\footnote{Studying the embedding of various probe branes within the geometry, in the spirit of \cite{Chen:2010jga,Bah:2013wda,Bah:2014dsa}, gives a diagnostic for the types of punctures that can be considered.} Furthermore, the general recipe that we explain in this paper can be applied more widely. For example, one could consider punctures for different theories, for example the recent work \cite{Apruzzi:2025znw} studying punctures in Riemann surface compactifications of the 6d $\mathcal{N}=(1,0)$ SCFTs in massive type IIA.
Similarly one can use our results to compute defect Weyl anomalies, for example Gukov--Witten-like defects in \cite{Gutperle:2023yrd,Capuozzo:2023fll,Bomans:2024vii}.

Before turning to an outline of the main results, we note that 
a previous series of works examined M5-brane punctures from the perspective of 
holographic anomaly inflow, 
computing the M5-brane anomaly polynomial from the 11d Chern--Simons interactions in M-theory 
\cite{Bah:2019jts, Bah:2019rgq}. Although conceptually different, 
it is a general fact that 
anomalies in SCFTs are related to central charges via the superconformal algebra
\cite{Anselmi:1997am},  
which should mean there is a relationship between anomaly inflow and equivariant localization. We comment briefly on this further in section~\ref{sec:conclusion}. 

\subsubsection*{The bulk geometry}

Our starting point is $N$ M5-branes wrapped over a \emph{smooth} Riemann surface $\Sigma_g$. To preserve supersymmetry, $\Sigma_g$ 
is embedded as a complex curve inside a Calabi--Yau manifold, implementing a partial topological twist of the M5-brane theory along the Riemann surface.
The Calabi--Yau geometry is 
\begin{align}\label{fibreoverSigma}
\mathcal{O}(-p_1^{\text{bulk}})\oplus 
\mathcal{O}(-p_2^{\text{bulk}})\rightarrow \Sigma_g\, , \qquad p_1^{\text{bulk}}+p_2^{\text{bulk}}=\chi(\Sigma_g)=2-2g\, ,
\end{align} 
where the latter equation is the Calabi--Yau condition. 
The near-horizon limit of the M5-branes is an $S^4$
bundle over $\Sigma_g$, where $S^4\subset \C_1\oplus\C_2\oplus\R$, with the complex planes $\C_i$ fibred over $\Sigma_g$ with Chern numbers $-p_i^{\text{bulk}}$, respectively. 
These are precisely the 
$\mathcal{N}=1$ AdS$_5$ solutions constructed in 
\cite{Bah:2012dg}. 
The latter reference  matched the $a$ central charge of the supergravity solutions to  a dual field theory computation by integrating the anomaly polynomial of the M5-branes over $\Sigma_g$ and 
using $a$-maximization \cite{Intriligator:2003jj}.
The \emph{off-shell} (trial) central charge function for these solutions was computed using equivariant localization in \cite{BenettiGenolini:2023ndb}, obtaining
\begin{equation}\label{abulk}
    a^{\text{bulk}}= -\frac{9}{8}b_1b_2(b_1p_2^{\text{bulk}}+b_2p_1^{\text{bulk}})N^3\, .
\end{equation}
Here the R-symmetry Killing vector $\xi$ has been written as $\xi=b_1\partial_{\varphi_1}+b_2\partial_{\varphi_2}$, 
with $\partial_{\varphi_i}$ generating rotations of the complex planes $\C_i$, so that $\xi$ rotates the $S^4$ normal to the M5-branes. 
The on-shell central charge is obtained from \eqref{abulk} by extremizing over $b_1,b_2$, subject to the constraint $b_1+b_2=1$ which correctly normalizes $\xi$ so that the Killing spinor on $M$ has R-charge $\tfrac{1}{2}$. 

Given such a smooth bulk geometry, we may choose a set of marked points $x^I\in\Sigma_g$, labelled by $I=1,\ldots,n$, and glue in local ``puncture geometries''. 
Equivariant localization will imply that the total (off-shell) central charge is
\begin{align}\label{bulkplusdelta}
a = \abulk - \sum_{I=1}^n \delta a^I\, ,
\end{align}
where $\abulk$ is given by \eqref{abulk} and each $\delta a^I$ is a local contribution at each puncture point.\footnote{The minus sign is due to an orientation reversal when we glue.}
In the remainder of this introduction we summarize how to compute $\delta a^I$ for $\mathcal{N}=2$ and $\mathcal{N}=1$ punctures, henceforth
dropping the superscript $I$ and focusing on a single puncture contribution.
 
\subsubsection*{\texorpdfstring{$\mathcal{N}=2$}{N=2} punctures }

A regular 
 $\mathcal{N}=2$ puncture
is given by wrapping  $N$ M5-branes over the local geometry $\C/\Z_\K\subset\C^2/\Z_\K$.\footnote{
Reducing the system of $N$ M5-branes wrapped on the $\C^2/\Z_\K$ singularity
to type IIA, along a circle inside $\C^2/\Z_\K$, 
leads to a type IIA configuration with $K$ D6-branes and $N$ D4-branes. 
The AdS$_5$ solution is
interpreted as a backreaction of the $N$ D4-branes, to which D6-branes are then added, which 
is expected to be a consistent description only for  $K\lesssim N$.}  
We will describe how to glue this into the bulk geometry below, where there are inequivalent ways to do this, but for now we will just focus on the local puncture geometry.
Geometrically, the data for
an $\mathcal{N}=2$ puncture 
is a partial resolution of the $\mathbb{C}^2/\mathbb{Z}_\K$ singularity, together with a choice of M5-brane fluxes through the blown up cycles.
This data is specified 
by a partition of
the positive integers $\K$ and $N$. That is, we write 
\begin{align}\label{KandN}
\K = \sum_{a=1}^d k_a\, , \qquad 
N = \sum_{a=1}^d k_a n_a\, ,
\end{align}
 with $k_a\in\mathbb{N}$, $n_a\in\mathbb{Z}_{\geq 0}$, for some choice of $d=1,\ldots,\K$.
  This is the well-known statement that an $\mathcal{N} = 2$ puncture is specified by a Young diagram of $\mathrm{SU}(N)$ \cite{Gaiotto:2009we}: the $k_a$ are the widths of the blocks whilst the $n_a$ are the heights.  
 It is  convenient to also introduce the  quantities
 \begin{align}\label{landy}
l_a = \sum_{b=a}^d k_b\, , \qquad \hat{y}_a = l_a n_a + \sum_{b=1}^{a-1}k_b n_b\, ,
 \end{align}
 where note that $l_a$ is a decreasing sequence with $l_1=\K$, $l_d=k_d$, and $\hat{y}_d = N$ is the number of M5-branes. 

Our main $\mathcal{N}=2$ result is that 
such a local puncture contributes to the (off-shell) central charge as 
\begin{align}\label{deltaaN2}
 \delta a=-\frac{9 b_1 b_2^2}{16}\bigg[\sG\Big(2-{3}\frac{n_d}{N}+\frac{1}{k_d}\Big)N^3- \sum_{a=1}^{d-1} \frac{k_a\hat{y}_a^3}{l_a l_{a+1}} \bigg]\,.
    \end{align}
In particular, the sum over $a=1,\ldots,d-1$ arises directly from fixed point contributions of the R-symmetry Killing vector 
$\xi$ in the puncture geometry, with the term 
$\hat{y}_a^3$ arising from the zero-form part of the equivariantly closed form constructed in \cite{BenettiGenolini:2023ndb}. 
More precisely, the result 
\eqref{deltaaN2} is valid for a ``$(1,0)$ puncture''. This means that the original
$\C^2/\Z_\K$ singularity is realized as the $\Z_\K$ quotient
$\C^2\ni (w,z_1)\mapsto (\ex^{2\pi \ii/\K}w,\ex^{-2\pi \ii/\K}z_1)$.  The $N$ M5-branes are wrapped on the locus $z_1=0$, with $z_1$ a coordinate on $\C_1$ that is rotated by $\partial_{\varphi_1}$, 
and 
with $w$ a local complex coordinate on the Riemann surface. Instead for a ``$(0,1)$ puncture'' we swap the roles of $z_1$ and $z_2$ when we glue, which simply swaps $b_1$ and $b_2$ in \eqref{deltaaN2}. 

Locally each puncture preserves 
 $\mathcal{N}=2$, but 
 unless $p_2^{\text{bulk}}=0$ and all punctures are type $(1,0)$ (or equivalently $p_1^{\text{bulk}}=0$ and all punctures are type $(0,1)$), globally  
 these will be
 $\mathcal{N}=1$ solutions. Using \eqref{bulkplusdelta}
 we may in general write
\begin{equation}
    a= -\frac{9}{8}b_1b_2(b_1\mathfrak{p}_2+b_2\mathfrak{p}_1)N^3\, ,
\end{equation}
where the $\mathfrak{p}_i$ are independent of the R-symmetry mixing parameters, $b_i$,  given by
\begin{equation}
    \mathfrak{p}_1=p_1^{\text{bulk}}-\frac{1}{2}\sum_{I\in(1,0) \text{ punct.}}\left (2-\frac{3 n^{I}_d}{N}+\frac{1}{k_d^I}-\sum_{a=1}^{d^I-1}\frac{k_a^I }{l_a^I l_{a+1}^I }\frac{\hat{y}_{a,I}^{3}}{N^3}\right)\, ,
\end{equation}
with  a similar expression for $\mathfrak{p}_2$, with $(1,0)\rightarrow (0,1)$ exchanged. 
In particular, the off-shell central charge takes the same functional form as for the smooth Riemann surface case \eqref{abulk}. This is a feature of the set-up and is not a generic feature for $\mathcal{N}=1$ punctures. 
Maximizing the central charge over the constrained $b_i$ one finds 
the on-shell central charge
\begin{equation}
    a_{\mathrm{on-shell}}=-\frac{3b_1 b_2(\mathfrak{p}_1+\mathfrak{p}_2-\sqrt{\mathfrak{p}_1^2- \mathfrak{p}_1\mathfrak{p}_2+\mathfrak{p}_2^2})}{8} N^3\, .
\end{equation}
We will also see that the scaling dimensions of operators dual to certain BPS wrapped M2-branes may be computed via localization, leading to the simple results \eqref{DeltaS2}, \eqref{Deltaspindle}. 

\subsubsection*{\texorpdfstring{$\mathcal{N}=1$}{N=1} punctures}

Our localization result \eqref{deltaaN2}
allows us to compute central charges (and other observables) for globally $\mathcal{N}=1$ solutions that are \emph{locally}
$\mathcal{N}=2$. However, where equivariant localization really comes into its own is in computing observables for 
local $\mathcal{N}=1$ punctures, assuming such solutions exist. 
$\mathcal{N}=1$ supersymmetry allows for considerably more freedom, and we will not attempt to be exhaustive, instead describing general classes and then illustrating with some explicit examples. 
We will find some interesting qualitative and conceptual differences compared to the $\mathcal{N}=2$ case.

The simplest way to introduce an ``$\mathcal{N}=1$ puncture'' is to wrap the $N$ M5-branes over the local $\mathcal{N}=1$ geometry $\C/\Z_\K\subset \C^3/\Z_\K$. Concretely, 
we take the latter orbifold action to act as 
$\C^3\ni (w,z_1,z_2)\mapsto (\ex^{2\pi \ii /\K}w,\ex^{2\pi\ii \alpha_1/ \K}z_1,\ex^{2\pi \ii\alpha_2 /\K}z_2)$ where $\alpha_i\in\Z$ satisfy
$\alpha_1+\alpha_2+1=\K$.\footnote{As discussed later in the paper, the local geometry clearly depends only on $\alpha_i$ mod $\K$, but to specify how this is glued in globally we fix integer lifts.} Recall that 
the smooth $\mathcal{N}=1$
solutions are near-horizon limits of $N$ M5-branes wrapped over a smooth Riemann surface $\Sigma_g$, with normal directions 
having complex coordinates $z_i$, $i=1,2$, and fibred as in \eqref{fibreoverSigma}. 
We may 
introduce this 
$\C^3/\Z_\K$ orbifold at 
an arbitrary point $x\in\Sigma_g$ (with local coordinate $w\in\C)$, with the near-horizon limit then being an $S^4$ orbibundle 
over an orbifold Riemann surface. The $S^4$ may be realized as $\{|z_1|^2+|z_2|^2+t^2=1\}\subset \C_1\oplus\C_2\oplus\R=\R^5$, with the north and south poles of the sphere at $z_1=z_2=0$, $t=\pm 1$, respectively, and these are $\Z_\K$ orbifold points. 
We show that the local contribution of such an orbifold point to the central charge is
\begin{align}
\delta a^{\mathrm{orb}}=-\frac{9}{8}b_1b_2\left(b_2\frac{\alpha_1}{K}+b_1\frac{\alpha_2}{K}\right)N^3\, .
\end{align}
However, more generally we may consider partial resolutions of the orbifold, and we will recover this result from the general formula we  describe next. 

The data of a local $\mathcal{N}=1$ puncture is by definition a choice of partial resolution of the local $\C^3/\Z_\K$ singularity (in principle different choices of resolution at the north pole and south pole of the four-sphere), together with a choice of M5-brane fluxes through the blown up four-cycles. 
We may conveniently describe (partial) 
resolutions of $\C^3/\Z_\K$ using toric geometry, where the latter is equipped with its obvious U$(1)^3$ action. As we will see, for a generic choice of R-symmetry vector $\xi$ the fixed points are precisely the fixed points of U$(1)^3$ in the resolved geometry, and this allows us to compute the quantized M5-brane fluxes using localization. Specifically, if $D_A$ denotes a torus-invariant four-cycle in the (partially) resolved geometry, we have the localization formula
\begin{equation}\label{generalfluxN1intro}
    N_A=\frac{1}{(2\pi\lp)^3}\int_{D_A}G=\frac{1}{2}{b_1b_2}\sum_{a\in\mathcal{I}_A} \frac{1}{k_a}\frac{\hat{y}_a}{\epsilon_{i_a}^a\epsilon_{j_a}^a}\, .
\end{equation}
Here $G$ is the closed M-theory four-form, $\lp$ is the 11d Planck length, and $N_A$ is the quantized flux, with the cycles labelled by an index $A$. On the right hand side of \eqref{generalfluxN1intro} the sum is over the fixed points (vertices in toric geometry) of $D_A$, labelled by $a\in \mathcal{I}_A$, 
with the weights of $\xi$ being $\epsilon_{i_a}^a$, and with the $a$'th point being an orbifold point of order $k_a\in\mathbb{N}$ (so for a fully resolved geometry all $k_a=1$). All of this may be computed very explicitly, for any choice of (partial) resolution, using toric geometry -- it is data that depends only on the topology of the space and the choice of vector field $\xi$ (parametrized by $b_1,b_2$). 
Finally, the variables $\hat{y}_a$ will enter as zero-form components of equivariantly closed forms, evaluated at the fixed points. 

There is similarly a localization formula for $\delta a$, 
generalizing  the $\mathcal{N}=2$ result \eqref{deltaaN2}: 
\begin{align}\label{N1punctintro}
        \delta a
        &=-\frac{9}{16} b_1 b_2\left[3\Big(b_1\pnut_1+b_2N^{[1]}_\epsilon\Big)N^2+\Big(2b_1\porb_2-b_2 \porb_1\Big)N^3
        +\frac{1}{2}\sum_{a} \frac{1}{k_a}\frac{b_1^2b_2^2}{\epsilon_1^a\epsilon_2^a\epsilon_3^a}\hat y_a^3\right]\,.
    \end{align}
The final sum here is over all isolated fixed points, for both partial resolution geometries at the north and south poles of $S^4$, and generalizes the sum in \eqref{deltaaN2}. In particular the $\epsilon^a_i$ are simply weights of $\xi$ at the $a$'th fixed point. The remaining terms in \eqref{N1punctintro} are associated to fluxes 
 that are present in the original orbifold geometry, after resolution. We shall define these more carefully in section~\ref{sec:N=1}, but in brief: the term $\mathfrak{n}_1$ is defined via a similar expression to 
\eqref{generalfluxN1intro}, and in particular is linear in $\hat{y}_a$, 
  $N^{[1]}_\epsilon$ 
is a flux associated to a bulk cycle, while $\porb_i\in\mathbb{Q}$ are certain fractional Chern numbers.\footnote{There is a similar expression instead involving $\mathfrak{n}_2$ and $N^{[2]}_\epsilon$, with a constraint holding between quantities labelled by 1 and~2.}

The $\hat{y}_a$ may be regarded as ``unknown'' variables which must be solved for, with \eqref{generalfluxN1intro} a set of linear constraints.
However, a key difference between local $\mathcal{N}=2$ and $\mathcal{N}=1$ punctures is that in the former case (where the $n_a$ are essentially the $N_A$) the linear flux constraints determine completely the $\hat{y}_a$, resulting in \eqref{landy} that solves for $\hat{y}_a$ in terms of the fluxes. 
Instead for $\mathcal{N}=1$ punctures the flux constraints \eqref{generalfluxN1intro} leave some $\hat{y}_a$ undetermined. The counting here is topological. 
Let us focus on (say) the north pole of the sphere, and for simplicity fix a choice of fully resolved geometry $\mathcal{X}$ (so $\mathcal{X}$ is a non-compact Calabi--Yau three-fold, defining the $\mathcal{N}=1$ puncture at the north pole of $S^4$). Then the number of fixed points  is $d\equiv\chi(\mathcal{X})=1+b_2(\mathcal{X})+b_4(\mathcal{X})$, 
where $b_i(\mathcal{X})$ denote Betti numbers. 
The equations \eqref{generalfluxN1intro} 
give $b_4(\mathcal{X})$ constraints on the $d$ variables $\hat{y}_a$, with $A$ running from $1$ to $b_4(\mathcal{X})$, and the flux through the generic $S^4$ fibre fixes (in our choice of labelling of fixed points) $\hat{y}_d=N$. This generically still leaves $b_2(\mathcal{X})$ degrees of freedom in the $\hat{y}_a$ at a fixed choice of pole of the $S^4$. 

On the other hand, to obtain  \eqref{generalfluxN1intro}, \eqref{N1punctintro}  we have used only two equivariantly closed forms, associated to $G$ and the central charge, respectively. In particular, the former imposes the Bianchi identity $\diff G=0$. There is another equivariantly closed form that imposes the equation of motion for $G$, which defines a closed two-form on $M$. 
We find that imposing this equation of motion gives $b_2(\mathcal{X})$ \emph{quadratic} constraints on the $\hat{y}_a$, but more remarkably we find that these quadratic constraints are precisely given by extremizing $\delta a$ in \eqref{N1punctintro}, subject to the linear constraints 
\eqref{generalfluxN1intro}! 
In other words, varying the off-shell central charge 
effectively imposes the equation of motion for $G$, at the level of the parameters $\hat{y}_a$. 
This is related to the following formula for $a$:
\begin{align}\label{aaction}
a = \frac{1}{48(2\pi)^6\lp^9}\int_{M}\Phi^G\wedge \Phi^{*G}\, .
\end{align}
Here $\Phi^G$, $\Phi^{*G}$ 
are equivariantly closed forms, and crucially \eqref{aaction} 
is also a \emph{partially on-shell action} \cite{BenettiGenolini:2023ndb}. It thus makes sense that the equation of motion for $G$ effectively arises by varying $\delta a$. While we do not have a first principles proof of this precise variational principle, we shall check explicitly that this holds in examples.

We conclude by giving closed formulas in a simple example of an $\mathcal{N}=1 $ puncture, namely the $\C^3/\Z_3$ orbifold where the action of $\Z_3$ is diagonal. 
This is well-known to resolve to 
$\mathcal{X}=\mathcal{O}(-3)\rightarrow \mathbb{CP}^2$, the total space of the canonical line bundle over $\mathbb{CP}^2$. Using this resolution and the same fluxes at both north and south poles, the contribution to the central charge is:
\begin{align}\label{deltaasymintro}
  \delta a=  \frac{3}{16} b_2 \Big[ b_1^2 \Big\{&N^3-3 N^2 \big(3 N_{\epsilon}^{[2]}+2N_{\mathbb{CP}^2}\big)+12 N
N_{\mathbb{CP}^2}^2-8 N_{\mathbb{CP}^2}^3\!\Big\}+{b_2^2} \big( N-2 N_{\mathbb{CP}^2}-\hat{y}_{1}\big)^3\nonumber\\
-2 b_1 b_2 \Big\{&N^3-6 N^2N_{\mathbb{CP}^2}- 8 N_{\mathbb{CP}^2}^3-6 N_{\mathbb{CP}^2}^2 \hat{y}_{1}+6 N N_{\mathbb{CP}^2} \big(2 N_{\mathbb{CP}^2}+\hat{y}_{1}\big)\Big\}\Big]\,.
\end{align}
Here $N_{\mathbb{CP}^2}$ is the flux through the single $\mathbb{CP}^2$ four-cycle in 
$\mathcal{X}$ (which has 
$b_4(\mathcal{X})=1$). 
This leaves one variable $\hat{y}_1$
unconstrained, as expected since 
$b_2(\mathcal{X})=1$ and $\mathcal{X}$ has one two-cycle. Extremizing $\delta a$ with respect to $\hat{y}_1$ gives 
\begin{align}\label{eatmyhat}
    \hat y_1^* & =N-2N_{\mathbb{CP}^2}\pm2\sqrt{\frac{b_1}{b_2}N_{\mathbb{CP}^2}(N_{\mathbb{CP}^2}-N)}\,,
\end{align}
which one can check is implied by imposing the equation of motion for $G$  through the two-cycle. 
One should substitute \eqref{eatmyhat} back into \eqref{deltaasymintro} and add to the bulk contribution to $a$, before extremizing over $b_1, b_2$ subject to the constraint $b_1+b_2=1$ to obtain the final on-shell central charge. 
Notice that \eqref{deltaasymintro} is a cubic polynomial in $b_i$ and $\hat{y}_1$, precisely as expected for a trial central charge function in field theory \cite{Intriligator:2003jj}! This seems to be a non-trivial result of our supergravity localization results, and is not immediate from \eqref{N1punctintro} due to the last term. 

The outline of the rest of the paper is as follows. 
In section~\ref{sec:AdS5} we review equivariant forms for AdS$_5$ solutions of M-theory, constructed in \cite{BenettiGenolini:2023ndb}, and how these may be applied to compute observables for solutions describing M5-branes wrapped over a smooth Riemann surface.  
Section~\ref{sec:orbifoldsection} introduces orbifold singularities. 
In section~\ref{sec:N=2} we study 
$\mathcal{N}=2$ punctures, carefully describing the local and global geometry before applying  equivariant localization. 
Section~\ref{sec:N=1} performs a similar analysis for $\mathcal{N}=1$ punctures, where we describe how to compute for a general such puncture, and then illustrate with some relatively simple examples. 
We conclude in section~\ref{sec:conclusion}. Some additional material is included in an appendix.


\section{\texorpdfstring{AdS$_5$}{AdS(5)} solutions from M5-branes}\label{sec:AdS5}

We begin by summarizing some geometric properties 
of supersymmetric AdS$_5$ solutions of 11d supergravity \cite{Gauntlett:2004zh}, 
including the set of  equivariantly closed forms constructed in
\cite{BenettiGenolini:2023ndb}. We then
 review how this may be used to compute physical observables for
 solutions 
describing the near-horizon 
limit of M5-branes wrapped 
over a smooth Riemann surface.

\subsection{\texorpdfstring{AdS$_5$}{AdS(5)} solutions in M-theory and localization}

The 11d metric is assumed to take the warped product form 
\begin{equation}
    \dd s^2_{11}= \me^{2\lambda}(\dd s^2_{\mathrm{AdS}_5}+\dd s^2_{M})\, .
\end{equation}
Here the AdS$_5$ metric is normalized to have unit radius, with
the six-dimensional internal space $M$ assumed to be compact without boundary. 
In order to preserve the symmetries of AdS$_5$ the  warp
factor function $\lambda$ and M-theory four-form $G$ are taken to be pull-backs from~$M$. 

Supersymmetry requires the existence of a Dirac spinor, $\epsilon$, on 
$M$, satisfying a certain Killing spinor equation \cite{Gauntlett:2004zh}. From $\epsilon$, following the conventions of \cite{BenettiGenolini:2023ndb}, one constructs the following 
real bilinear differential forms:
\begin{align}\label{bildef}
 & y \equiv \frac{\ii}{2}\ex^{3\lambda}\bar\epsilon\gamma_7\epsilon\, , \qquad  
\xi^\flat \equiv \frac{1}{3}\bar\epsilon \gamma_{(1)}\gamma_7\epsilon\, , 
\qquad Y\equiv -\ii \bar\epsilon \gamma_{(2)}\epsilon\, , \qquad  Y' \equiv \bar\epsilon\gamma_{(2)}\gamma_7\epsilon\, .
\end{align}
Here $\gamma_7\equiv \gamma_{123456}$ and we have defined $\gamma_{(r)}\equiv \tfrac{1}{r!}\gamma_{\mu_1\cdots \mu_r}\diff x^{\mu_1}\wedge\cdots\wedge \diff x^{\mu_r}$. 
The Killing spinor equation implies that $\bar\epsilon\epsilon$ is a constant, 
which we normalize to 1, and the vector field 
$\xi$ dual to the one-form bilinear $\xi^\flat$ is Killing. 
The Killing spinor is charged under $\xi$, satisfying 
$\mathcal{L}_\xi\epsilon=\frac{\ii}{2}\epsilon$. 
The function $y$  was used as a canonical coordinate in \cite{Gauntlett:2004zh}, but together with the two-forms $Y$, $Y'$ also plays an important role in equivariant localization below.

We next introduce the equivariant exterior derivative
\begin{align}\label{dxi}\diff_\xi\equiv 
    \diff - \xi \hook\, ,
\end{align}
where $\diff$ is the usual exterior derivative acting on differential forms, and $\xi\hook$ denotes contraction with the vector field $\xi$. 
Reference \cite{BenettiGenolini:2023ndb}
showed that supersymmetry  implies the following polyforms are equivariantly closed under $\diff_\xi$:
\begin{align}
\label{equivariantforms}
\Phi^a  &\equiv \ex^{9\lambda} \vol + \frac{1}{12} \ex^{9\lambda} * Y - \frac{1}{36} y\mskip 2mu\ex^{6\lambda} Y - \frac{1}{162 } y^3 \, ,\nn \\
\Phi^G  &\equiv G - \frac{1}{3} \ex^{3\lambda} Y' + \frac{1}{9} y \,,\nn \\ \Phi^Y  &\equiv \ex^{6\lambda}Y + \frac{1}{3} y^2\, .
\end{align}
Here $\vol$ is the Riemannian volume form on $M$ and 
$*$ is the Hodge dual operator.
The Cartan formula implies $\diff_\xi^2=-\mathcal{L}_\xi$, 
and again supersymmetry implies that the polyforms \eqref{equivariantforms}, 
together with $\lambda$ and the four-form $G$, are 
invariant under $\mathcal{L}_\xi$. 
Integrals of these polyforms 
give the $a$ central charge, quantized four-form fluxes, and conformal dimensions of BPS operators dual to wrapped M2-branes:
\begin{align}\label{integrals}
   a & = \frac{1}{2(2\pi)^6\lp^9}\int_{M}\Phi^a\, ,\qquad   N_{D}=\frac{1}{(2\pi\lp)^3}\int_{D}\Phi^G\in\mathbb{Z}\,, \qquad  
    \Delta[\Sigma]  =-\frac{3}{(2\pi)^2\lp^3}\int_{\Sigma}\Phi^G\,.
\end{align}
Here $D\subset M$ is any four-cycle, while  $\Sigma\subset M$ 
wrapped by M2-branes needs to be a calibrated two-dimensional submanifold with 
$\vol_{\Sigma}= Y' |_{\Sigma}$ in order to preserve supersymmetry. The integrals 
in \eqref{integrals} mean 
that one integrates the 
appropriate degree part of the polyform. 
$\Phi^Y$ is related physically to an equivariantly closed polyform associated to $*G$, with 
\begin{align}
\Phi^{*G}\equiv 
\ex^{3\lambda}*G - \frac{1}{3}\ex^{6\lambda}=- 4\Phi^Y + 3\diff_\xi(\ex^{6\lambda}\xi^\flat)\, ,
\end{align} 
the latter following from supersymmetry.

Let $F\equiv\{\xi=0\}\subset M$ denote the fixed point set for $\xi$. 
One can show 
\cite{BenettiGenolini:2023ndb} 
that $\ex^{6\lambda} |_F = 4y^2 |_F$, together with
\begin{align}
\fleft\ex^{6\lambda} Y \fright = \omega\, , \qquad 
\fleft\ex^{3\lambda}Y' \fright = -\frac{1}{2y}\omega\, , \qquad \fleft
 \ex^{9\lambda}* Y \fright =-\frac{1}{4y}\omega\wedge\omega\, .
\end{align}
Here $\omega$ is a global two-form on $M$ which is closed when pulled back to $F$, so $\diff\omega |_F=0$. 
A global expression for $\omega$ may be found in \cite{BenettiGenolini:2023ndb}, but this will not be needed in what follows. We also note that $\omega$ restricted to $F$ is proportional to the K\"ahler form introduced in \cite{Gauntlett:2004zh}, and we hence refer to integrals $\int\omega$ as  ``K\"ahler classes'', but again we will not need any details of the K\"ahler geometry of \cite{Gauntlett:2004zh} in what follows. Similarly 
$\diff y |_F=0$ so that $y$ is constant over a connected component of $F$. 
We may  then evaluate the quantities in 
\eqref{integrals} using equivariant localization.   
We refer to 
\cite{BenettiGenolini:2023ndb} for further details, here just recording the final formulae:
\begin{align}\label{fixedstuff}
 a & = \frac{1}{2(2\pi)^6\lp^9}\Bigg\{\sum_{\Ftwo}\int_{\Ftwo}\Big[-\frac{(2\pi)^2}{\epsilon_1\epsilon_2}\frac{y}{36}\omega  +\frac{(2\pi)^3}{\epsilon_1\epsilon_2}\frac{y^3}{162}\left(\frac{c_1(\mathcal{L}_1)}{\epsilon_1}+ \frac{c_1(\mathcal{L}_2)}{\epsilon_2}\right)\Big] \nonumber \\ & \qquad \qquad \quad 
 - \sum_{\Fzero} \frac{1}{d_0}\frac{(2\pi)^3}{\epsilon_1\epsilon_2\epsilon_3}\frac{y^3}{162}\Bigg\}\, , \nonumber\\
 N_{D} & = \frac{1}{(2\pi\lp)^3}\Bigg\{  
\sum_{\Dtwo}\int_{\Dtwo}\Big[ \frac{2\pi}{\epsilon_1}\frac{1}{6y}\omega
-\left(\frac{2\pi}{\epsilon_1}\right)^2\frac{y}{9}c_1(\mathcal{L})\Big]
 +\sum_{\Dzero}\frac{1}{d_0}\frac{(2\pi)^2}{\epsilon_1\epsilon_2}\frac{y}{9}\Bigg\}\, ,\nonumber\\ 
 \Delta[\Sigma] & = 
  \frac{1}{(2\pi)^2\lp^3}\Bigg\{-\int_{\Sigma}\frac{1}{2y}\omega
-\sum_{\Sigmazero}\frac{1}{d_0}\frac{2\pi}{\epsilon_1}\frac{y}{3}\Bigg\}\, ,\nonumber\\
\int_{\Sigma}\Phi^Y & = \int_{\Sigma}\omega +\sum_{\Sigmazero} \frac{1}{d_0}\frac{2\pi}{\epsilon_1}\frac{y^2}{3}\, .
\end{align}
Since in this paper the 
fixed point set $F$ will 
have connected components $\Fzero$ of dimension zero (isolated fixed points, which we refer to as \emph{nuts}) and $\Ftwo$ of dimension two (fixed Riemann surfaces, which we refer to as \emph{bolts}), we have presented the above formulae only for these cases. 
In particular:
\begin{itemize}
\item A connected component of $\Ftwo$ (a Riemann surface) has normal bundle $\mathcal{L}_1\oplus\mathcal{L}_2$ a sum of two complex line bundles, with $c_1$ denoting first Chern class. 
\item $\Dzero, \Dtwo\subset D$ 
are fixed submanifolds inside $D$,
where note $\Dzero=D\cap \Fzero$, and $\Dtwo=D\cap \Ftwo$. 
 $\mathcal{L}$ is the 
normal bundle of $\Dtwo$ inside $D$. Note that since $G$ is closed the 
integer $N_{D}$ 
depends only on the homology class $[D]\in H_4(M,\Z)$.
\item $\Sigmazero\subset \Sigma$ are the fixed points inside $\Sigma$, 
so $\Sigmazero=\Sigma\cap\Fzero$, where note the first
term on the right hand side is only present (and by itself) when the entire $\Sigma$ is fixed, while the second term
is present (and by itself) when the fixed point set is $\Sigmazero\subset \Sigma$.
\end{itemize}
The $\epsilon_i$ 
are the weights of the vector field $\xi$ 
on the normal directions to the fixed point sets, while 
$d_0\in\mathbb{N}$ denote dimensions of orbifold groups (in the case where there are orbifold singularities at nuts). 
Notice that the quantities in \eqref{fixedstuff} depend on integrals 
of the two-form~$\omega$, together with $y$ which is locally constant on $F$.

In  \cite{BenettiGenolini:2023ndb} these observables have been computed with $M$ having various topologies.
In the next subsection we review the case in which  
$M$ is an $S^4$ bundle over a smooth Riemann surface. The aim of this work is to extend these results to include punctures on the Riemann surface, which modifies the $S^4$ bundle locally over a neighbourhood of a puncture. 


\subsection{M5-branes on a smooth Riemann surface}\label{sec:M5review}

We start with the set-up without punctures, where the general 
supergravity solutions were originally constructed in \cite{Bah:2012dg}, 
generalizing \cite{Maldacena:2000mw}. 
The localization is described 
in \cite{BenettiGenolini:2023ndb}, which we follow closely. 

Consider $N$ M5-branes wrapped on a smooth Riemann surface 
$\Sigma_g$ inside a local Calabi--Yau three-fold $X$.
The latter is the total space of the bundle 
$\mathcal{O}(-p_1)\oplus\mathcal{O}(-p_2)\rightarrow \Sigma_g$, where 
this is Calabi--Yau provided 
\begin{align}\label{CY3}
p_1+p_2= 2-2g = \chi(\Sigma_g)\, .
\end{align}
The latter is the Calabi--Yau condition $c_1(X)=0$. 
The near-horizon limit of the wrapped M5-branes is an $S^4$ bundle over 
$\Sigma_g$
\begin{align}
S^4\hookrightarrow M \rightarrow \Sigma_g\, . 
\end{align}
Here $S^4\subset \C_1\oplus\C_2\oplus\R$, where the two copies of $\C_i$ 
are twisted using the complex line bundles $\mathcal{O}(-p_i)$, respectively. 
The unit sphere may be realized as $S^4=\{|z_1|^2+|z_2|^2+t^2=1\}$, 
where $z_i$ are complex fibre coordinates on $\mathcal{O}(-p_i)$, and 
$t\in\R$ is a coordinate on $\R$. The poles of $S^4$ are then  
$N,S=\{z_1=z_2=0,t=\pm1\}$.
 Let $\partial_{\varphi_i}$ rotate the $\mathbb{C}_i$ above, so 
 $z_i=|z_i|\ex^{\ii\varphi_i}$, and assume that these are Killing vectors in the full solution. 
We take the R-symmetry Killing vector $\xi$ to be a general element in the Cartan subalgebra 
of $\mathrm{so}(5)_R$ associated to $S^4$
\begin{equation}
    \xi=\sum_{i=1}^{2} b_i \partial_{\varphi_i}\, ,
\end{equation}
where the coefficients $b_i\in\R$ are
subject to the constraint $b_1+b_2=1$. The latter follows from 
regularity of the Killing spinor $\epsilon$, which recall satisfies $\mathcal{L}_\xi\epsilon=\tfrac{\ii}{2}\epsilon$ \cite{BenettiGenolini:2023ndb}. 
Then, for $b_1 b_2\neq0$ as we assume throughout, the fixed point set consists of two disjoint copies of the Riemann surface $\Sigma_g$ (which are ``bolt'' fixed point sets), namely $F=\Ftwo=\Sigma_g^N\cup\Sigma_g^S$ 
 at the poles $N, S$ of the $S^4$ fibre.

To compute the observables in \eqref{fixedstuff} in this set-up, it is helpful to first note that there are two linearly embedded $S^2$'s inside the $S^4$, given by
$S^2_i\subset \C_i\oplus\R\subset \R^5$. These are invariant under the action of $\xi$, with fixed points at the poles $N,S$ above. Because 
 $H_2(S^4,\Z)=0$ 
 the homology classes are trivial, and using localization for $\Phi^Y$  we deduce
 \begin{equation}\label{eq:locY}
    0=\int_{S^2_i}\Phi^Y=\frac{2\pi}{3b_i}(y^2_N-y_S^2)\,.
\end{equation}
Here we have used the last equation in \eqref{fixedstuff}, with weights $\epsilon_1^N=b_i=-\epsilon_1^S$. 
 This implies that $|y_N|=|y_S|$. We may then quantize the flux through the $S^4$ at a point on the Riemann surface,
\begin{equation}\label{eq:S4flux}
    N\equiv \frac{1}{(2\pi\lp)^3}\int_{S^4}G=\frac{y_N-y_S}{18\pi\lp^3b_1b_2} = \frac{y_N}{9\pi\lp^3b_1b_2}\,.
\end{equation}
Here we have used the localization equation in \eqref{fixedstuff} with $\epsilon_1^N\epsilon_2^N=b_1b_2=-\epsilon_1^S\epsilon_2^S$. For $N\neq 0$ M5-brane flux through the $S^4$, in the last equality in \eqref{eq:S4flux} we have deduced that  necessarily $y_S=-y_N$, and then solve this equation for $y_N=9\pi \lp^3 b_1b_2 N$, where we take $y_N>0$. 

There are further four-cycles that are helpful to consider in order to continue the computation. 
Here we proceed slightly differently to the 
discussion in \cite{BenettiGenolini:2023ndb}, since what follows will generalize more straightforwardly later. We may introduce the following round metric on the $S^4$ fibres
\begin{align}
\diff s^2_{S^4} = 
 \diff\psi^2 + \sin^2\psi \left[\diff \theta^2 + \sin^2\theta (\diff\varphi_1 - A_1)^2
+\cos^2\theta (\diff\varphi_2-A_2)^2\right]\, .
\end{align}
Here $\psi\in[0,\pi]$ is a polar coordinate with $S^3$ slices for $\psi\in (0,\pi)$
with poles $N=\{\psi=0\}$,
$S=\{\psi=\pi\}$, 
while $\theta\in[0,\pi/2]$. 
The periodic angular coordinates $\varphi_i$ are fibred over the Riemann surface $\Sigma_{g}$ with 
connection forms $A_i$ on the line bundles $L_i\equiv\mathcal{O}(p_i)$, respectively, 
where as in 
\cite{BenettiGenolini:2023ndb} we choose an orientation convention such that 
the normal bundle is given by $\mathcal{N}(\Sigma_g^N\hookrightarrow M)=\mathcal{L}_1^N\oplus\mathcal{L}^N_2=\mathcal{O}(p_1)\oplus \mathcal{O}(p_2)$.\footnote{Instead at the south pole we have normal bundle $\mathcal{N}(\Sigma_g^S\hookrightarrow M)=\mathcal{L}_1^S\oplus\mathcal{L}^S_2=\mathcal{O}(\mp p_1)\oplus \mathcal{O}(\pm p_2)$, where 
the choice of upper/lower signs are a convention.
 Throughout the paper $L_i$ will have a fixed orientation, but the orientations of the the normal bundles $\mathcal{L}_i=\mathcal{O}(\pm p_i)$ depend on the pole and choices of conventions.}
By definition we then have 
\begin{align}\label{c1}
\int_{\Sigma_g}c_1(\mathcal{O}(p_i))=\int_{\Sigma_g} \frac{F_i}{2\pi} = p_i\, ,
\end{align}
where $F_i=\diff A_i$ is the curvature. One then verifies that the following four-form is globally well-defined on $M$, closed ($\diff G=0$), and has flux $N$ through the $S^4$, as in \eqref{eq:S4flux}:
\begin{align}\label{Gformula}
G & = 3\pi \lp^3 N \sin^3\psi\, \diff\psi\wedge\bigg[ \sin\theta\cos\theta\diff\theta\wedge (\diff\varphi_1-A_1)\wedge (\diff\varphi_2-A_2)\nonumber\\
& \qquad + \frac{1}{2}\cos^2\theta F_1 \wedge (\diff\varphi_2-A_2) +\frac{1}{2}\sin^2\theta F_2\wedge (\diff\varphi_1 -A_1)\bigg]\, .
\end{align}
The four-cycles of interest are $D_{[i]}$, which are defined to be 
the submanifolds where $\del_{\varphi_i}$ vanish, for $i=1,2$, corresponding to the total spaces of the $S^2_{j\neq i}$ bundles over $\Sigma_g$.\footnote{Note that we change convention compared to \cite{BenettiGenolini:2023ndb, Couzens:2024vbn}. There the cycle was labelled by the $S^2_i$ fibre, such that  $D_{[1]}^\mathrm{here}\equiv C_4^{(2)\mathrm{there}}$ and similarly the labels on $N^{[i]}$ are exchanged.\label{footnote:4cycle}}
These are located at $\theta=0, \,\theta=\pi/2$, respectively.
By direct computation
from \eqref{Gformula}:
\begin{align}
N^{[i]}\equiv\frac{1}{(2\pi \lp)^3}\int_{D_{[i]}} G = N \int_{\Sigma_g} \frac{F_i}{2\pi} = p_iN\, .
\end{align}
Here in the first equality we have integrated over $\psi$ and $\varphi_{j\neq i}$. 
On the other hand, the same integral may be computed using the localization formula in \eqref{fixedstuff}:
\begin{equation}  \label{localizeG} p_1N=\frac{1}{(2\pi\lp)^3}\int_{D_{[1]}}G=\frac{2}{(2\pi\lp)^3}\frac{2\pi}{b_2}\bigg[\int_{\Sigma_{g}^N}\frac{\omega}{6y_N}-\frac{2\pi p_2}{b_2}\frac{y_N}{9}\bigg]\,,
\end{equation}
and similarly for $1\leftrightarrow 2$.
Notice that on the right hand side the fixed point set $\Dtwo_{[i]}\subset D_{[i]}$ is $\Dtwo_{[i]}=\Sigma_g^N \cup \Sigma_g^S$, with the normal bundles being 
$\mathcal{N}(\Sigma_g^N\hookrightarrow D_{[i]})=\mathcal{L}^N=\mathcal{O}(p_{j\neq i})$, and $\mathcal{N}(\Sigma_g^S\hookrightarrow D_{[i]})=\mathcal{L}^S=\mathcal{O}(-p_{j\neq i})$, weights $\epsilon_i^N=b_{j\neq i}=-\epsilon_i^S$, and 
we have used $\int_{\Sigma_g^N}\omega=\int_{\Sigma_g^S}\omega$, which follows since 
$[\Sigma_g^N]=[\Sigma_g^S]\in H_2(M,\Z)$. The $\Sigma_g^N$ and $\Sigma_g^S$ contributions are hence equal, giving an overall factor of 2 in \eqref{localizeG}. 
Combining the formulae derived so far implies that
\begin{equation}\label{eq:Jbulk}
    \int_{\Sigma_{g}^N} \omega=108\pi^3\lp^6b_1b_2(b_1p_2+b_2p_1)N^2\,.
\end{equation}

We now have all the ingredients needed to compute the central charge $a$ using the localization formula \eqref{fixedstuff}. 
The fixed point set is $F=\Ftwo=\Sigma_g^N \cup \Sigma_g^S$, 
and as above the north and south poles give an equal contribution, so that with this factor of 2 we have
\begin{align}\label{eq:abulk1}
 a & = \frac{2}{2(2\pi)^6\lp^9}\int_{\Sigma_g^N}\Big[-\frac{(2\pi)^2}{b_1b_2}\frac{y_N^{}}{36}\omega  +\frac{(2\pi)^3}{b_1b_2}\frac{y_N^3}{162}\left(\frac{p_1}{b_1}+ \frac{p_2}{b_2}\right)\Big] \nonumber\\
 & = -\frac{9}{8}b_1b_2(b_1p_2+b_2p_1)N^3\,.
\end{align}
This is an off-shell result that requires extremization over the weights $b_i$ subject to the constraint $b_1+b_2=1$.

The special case where one of the $p_i$ vanishes, say $p_2=0$, leads to a 
 $d=4$ SCFT with $\mathcal{N}=2$ supersymmetry. 
In this case the Calabi--Yau geometry is $X=T^*\Sigma_g\times \C_2$, 
where the fibre of $T^*\Sigma_g=\mathcal{O}(-2+2g)$ is rotated by SO(2)$_R\subset \mathrm{SO}(2)_R\times \mathrm{SO}(3)_R\subset \mathrm{SO}(5)_R$. 
The total normal bundle of the M5-branes has an additional $\R$ direction (with coordinate $t\in\R$), so that the fibre is $\C_1\oplus \R^3$ 
 where 
the $\R^3=\C_2\oplus\R$ factor is rotated by SO$(3)_R$ and this symmetry is later
left unbroken by the addition of 
$\mathcal{N}=2$ punctures.
Note that the special case $p_2=0$ fixes $b_1=1/3$, $b_2=2/3$ and $p_1=2(1-g)$, such that 
\begin{equation}\label{eq:aN2}
    a=\frac{1}{3}(g-1)N^3\,,
\end{equation}
the central charge of the $d=4$, $\mathcal{N}=2$ SCFT obtained by compactifying the 6d $\mathcal{N}=(2,0)$ theory on a smooth Riemann surface.


\section{Getting orbifoldy}\label{sec:orbifoldsection}

A richer set of theories may be constructed by wrapping M5-branes on a punctured Riemann surface.
We may add punctures into the wrapped M5-brane configuration just described by introducing conical defects on the Riemann surface, and then further (partially) resolving these orbifold singularities. In this section 
we introduce a general class of orbifold singularities preserving $\mathcal{N}=1$ supersymmetry, before then studying the associated puncture geometries in the remainder of the paper. Some useful 
references for the material in this section are \cite{Closset:2018ghr, Ferrero:2021etw}.

\subsection{Orbifold Riemann surfaces}\label{sec:orbifolds}

We may replace the smooth Riemann surface $\Sigma_g$ that the M5-branes wrap by a general two-dimensional orbifold $\Sigma_{g,n}$. Topologically this is a 
closed genus $g$ Riemann surface $\Sigma_g$ together with 
$n$ marked points $\pI\in\Sigma_g$, $I=1,\ldots,n$. 
A neighbourhood of $\pI$ is  modelled on the quotient space 
$\C/\Z_{\KI}$, with $\KI\in \mathbb{N}$. Introducing a local complex coordinate $\wI\in\C$ 
in a neighbourhood of $\pI$, with the point $\pI=\{\wI=0\}$, the quotient means that we identify points under the map 
$\wI \mapsto\ex^{2\pi \ii /\KI} \wI$. 
 Equivalently, 
$\wI= |\wI|\ex^{\ii \wvarphiI}$ where after taking the quotient the angular coordinate
$\wvarphiI$ is identified with period 
$\Delta \wvarphiI=2\pi/\KI$. 
When $\KI=1$ the marked point $\pI$ is a smooth point.

The orbifold Euler characteristic of $\Sigma_{g,n}$ is 
\begin{align}\label{chiorb}
\chi(\Sigma_{g,n})&= 
\chi(\Sigma_g) - \sum_{I=1}^n \Big(1-\frac{1}{\KI}\Big)\, ,
\end{align}
where $\chi(\Sigma_g)=2-2g$ 
is the Euler characteristic of the underlying 
smooth Riemann surface. 
Notice that  since 
$\Delta \wvarphiI=2\pi/\KI$,
the conical deficit angle 
at $\pI$ is 
$2\pi (1-1/\KI)$, 
and the right hand side 
of \eqref{chiorb} 
is what one obtains from the Gauss--Bonnet formula by integrating $R/4\pi$ 
over $\Sigma_{g,n}$, 
where $R$ is the Ricci scalar of any
orbifold metric on $\Sigma_{g,n}$. 

Recall that the smooth Riemann surface $\Sigma_g$
was embedded inside a local Calabi--Yau three-fold geometry. When introducing 
the orbifold points 
$\pI$ as a quotient by 
$\Z_{\KI}$,
we need to specify how this group
acts in the full geometry, preserving the Calabi--Yau condition and hence $\mathcal{N}=1$ supersymmetry for the wrapped M5-branes. 
Local complex coordinates for the Calabi--Yau
near to $\pI$ are 
$(\wI,z_1,z_2)$, 
where recall $z_i$, $i=1,2$, 
are complex fibre coordinates. 
We may then identify points under
the $\Z_{\KI}$ action
\begin{align}\label{wIlift}
(\wI,z_1,z_2)\mapsto(\ex^{2\pi \ii /\KI} \wI, \ex^{2\pi \ii{\alphaI_1} /\KI} z_1,\ex^{2\pi \ii{\alphaI_2} /\KI} z_2)\, ,
\end{align}
where $\alphaI_i\in\Z$, 
$i=1,2$. 
Notice here that $\alphaI_i$
and $\alphaI_i + m_i^I\KI$ 
give the same action, for any $m_i^I\in\Z$, so really $\alphaI_i\in\Z_{\KI}$; but 
as explained below, globally we pick 
particular integer lifts for each $\pI$
and regard $\alphaI_i\in\Z$. 
The action \eqref{wIlift} preserves the holomorphic volume form 
$\Omega_{(3,0)}=\diff \wI\wedge \diff z_1\wedge \diff z_2$, so $\Z_{\KI}\subset \mathrm{SU}(3)$ with $\C^3/\Z_{\KI}$ then being Calabi--Yau, if and only if 
$1+\alphaI_1+\alphaI_2=0$ mod 
$\KI$. In the special case that $\alphaI_2=0$ (or $\alphaI_1=0$)
the action \eqref{wIlift} has $\Z_{\KI}\subset SU(2)$
and (locally) preserves $\mathcal{N}=2$ supersymmetry. In this case the geometry is locally $T^*\Sigma_{g,n}\times \C_2$, and $\alphaI_1=-1$ 
gives the natural lift of the $\Z_{\KI}$ action on the Riemann surface to its cotangent bundle. 

Globally the Calabi--Yau 
$\mathcal{O}(-p_1^\mathrm{bulk})\oplus 
\mathcal{O}(-p_2^\mathrm{bulk})\rightarrow \Sigma_g$ 
is now replaced by a sum of two orbifold 
complex line bundles 
$L_1^{-1}\oplus L_2^{-1}\rightarrow \Sigma_{g,n}$, 
where the curvature forms $F_i= \diff A_i$ for $L_i$ introduced in section \ref{sec:M5review} may be written
\begin{align}\label{fluxorb}
\frac{F_i}{2\pi} = p_i^\mathrm{bulk} \vol_{\Sigma_{g,n}} - 
\sum_{I=1}^n \frac{\alphaI_i}{\KI}\delta_{\pI}\,.
\end{align}
Here $\vol_{\Sigma_{g,n}}$ 
is any two-form that integrates to 1 
over $\Sigma_{g,n}$, while 
$\delta_{\pI}$ is 
a two-form current with a Dirac delta-function supported 
at the point $\pI$. 
Similarly to \eqref{c1}, it follows that  
\begin{align}\label{orbi}
p_i^\mathrm{tot} = 
\int_{\Sigma_{g,n}} 
c_1(L_i) = 
\int_{\Sigma_{g,n}}
\frac{F_i}{2\pi} = 
p_i^\mathrm{bulk} - \sum_{I=1}^n 
\frac{\alphaI_i}{\KI}\in\mathbb{Q}\, ,
\end{align}
with $p_i^\mathrm{bulk}\in\Z$ being the 
``bulk'' first Chern class (mathematically called the \emph{degree} of $L_i$). 
We  may correspondingly split the Riemann surface as
\begin{align}\label{splitSigma}
\Sigma_{g,n} = \Sigmabulk \bigcup_{I=1}^n (-\Sigma_{\epsilon}^I)\, .
\end{align}
Here $\Sigmabulk$ 
is the smooth Riemann surface with boundary
obtained from $\Sigma_{g,n}$ 
by removing 
 small  neighbourhoods $\Sigma^I_\epsilon$ around 
 each orbifold point $x^I$. 
The latter may be written locally as 
$\Sigma_{\epsilon}^I\equiv \{|\wI|\leq \epsilon\}$, for some small $\epsilon>0$. 
In \eqref{splitSigma} the $\Sigma^I_\epsilon$ 
may be glued back on to 
 the bulk 
$\Sigmabulk$ along their common $S^1$ boundaries, 
after reversing orientations which give rise to the minus sign above
so that the common orientations agree.
We may then 
write \eqref{orbi} as (assuming the support for 
$\vol_{\Sigma_{g,n}}$ lies inside $\Sigmabulk$)
\begin{align}\label{bobthebuilder}
    p_i^\mathrm{tot} = \int_{\Sigmabulk}c_1(L_i) -\sum_{I=1}^n \int_{\Sigma^I_\epsilon} c_1(L_i) = p_i^\mathrm{bulk} - \sum_{I=1}^n\frac{\alphaI_i}{\KI}\, ,
\end{align}
and in particular we have the discrete fluxes 
\begin{align}\label{intc1Li}
\frac{\alphaI_i}{\KI}    = \int_{\Sigma_{\epsilon}^I} c_1(L_i) = 
\int_{\partial \Sigma_{\epsilon}^I} \frac{A_i}{2\pi}\, .
\end{align}
Here $\partial 
\Sigma_{\epsilon}^I$
is a small circle in the Riemann surface around the point $\pI$, and we have 
written $F_i=\diff A_i$ 
locally near to $\pI$
and used Stokes' theorem. 
Indeed, the Dirac delta-function 
in \eqref{fluxorb} 
is obtained by simply 
writing $A_i=\frac{\alphaI_i}{\KI} \diff \hat\wvarphi^I$ where 
$\hat\wvarphi^I$ is a 
$2\pi$-periodic 
coordinate on the angular direction in 
$\Sigma_{\epsilon}^I$ (so $\wvarphiI=\hat\wvarphi^I/\KI$). 

Notice 
that we may shift 
\begin{align}\label{banishtothebulk}
\alphaI_i\mapsto \alphaI_i + \mI_i \KI\, , \quad 
p_i^\mathrm{bulk} \mapsto p_i^\mathrm{bulk}+\sum_{I=1}^n \mI_i\, ,
\end{align}
for any choice of $\mI_i\in\Z$, leaving 
the total orbifold first Chern number  $p_i^\mathrm{tot}$ in \eqref{orbi} (and hence isomorphism class of $L_i$) invariant. 
As commented earlier, 
the shift \eqref{banishtothebulk} leaves the local 
$\C^3/\Z_{\KI}$ 
spaces invariant, 
shifting integer parts of the ``discrete fluxes'' 
$\alphaI_i/\KI$ into 
the ``bulk flux'' $p_i^\mathrm{bulk}\in\Z$. 
These are simply large gauge transformations of the $A_i$ in each neighbourhood $\Sigma^I_\epsilon$, 
or equivalently  different choices of trivialization of the line bundle $L_i$ 
over each $\partial \Sigma_{\epsilon}^I$.  
Note that intermediate steps for calculations can depend on this choice of gauge, as we shall illustrate in the next subsection, but necessarily this is not the case for final (global) formulae,
 which must be invariant under \eqref{banishtothebulk}.

Finally, the Calabi--Yau condition \eqref{CY3} now reads
\begin{align}\label{CY3orb}
p_1^\mathrm{tot}+p_2^\mathrm{tot} 
= \chi(\Sigma_{g,n}) &= 
2-2g - \sum_{I=1}^n 
\Big(1-\frac{1}{\KI}\Big)\, . 
\end{align}
A canonical way to 
impose this is to 
fix $p_1^\mathrm{bulk}+p_2^\mathrm{bulk}=2-2g$, 
so that the bulk 
satisfies the same 
condition as for the smooth Riemann surface 
\eqref{CY3}, and then 
for each $I$ also fix 
\begin{equation}
  \alphaI_1+\alphaI_2=\KI-1\,.  
\end{equation}
In particular for an $\mathcal{N}=2$ orbifold singularity $\alphaI_2=0$ (or $\alphaI_1=0$), which fixes 
$\alphaI_1=\KI-1$ (or $\alphaI_2=\KI-1$), such that
\begin{equation}\label{N2twists}
    \int_{\Sigma_\epsilon^I} c_1(L_1)=1-\frac{1}{K^I}\,, \quad  
    \int_{\Sigma_\epsilon^I} c_1(L_2)=0\,.
\end{equation}

\subsection{M5-branes 
on an orbifold Riemann surface}\label{sec:orbi2}

It is straightforward to 
run through the analysis of section 
\ref{sec:M5review}, 
where the M5-branes 
are now assumed to wrap the orbifold Riemann surface zero-section of $X=L_1^{-1}\oplus L_2^{-1}\rightarrow \Sigma_{g,n}$, 
with near-horizon 
limit being the corresponding $S^4$
 orbibundle over $\Sigma_{g,n}$. 

Firstly, we note that the fibre over the orbifold point $\pI=\{\wI=0\}\in \Sigma_{g,n}$ is  $S^4/\Z_{\KI}$, with the 
$\Z_{\KI}$ action 
on $\C_1\oplus\C_2\subset \R^5\supset S^4$ given by \eqref{wIlift}. 
The flux through this $S^4/\Z_{\KI}$ is correspondingly 
$N/\KI$. 
Next, we note that the expression 
\eqref{Gformula} 
still holds for a representative of the cohomology class of the four-form $G$ on $M$, with 
$F_i=\diff A_i$ now 
replaced by \eqref{fluxorb}.
We may then define $D^{[i]}$ 
to be the total space of the $S^2_{j\neq i}$ orbibundle over
$\Sigma_{g,n}$, with $D^{[i]}_{\epsilon,\,I}$ 
being the $S^2_{j\neq i}$ bundle 
over the neighourhood 
$\Sigma_{\epsilon}^I$. 
Then
\begin{align}\label{Gfluxorb}
    N^{[i]}\equiv\frac{1}{(2\pi\lp)^3}\int_{D^{[i]}}G = 
    N\int_{\Sigma_{g,n}}\frac{F_i}{2\pi} 
    = p_i^\mathrm{tot} N \, ,
    \end{align}
where $ p_i^\mathrm{tot}$ decomposes as in \eqref{bobthebuilder}. In particular this means that  the
four-form $G$ 
\emph{necessarily} has a 
Dirac delta-function
contribution localized 
near each orbifold point, with flux  
\begin{align}\label{GfluxDirac}
N^{[i]}_{\epsilon,\,I}\equiv\frac{1}{(2\pi\lp)^3} \int_{D^{[i]}_{\epsilon,\,I}} 
G = 
\frac{\alphaI_i}{\KI}N\,.
\end{align}
Note that the flux \eqref{GfluxDirac} 
is \emph{not} invariant under the large gauge transformations \eqref{banishtothebulk}, but rather shifts by 
the integer $m_i^I N$. 
On the other hand $D^{[i]}_{\epsilon,\,I}$ is a four-cycle with boundary, and physically this integral need not be gauge-invariant.
Recall that specifying 
$\alphaI_i\in\Z$ 
(rather than $\alphaI_i\in\Z_{\KI}$) specifies not only the 
$\C^3/\Z_{\KI}$ local geometry, but also the large gauge transformation that it is glued in with. Instead, fluxes on the total space such as \eqref{Gfluxorb} are gauge-invariant, as expected.
 
Finally the upshot of the computation is that the 
central charge now reads
\begin{align}\label{eq:aorbtot}
a = -\frac{9}{8}b_1b_2(b_1 p_2^\mathrm{tot}+b_2 p_1^\mathrm{tot})N^3\,, 
\end{align}
as in \eqref{eq:abulk} but with orbifold Chern numbers $p_i^\mathrm{tot}$ given by \eqref{orbi}, satisfying the constraint \eqref{CY3orb} and $N$ being the flux $G/(2\pi \lp)^3$ through a copy of 
$S^4$ over a generic smooth point in $\Sigma_{g,n}$.
Therefore it can be decomposed as 
\begin{equation}\label{eq:asplit}
    a=\abulk-\sum_{I=1}^n\delta a^I\,,
\end{equation}
with 
\begin{equation}\label{eq:abulk}
    \abulk=-\frac{9}{8}b_1b_2(b_1p_2^\mathrm{bulk}+b_2p_1^\mathrm{bulk})N^3\,,
\end{equation}
the central charge of a smooth Riemann surface computed in \eqref{eq:abulk1}, and
\begin{equation}\label{eq:aorb}
    \delta a^I=-\frac{9}{8}b_1b_2\Big(b_1\frac{\alpha_2^I}{K^I}+b_2\frac{\alpha_1^I}{K^I}\Big)N^3\,.
\end{equation}
Notice here that as in 
\eqref{splitSigma}  we have separated the contributions from the bulk and the orbifold points, where the latter are subtracted due to the change in orientation. Notice also that we do not consider any boundary contributions since they cancel when gluing.

As a quick application of \eqref{eq:aorbtot}, let us derive the central charge for $N$ M5-branes wrapped on a  spindle \cite{Ferrero:2021wvk, BenettiGenolini:2023ndb}, with the equivariant parameter mixing the R-symmetry with the spindle isometry set to vanish. This leaves a sphere with two punctures, one at each pole. 
At the north and south pole punctures the orbifold acts with $\mathbb{Z}_{n_+}$ and $\mathbb{Z}_{n_-}$ action as in \eqref{wIlift} respectively. Then from \eqref{eq:aorbtot} we have
\begin{equation}
    a=-\frac{9}{8}b_1 b_2 \Big[b_1\Big(p_2^\mathrm{bulk}-\frac{\alpha_2^{+}}{n_+}-\frac{\alpha_2^{-}}{n_-}\Big)+b_2\Big(p_1^\mathrm{bulk}-\frac{\alpha_1^{+}}{n_+}-\frac{\alpha_1^{-}}{n_-}\Big)\Big]N^3\,,
\end{equation}
where $p_1^\mathrm{bulk}+p_2^\mathrm{bulk}=2$ and $\alpha_1^{\pm}+\alpha_{2}^{\pm}=n_{\pm}-1$.
To compare with \cite{BenettiGenolini:2023ndb} we identify\footnote{Note that the quantization of the $p_{i}^{\text{there}}$ was chosen such that they are integer which accounts for the appearance of $n_+ n_-$.}
\begin{equation}
    \frac{p_i^{\text{there}}}{n_+ n_-}=p_i^\mathrm{bulk}-\frac{\alpha_i^{+}}{n_+}-\frac{\alpha_i^{-}}{n_-}\, ,
\end{equation}
which satisfies $p_1^{\text{there}}+p_2^{\text{there}}=n_++n_-$ as required. We find exact agreement upon setting the spindle equivariant parameter to vanish. The conical singularity of the spindle therefore has an interpretation as an $\mathcal{N}=1$ puncture \cite{Bomans:2024mrf}.

In this section we studied the full orbifold Riemann surface geometry and noticed that
the formulae, such as the central charge in \eqref{eq:asplit}, consist of a ``bulk part'', which is 
the result for a smooth Riemann surface with $p_1^\mathrm{bulk}+p_2^\mathrm{bulk}=2-2g$, 
together with localized 
contributions from a small neighbourhood of each 
orbifold singularity.
Therefore these results can equally be obtained by studying the local contributions around an orbifold point and subtracting them from the smooth Riemann surface results. This local analysis can be performed using equivariant localization and is more suited to obtain the contributions from a puncture rather than a simple orbifold point.
Since equivariant localization will be entirely local at each such orbifold/puncture point, we henceforth drop the label $I$ and focus on the contribution of a single puncture. 


\section{\texorpdfstring{$\mathcal{N}=2$}{N=2} punctures}\label{sec:N=2}

In this section we analyse $\mathcal{N}=2$ 
punctures. The local orbifold action \eqref{wIlift} is taken to be
\begin{align}\label{Ak}
(w,z_1,z_2)\mapsto (\ex^{2\pi \ii/\K} w, \ex^{2\pi \ii(K-1)/\K} z_1,z_2)\, .
\end{align}
Note here we have chosen an embedding $\Z_\K\subset \mathrm{SU}(2)\subset \mathrm{SU}(3)$ into a particular SU$(2)$ subgroup of 
SU$(3)$, and that  
with this choice the SO$(2)_R\cong\mathrm{U}(1)_R$
of the $\mathcal{N}=2$ R-symmetry rotates $\mathbb{C}_1$, 
with generator $\partial_{\varphi_1}$, and the two-sphere $S^2_R$ rotated by SO$(3)_R$ is embedded as $S^2_R=S^2_2\subset \mathbb{C}_2\oplus \mathbb{R}$.
If there is a single orbifold point/puncture, we may always choose conventions so that the $\mathcal{N}=2$
quotient takes the form \eqref{Ak},  but we will later comment on choosing different 
embeddings
at different points (which 
locally preserve $\mathcal{N}=2$, but 
globally only $\mathcal{N}=1$ supersymmetry). 

The $\Z_K$ action on the 
coordinates $(w,z_1)\in\C^2$ 
in \eqref{Ak} by definition gives 
an $A_{\K-1}$ singularity $\C^2/\Z_\K$ on taking the quotient. 
This is naturally the cotangent space of $\C/\Z_\K$, where $\C/\Z_\K$ (with coordinate $w$) is the 
space locally wrapped by the $N$ M5-branes. 
We may then consider 
(partially) resolving the $A_{\K-1}$ singularity inside the full geometry, which we refer to 
as a local \emph{puncture geometry}. Physically the presence of an $A_{\K-1}$ singularity leads to an SU$(\K)$ flavour symmetry in the dual field theory, and we can understand a (partial) resolution as a (partial) nilpotent Higgsing of this flavour symmetry.  

In the remainder of this section we describe this local puncture 
geometry,  and then analyse flux quantization and compute the local 
contribution to the $a$ central charge  using equivariant localization.
A key point is that our analysis is purely topological, and does not require any explicit form of (locally) $\mathcal{N}=2$ M5-brane solutions. It will
therefore extend to $\mathcal{N}=1$ punctures.

\subsection{Geometry of \texorpdfstring{$A_{\K-1}$}{A(K-1)} singularities}\label{sec:N=2theory}

We begin by describing the 
 toric geometry of $\mathbb{C}^2/\mathbb{Z}_\K$. Recall that 
 we write the 
 covering space coordinates $(w,z_1)\in \mathbb{C}^2$  as  $w=|w|\ex^{\ii\wvarphi}$, $z_1=|z_1|\ex^{\ii\varphi_1}$. 
The quotient space admits a 
U$(1)^2$ action with basis of generating  vector fields
\begin{align}
e_1 \equiv\partial_{\varphi_1}\, , \qquad e_2\equiv \frac{1}{\K}(\partial_{\wvarphi}-\partial_{\varphi_1})\, .
\end{align}
These are normalized so that exponentiating by $2\pi$ generates an effective torus action, and the 
$(2,0)$-form $\Omega_{(2,0)}\equiv \diff w\wedge \diff z_1$ 
has charge 1 under $e_1$, and is uncharged under $e_2$ -- this is the canonical 
basis for toric Calabi--Yau geometries. 
Notice that the 
U$(1)_R$ generator is precisely $e_1=\partial_{\varphi_1}$. 

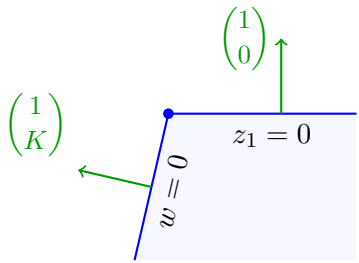
\begin{figure}
\centering
\begin{tikzpicture}

\coordinate (B) at (6.5,5.0);     
\coordinate (A) at (6.05,3.05);   
\coordinate (C) at (9.0,5.0);     

\fill[blue, opacity=0.035]
  (A) -- (B) -- (C) -- (9.0,3.05) -- (6.05,3.05) -- cycle;

\draw[thick, blue] (A) -- (B)
  node[pos=0.50, below, sloped, black] {$w=0$};
\draw[thick, blue] (B) -- (C)
  node[pos=0.55, below, black] {$z_1=0$};

\fill[blue] (B) circle(2pt); 

\draw[->, thick, green!60!black] (8,5) -- (8,6) node[anchor=east] {\small $\begin{pmatrix}1\\0\end{pmatrix}$};

\draw[->, thick, green!60!black] (6.275,4.025) -- ++(-0.975,0.225)
  node[above left] {\small $\begin{pmatrix}1\\\K\end{pmatrix}$};

\end{tikzpicture}
\caption{Toric diagram for $\C^2/\Z_\K$.}
\label{figjam}
\end{figure}

The toric diagram for $\C^2/\Z_\K$ is  shown in figure~\ref{figjam}. 
The space fibres over this 
wedge region, with generic 
fibre over an interior 
point being U$(1)^2$. 
The subspaces $\{z_1=0\}$,
$\{w=0\}$ are (non-compact) toric divisors $\C/\Z_\K$, fixed by
the U$(1)\subset\mathrm{U}(1)^2$ subgroups generated by
$\partial_{\varphi_1}=(1,0)$ and 
$\partial_{\wvarphi}=(1,K)$, respectively. 
These meet at the $\Z_\K$ orbifold singularity 
at the origin, shown as a blue dot, which is 
fixed under 
the entire U$(1)^2$.

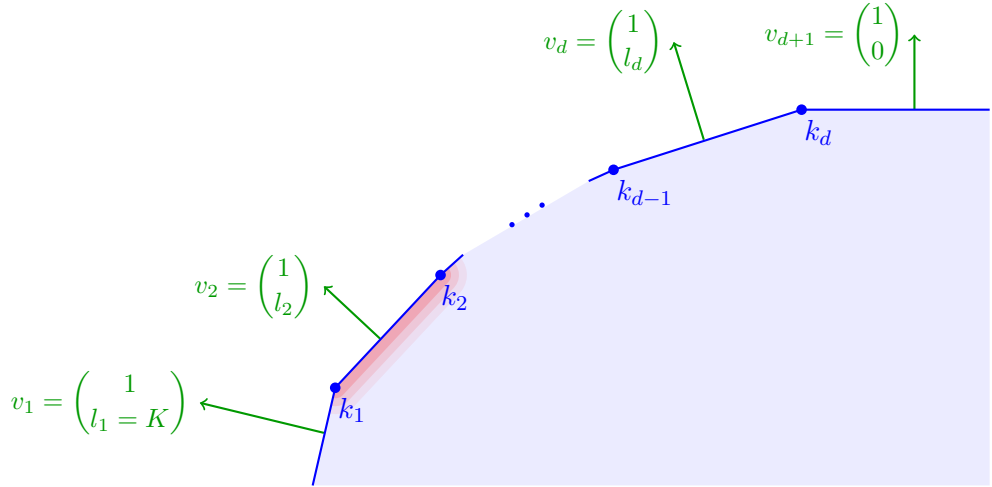
\begin{figure}
\centering
\begin{tikzpicture}

\draw[->, thick, green!60!black] (8,5) -- (8,6.)
  node[anchor=east] {\small $v_{d+1}=\begin{pmatrix}1\\0 \end{pmatrix}$};
\draw[->, thick, green!60!black] (5.2,4.6) -- (4.8,5.9)
  node[anchor=east] {\small $v_d=\begin{pmatrix}1\\l_d\end{pmatrix}$};
\draw[->, thick, green!60!black] (0.9,1.95) -- ++(-0.75,0.7)
  node[anchor=east] {\small $v_2=\begin{pmatrix}1\\l_2 \end{pmatrix}$};
\draw[->, thick, green!60!black] (0.15,.7) -- (-1.5,1.1)
  node[anchor=east] {\small $v_1=\begin{pmatrix}1\\l_1=\K \end{pmatrix}$};

\fill[blue, opacity=0.08]
  (0,0) -- (.3,1.3) -- (1.7,2.8) -- (2.,3.07) --
  (3.67,4.05) -- (4.,4.2) -- (6.5,5.) -- (9,5.) -- (9,0) -- cycle;

\begin{scope}
  \clip (0,0) -- (.3,1.3) -- (1.7,2.8) -- (2.,3.07) --
        (3.67,4.05) -- (4.,4.2) -- (6.5,5.) -- (9,5.) -- (9,0) -- cycle;
  \foreach \w/\op in {20/0.06, 14/0.10, 8/0.16}{
    \draw[red, line cap=round, line join=round, line width=\w, opacity=\op]
      (.3,1.3) -- (1.7,2.8);
  }
\end{scope}

\draw[thick, blue]  (0,0) -- (.3,1.3) -- (1.7,2.8) -- (2.,3.07);
\draw[thick, blue] (3.67,4.05) -- (4.,4.2) -- (6.5,5.) -- (9,5.);

\fill[blue] (.3,1.3) circle(2pt) node[below] {$\quad\, k_1$};
\fill[blue] (1.7,2.8) circle(2pt) node[below] {$\quad k_2$};
\fill[blue] (4,4.2) circle(2pt) node[below] {$\quad\quad\, k_{d-1}$};
\fill[blue] (6.5,5) circle(2pt) node[below] {$\quad\, k_d$};

\fill[blue] (2.65,3.47) circle(1pt);
\fill[blue] (2.85,3.6) circle(1pt);
\fill[blue] (3.05,3.73) circle(1pt);

\end{tikzpicture}
\caption{Partial resolution of the $\C^2/\Z_\K$ singularity. To each compact face (blue line between two fixed points) one can associate a compact divisor. The Poincar\'e dual to these divisors (discussed further in appendix~\ref{app:explicitG}) has support in a small neighbourhood of the blue line. For the divisor $\D_2$ we have plotted the support in red for illustration.}
\label{figpartial}
\end{figure}

With this toric description of the $A_{\K-1}$ singularity, we may next consider its (partial) resolutions by performing blow-ups. Observe that the vectors $(1,0)$ and $(1,\K)$ do not span $\mathbb{Z}^2$ over $\mathbb{Z}$, exemplifying the $\mathbb{Z}_\K$ singularity at the origin. 
To resolve we introduce additional vectors which subdivide the toric fan. For any choice of 
$d\in\mathbb{N}$ with $1\leq d\leq \K$, we  
choose $k_a\in\mathbb{N}$, $a=1,\ldots,d$,  
satisfying  
$\sum_{a=1}^d k_a = \K$, and further define
\begin{align}  
l_{a}\equiv\sum_{b=a}^{d}k_b\, ,
\end{align}
with $l_{d+1}\equiv 0$. 
Notice that the sequence $l_a$ is strictly decreasing, with $l_1=\K$ and $l_d=k_d$. 
We then consider the partial resolution with normal vectors
\begin{equation}
    v_1\equiv(1,l_1)\, ,\quad \ldots \, , \quad v_a\equiv (1,l_a)\, ,\quad \ldots\, , \quad v_d\equiv (1,l_d)\, ,\quad v_{d+1}\equiv (1,0)\, .
\end{equation}
The toric diagram is shown in figure~\ref{figpartial}. 
Each of the $d$ blue dots is a 
$\mathbb{C}^2/\mathbb{Z}_{k_a}$ singularity, 
where a full resolution then necessarily has
$d=\K$ with $k_a=1$ for all $a=1,\ldots,\K$. 
Each of the $d-1$ finite blue line segments is the image 
of a compact toric divisor $\D_a$, fixed by the U$(1)$ subgroup generated by 
$v_a=(1,l_a)$, $a=2,\ldots,d$, and is topologically 
a weighted projective space (or spindle) 
$\D_a\cong\mathbb{WCP}^1_{[k_a,k_{a-1}]}$. 
The non-compact toric divisors 
are now $\D_1\cong \C/\Z_{k_1}$ and $\D_{d+1}\cong \C/\Z_{k_d}$, which thus 
in general still have orbifold singularities. 
Note that the original $\C^2/\Z_\K$ 
singularity has $d=1$, $k_1=\K$, with $\D_1=\{w=0\}$, $\D_{d+1}=\{z_1=0\}$. 

Finally, recall that the U$(1)_R$ generator 
is $e_1=\partial_{\varphi_1}$. In the next subsection we will want to analyse the fixed points of this action in the partially resolved puncture geometry. The non-compact toric divisor 
$\D_{d+1}$ (the horizontal blue line segment in figure~\ref{figpartial}) is a fixed bolt 
for $e_1$, while the remaining fixed points are the blue dots
$a=1,\ldots,d-1$, which are \emph{nuts}. 
The weights of U$(1)_R$ 
at the $a$'th nut, which recall is locally $\C^2/\Z_{k_a}$,  are then given by the
toric geometry formulae
\begin{align}\label{4dRweights}
(\epsilon_1^a,\epsilon_2^a)= 
\left(\frac{\det (e_1,v_{a+1})}{\det(v_a,v_{a+1})},\frac{\det (v_a,e_1)}{\det(v_a,v_{a+1})}\right)=
\left(-\frac{l_{a+1}}{k_a},\frac{l_{a}}{k_a}\right)\, ,
\end{align}
for $a=1,\ldots,d-1$, where we have used 
$\det(v_a,v_{a+1}) = l_{a+1}-l_a = -k_a$. 
Notice here that $\epsilon^{d}_1=0$, 
signifying the fact that $\D_{d+1}$ is a bolt with its tangent direction fixed by $e_1$.

\subsection{Local \texorpdfstring{$\mathcal{N}=2$}{N=2}
punctures}
\label{sec:N2}

We now want to glue in our puncture geometry. 
Recall that the $N$ M5-branes 
wrap the orbifold Riemann surface zero-section
 of $L_1^{-1}\oplus L_2^{-1}\rightarrow\Sigma_{g,n}$. Near to an orbifold point $x\in\Sigma_{g,n}$, 
 we have local coordinates $(w,z_1,z_2,t)\in \C^3\times\R$, 
with $w$ a coordinate on the Riemann surface with $x=\{w=0\}$, 
 $z_i$ are coordinates on the fibres of $L_i^{-1}$, and $t\in\R$. In the near-horizon limit 
the normal directions
to the Riemann surface/M5-branes
satisfy 
$\{|z_1|^2+|z_2|^2+t^2=1\}$, which over a smooth point of $\Sigma_{g,n}$ 
($w\neq 0$) is an $S^4$ fibre. 

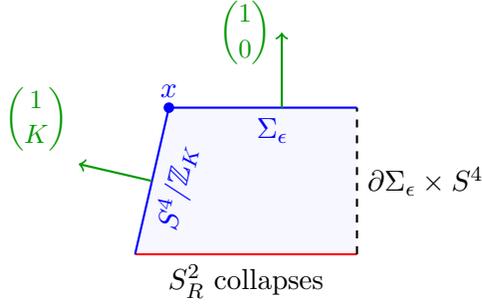
\begin{figure}
\centering
\begin{tikzpicture}

\coordinate (B) at (6.5,5.0);     %
junction (blue dot)
\coordinate (A) at (6.05,3.05);  
\coordinate (C) at (9.0,5.0);     

\fill[blue, opacity=0.035]
  (A) -- (B) -- (C) -- (9.0,3.05) -- (6.05,3.05) -- cycle;
  
\draw[thick, red] (6.05,3.05) -- (9.0,3.05)
  node[midway, below, text=black] {$S^2_R$ collapses};

\draw[thick, dashed] (9.0,3.05) -- (9.0,5.0)
  node[midway, right] {$\partial\Sigma_\epsilon\times S^4$};

\draw[thick, blue] (A) -- (B)
  node[pos=0.50, below, sloped, blue] {$S^4/\Z_\K$};
\draw[thick, blue] (B) -- (C)
  node[pos=0.55, below, blue] {$\Sigma_\epsilon$};

\fill[blue] (B) circle(2pt)
node[above] {$x$};;

\draw[->, thick, green!60!black] (8,5) -- (8,6)
  node[anchor=east] {\small $\begin{pmatrix}1\\0\end{pmatrix}$};

\draw[->, thick, green!60!black] (6.275,4.025) -- ++(-0.975,0.225)
  node[above left] {\small $\begin{pmatrix}1\\\K\end{pmatrix}$};

\end{tikzpicture}
\caption{Embedding of the $A_{\K-1}$ singularity into $M$. Over each interior point of the shaded region we have a copy of $\mathrm{U}(1)\times \mathrm{U}(1)_R\times S^2_R$. The first U$(1)$ rotates the Riemann surface coordinate $w$, with $\Sigma_\epsilon=\{|w|\leq \epsilon\}$,  and is not an isometry of the full solution. Note: (i) the U$(1)$'s collapse along the blue edges, as per the toric diagram, (ii) $S^2_R$ collapses along the red edge, which is $|z_1|=1$, (iii) the dotted black 
line is the locus $|w|=\epsilon$, which is a copy 
of $\partial\Sigma_\epsilon\times S^4$, with the first factor being a small $S^1$ around the orbifold point $x\in\Sigma_{g,n}$. The region shown is hence 
the $S^4$ orbibundle over $\Sigma_\epsilon$, and
this geometry is glued into $\Sigmabulk$
along the dotted black line.}
\label{figM}
\end{figure}

The $\mathcal{N}=2$ 
quotient by $\Z_\K$ acts on the coordinates $(w,z_1)$, fixing $w=z_1=0$. The latter is a copy of $S^2_R=S^2_2\subset \C_2\oplus\R$ rotated by the SO$(3)_R$ symmetry of $\mathcal{N}=2$ solutions. A neighbourhood of this 
$A_{\K-1}$ singularity in the full geometry is hence (for $|w|,|z_1|\ll 1$ both small) $\C^2/\Z_\K\times S^2_R$. On the other hand, if we fix a smooth point $\{w\neq 0\}\in \Sigma_{g,n}$ near to $x$, this neighbourhood is simply $\C_1\times S^2_R\subset S^4$. Here $z_1$ is a coordinate 
on $\C_1$, and increasing $|z_1|$ to
 $|z_1|=1$ one then has 
$z_2=t=0$.
Thus at $|z_1|=1$
the 
 $S^2_R$ collapses 
to a point, 
and $\C_1\times S^2_R$ smoothly caps off to give the $S^4$ fibre. 
This is shown in figure~\ref{figM}, which 
depicts the $S^4$ orbibundle over the neighbourhood $\Sigma_\epsilon=\{|w|\leq \epsilon\}$ 
of the point $x\in\Sigma_{g,n}$. 
This region is glued into the bulk of $M$ along 
the copy of $\partial\Sigma_\epsilon\times S^4\cong S^1\times S^4$ depicted by the dotted black line. 

\begin{figure}
    \centering
\begin{tikzpicture}

\draw[-stealth] (0,0) -- (10,0) node[anchor=west] {$n$};
\draw	
        (0,1.275) node {$y_1\,-$$\quad\,$}
		(0,2.775) node {$y_2\,-$$\quad\,$}
		(0,4.175) node {$y_{d-1}\,-$$\quad\quad\;\,$}
		(0,4.975) node {$y_d\,-$$\quad\;\,$};

\draw[-stealth] (0,0) -- (0,6.) node[anchor=east] {$y$};

\draw[thick, dash pattern=on 6pt off 2pt] (9,0) -- (9,5)
  node[midway, anchor=west] {$\partial\Sigma_\epsilon\times S^4$};

\draw[dotted] (0,1.3) -- (.3,1.3);
\draw[dotted] (0,2.8) -- (1.7,2.8);
\draw[dotted] (0,4.2) -- (4,4.2);
\draw[dotted] (0,5) -- (6.5,5);

\draw[dotted] (0.3,0) -- (.3,1.3);
\draw[dotted] (1.7,0) -- (1.7,2.8);
\draw[dotted] (4,0) -- (4,4.2);
\draw[dotted] (6.5,0) -- (6.5,5);

\draw	
        (0.3,0) node[below] {$n_1$}
		(1.7,0) node[below] {$n_2$}
		(4,0) node[below] {$n_{d-1}$}
		(6.5,0) node[below] {$n_d$};

\fill[blue, opacity=0.08]
  (0,0) -- (.3,1.3) -- (1.7,2.8) -- (2.,3.07) --
  (3.67,4.05) -- (4.,4.2) -- (6.5,5.) -- (9,5.) -- (9,0) -- cycle;

\begin{scope}
  \clip (0,0) -- (.3,1.3) -- (1.7,2.8) -- (2.,3.07) --
        (3.67,4.05) -- (4.,4.2) -- (6.5,5.) -- (9,5.) -- (9,0) -- cycle;
  \foreach \w/\op in {20/0.06, 14/0.10, 8/0.16}{
    \draw[red, line cap=round, line join=round, line width=\w, opacity=\op]
      (6.5,5) -- (9,5);
  }
\end{scope}

\draw[-stealth, thick, Green] (8,5) -- (8,6.) node[anchor=east] {\small $\begin{pmatrix}1\\0 \end{pmatrix}$};
\draw[-stealth, thick, Green] (5.2,4.6) -- (4.8,5.9) node[anchor=east] {\small $\begin{pmatrix}1\\l_d\end{pmatrix}$};
\draw[-stealth, thick, Green] (.9,1.95) -- (-1.,3.5) node[anchor=east] {\small $\begin{pmatrix}1\\l_2 \end{pmatrix}$};
\draw[-stealth, thick, Green] (0.15,.7) -- (-1.5,1.1) node[anchor=east] {\small $\begin{pmatrix}1\\l_1=\K \end{pmatrix}$};

\draw[red, thick] (.3,1.3) to[out=-10,in=180+60] (1.4,1.7) to[out=60,in=-125] (1.7,2.8);
\draw[red, thick] (.3,1.3) to[out=90+10,in=190] (.9,2.5) to[out=-170,in=-140] (1.7,2.8);
\draw[red] (1.6,1.6) node[below] {$D_2$};
\draw[blue] (0.2,0.4) node[ anchor=west] {$S^4/\mathbb{Z}_{k_1}$};

\draw[thick, blue]  (0,0) -- (.3,1.3)  -- (1.7,2.8)--(2.,3.07);
\draw[thick, blue] (3.67,4.05) -- (4.,4.2) -- (6.5,5.)--(9,5.)
  node[midway, anchor=north] {$\Sigma_\epsilon$};

\draw[thick, red] (0,0) -- (9,0) node[midway, above, text=black] {$S^2_R$ collapses};

\fill[blue] (.3,1.3) circle(2pt) node[below] {$\quad\, k_1$};
\fill[blue] (1.7,2.8) circle(2pt) node[below] {$\quad k_2$};
\fill[blue] (4,4.2) circle(2pt) node[below] {$\quad\quad\, k_{d-1}$};
\fill[blue] (6.5,5) circle(2pt) node[below] {$\quad\, k_d$};

\fill[blue] (2.65,3.47) circle(1pt);
\fill[blue] (2.85,3.6) circle(1pt);
\fill[blue] (3.05,3.73) circle(1pt);

\end{tikzpicture}
    \caption{Local $\mathcal{N}=2$ puncture geometry. Again, over each interior point of the shaded region is a copy of $\mathrm{U}(1)\times \mathrm{U}(1)_R\times S^2_R$, with the U$(1)$'s collapsing along the blue edges as per the toric diagram. The finite blue line segments now represent four-cycles $D_a \equiv\D_a\times S^2_R\cong \mathbb{WCP}^1_{[k_a,k_{a-1}]}\times S^2_R$, associated to new fluxes $n_a\in\Z$ for the M-theory four-form $G$ as described in the main text. 
    The shaded red region is the support of the bulk part of the flux, as discussed in appendix \ref{app:explicitG}. 
    }
    \label{fig:toric} 
\end{figure}
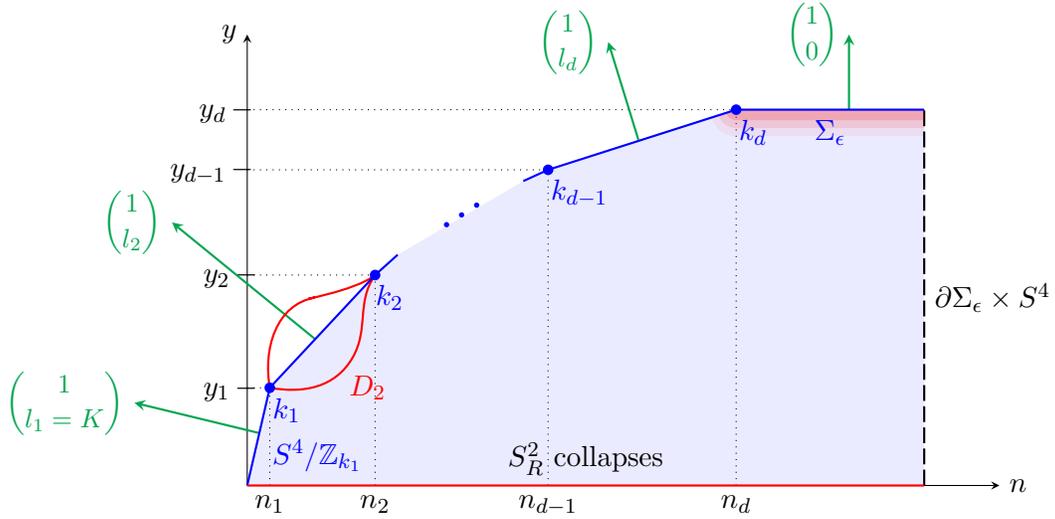

With this neighbourhood clearly described, we may now partially resolve the $A_{\K-1}$ singularity to obtain the local $\mathcal{N}=2$ puncture geometry. This is shown in figure~\ref{fig:toric}.
Our aim in the remainder of this section is to apply equivariant localization to this neighbourhood of a puncture, to compute its contribution to the $a$ central charge. To begin, we recall that the R-symmetry Killing vector is
\begin{align}\label{Rpuncture}
\xi = b_1 \partial_{\varphi_1} + b_2 \partial_{\varphi_2}\, ,
\end{align}
where $\partial_{\varphi_1}=(1,0)$ generates U$(1)_R$ and
is one of the two toric U$(1)$'s, and $\partial_{\varphi_2}$ 
rotates $S^2_R=S^2_2\subset\C_2\times \R$, fixing the poles $N, S$ of this two-sphere. The fixed point set 
of $\xi$ in this region thus 
consists of the two bolts $\Sigma_\epsilon^N, \Sigma_\epsilon^S$, as per the 
original M5-brane geometry without any 
partial resolution, together with 
$2(d-1)$ nuts: these are the blue dots 
labelled by $a=1,\ldots,d-1$, at either the 
north $N$ or south $S$ pole of the $S^2_R$. 
The weights of $e_1$ at the fixed points 
of the 4d geometry were computed in 
\eqref{4dRweights}, so from \eqref{Rpuncture} 
we can write down the weights 
at the $2(d-1)$ nuts in the 6d geometry $M$:
\begin{equation}\label{weightsN2}
    (\epsilon_1^a,\epsilon_2^a,\epsilon_3^a)=\left(-\frac{l_{a+1}}{k_a}b_1,\frac{l_{a}}{k_a}b_1,\pm b_2\right)\,,
\end{equation}
$a=1,\ldots, d-1$, 
with $\pm b_2$ being the weight of $\xi$ 
on the tangent spaces to $S^2_R$ at its north and south poles, respectively.

We may then proceed as in section 
\ref{sec:M5review}. Consider first 
the copy of $S^2_R$ over the $a$'th blue dot, which we denote by  $S^2_{R,a}$.
This has trivial homology class: 
we may connect any point in the partially 
resolved $A_{\K-1}$ space to a point 
on the red edge $\{|z_1|=1\}$ where $S^2_R$ collapses by a 
curve, where this curve together with $S^2_R$ is topologically 
a ball $B_3$ with boundary $\partial B_3=S^2_R$. Thus as in~\eqref{eq:locY}
we may write down
\begin{align}\label{eq:intYN2}
0 = \int_{S^2_{R,a}}\Phi^Y = \frac{2\pi}{3b_2}\left[(y_a^N)^2-(y_a^S)^2\right]\, ,
    \end{align}
 where $y_a^N, y_a^S$ are the values of the function $y$ at these nuts. 
In order to have non-zero fluxes in what follows 
we again solve this by taking $y_a\equiv y_a^N=-y_a^S>0$, where the $y_a$ are shown in figure~\ref{fig:toric}, and described in more detail below.\footnote{One could in principle instead take $y_a^N=y_a^S$. However, this would require the fluxes we are about to introduce to vanish. Moreover, the conformal dimension associated to an M2-brane wrapping the $S^2_R$ would vanish despite the cycle having finite size, signifying a pathology in this choice.}  
Notice that necessarily $y_d=y_N=9\pi \lp^3 b_1b_2N$, 
where $y_N$ was the value of $y$ at the north pole of the $S^4$ fibre over a smooth point of the Riemann surface in section~\ref{sec:M5review} 
(with this value of $y$ independent of the point on the Riemann surface, which we can then slide towards the blue dot at the left end of $\Sigma_\epsilon$ in figure~\ref{fig:toric}). Indeed since we want 
to glue this local puncture geometry into the bulk considered earlier, we need to 
identify $y_d=y_N$.

\subsection{Flux quantization}\label{sec:fluxN2}

We now turn to quantizing the four-form flux $G$. 
We begin with the four-cycle $D_1$, which is given by the blue line segment connecting the first blue dot to the red line where the $S^2_R$ collapses.
Topologically this space is $S^4/\Z_{k_1}$, in the 4d geometry recall this line segment was the non-compact toric divisor $\D_1\cong\C/\Z_{k_1}$, and $\D_1\times S^2_R$ then caps off to $S^4/\Z_{k_1}$. 
Localization immediately gives the flux
\begin{align}\label{n1def}
    n_1\equiv \frac{1}{(2\pi\lp)^3} \int_{D_1=S^4/\mathbb{Z}_{k_1}}G & =  \frac{1}{(2\pi\lp)^3} \frac{1}{k_1}\left[\frac{(2\pi)^2}{\epsilon_2^1\epsilon_3^1}\frac{y}{9}\Big|_{y_1^N} + \frac{(2\pi)^2}{\epsilon_2^1\epsilon_3^1}\frac{y}{9}\Big|_{y_1^S}\right]
    \\ \nn &
    =\frac{1}{9\pi \lp^3 b_1 b_2}\frac{y_1}{l_1}\equiv \frac{\hat{y}_1}{l_1}\,.
\end{align}
Here \eqref{n1def} defines the flux number $n_1$, 
we have used that $y_1\equiv y_1^N=-y_1^S$, the weights 
$\epsilon^a_3 \mid_{y_a^N}=b_2=-\epsilon^a_3 \mid_{y_a^S}$ (in a slight abuse of notation using the value of $y$ at a fixed point to label also the fixed point), 
and 
we have introduced the convenient rescaled
variable
\begin{equation}
    \hat{y}_a\equiv \frac{y_a}{9 \pi \lp^3 b_1 b_2}\, . .
\end{equation}

Next we have the four-cycles $D_a\equiv \D_a\times S^2_R$ described in figure~\ref{fig:toric}. These consist of the compact toric divisors 
$\D_a\cong \mathbb{WCP}^1_{[k_a,k_{a-1}]}$ 
in the 4d toric geometry, together with $S^2_R$, and are represented by the finite blue edges. 
Quantizing the flux over these four-cycles for $a=2,\dots,d$ we have 
\begin{align}\label{eq:fluxN2}
    n_a-n_{a-1}&\equiv \frac{1}{(2\pi\lp)^3} \int_{D_a}G=  \frac{2}{(2\pi\lp)^3}\left[ \frac{1}{k_a}\frac{(2\pi)^2}{\epsilon_2^a\epsilon_3^a}\frac{y_a}{9}+\frac{1}{k_{a-1}}\frac{(2\pi)^2}{\epsilon_1^{a-1}\epsilon_3^{a-1}}\frac{y_{a-1}}{9}\right]
    \\ \nn &
    = \frac{\hat{y}_a-\hat{y}_{a-1}}{l_a}\,.
\end{align}
These equations then define $n_a$ iteratively, starting
with $n_1$ defined in \eqref{n1def}. 
Notice that the weights tangent to $\D_a=\mathbb{WCP}^1_{[k_a,k_{a-1}]}$ satisfy 
$k_{a-1}\epsilon_1^{a-1}=-l_ab_1=-k_a\epsilon_2^a$. 
Finally note that there are in fact four 
fixed points contributing to the flux integral \eqref{eq:fluxN2}, and 
we have used 
$y_a\equiv y_a^N=-y_a^S$. As in \eqref{n1def} these come in two equal pairs which we have accounted for with the factor of 2 in front of the square bracket. 
This comment applies also to other integrals below.

\begin{figure}
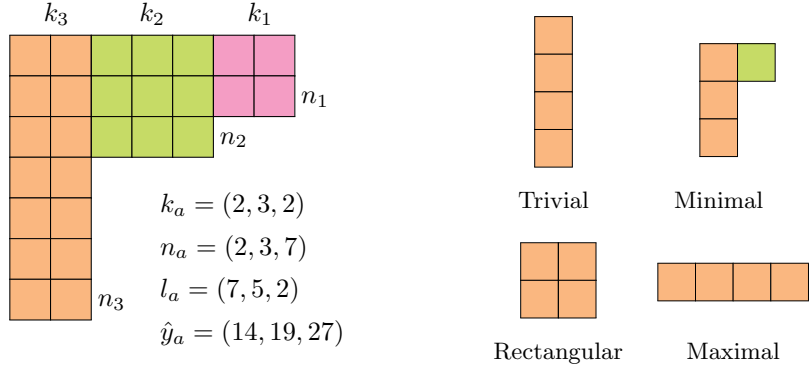

    \centering
      \begin{subfigure}[b]{0.45\textwidth}
        \centering
        \small{
        \begin{ytableau}
        \none[\qquad k_3] & \none & \none & \none[k_2] & \none & \none[\qquad k_1]  \\
*(Apricot) & *(Apricot)  &*(SpringGreen) & *(SpringGreen)  &*(SpringGreen) & *(Lavender)  &*(Lavender)   \\
*(Apricot) & *(Apricot)  &*(SpringGreen) & *(SpringGreen)  &*(SpringGreen)  & *(Lavender)  &*(Lavender) & \none[n_1]  \\
*(Apricot) & *(Apricot)  &*(SpringGreen) & *(SpringGreen)  &*(SpringGreen) & \none[n_2]  \\
*(Apricot) & *(Apricot) & \none & \none[]   \\
*(Apricot) & *(Apricot) \\
*(Apricot) & *(Apricot) \\
*(Apricot)  & *(Apricot)  & \none[n_3]  \\
 \none[\qquad ]
\end{ytableau}
\vspace{-2.7cm}
\hspace{2cm}
\begin{align*}
    &k_a=(2,3,2) \\
    &n_a=(2,3,7)\\
    &l_a=(7,5,2)\\
    &\hat y_a=(14,19,27)
\end{align*}
}
\\[-8pt]
        \caption{A general puncture \\ with $d=3$ and $N=27$}
    \end{subfigure}%
    \hspace{-1.5cm}
    ~ 
    \begin{subfigure}[b]{0.35\textwidth}
            \centering
     \begin{minipage}[b]{0.38\linewidth}
      \centering
        \footnotesize{
      \begin{ytableau}
    *(Apricot) ~ \\
  *(Apricot)   ~  \\
  *(Apricot)  ~  \\
  *(Apricot)  ~ \\
\end{ytableau}}
\\[8pt]
Trivial
    \end{minipage}
 \begin{minipage}[b]{0.38\linewidth}
 \centering
\footnotesize{
      \begin{ytableau}
    \none & *(Apricot)~  & *(SpringGreen) \\
  \none & *(Apricot)  ~ & \none \\
  \none & *(Apricot) ~ & \none \\ 
\end{ytableau}}
\\[12pt]
Minimal
   \end{minipage}
 \\[12pt]~
  \begin{minipage}[b]{0.38\linewidth}
 \centering
 \footnotesize{
      \begin{ytableau}
   *(Apricot)  ~ &  *(Apricot)  \\
  *(Apricot)  ~ &   *(Apricot)
\end{ytableau}}
\\[8pt]
Rectangular
\\ $\quad$
\end{minipage}~
 \begin{minipage}[b]{0.38\linewidth}
 \centering
 \footnotesize{
      \begin{ytableau}
  *(Apricot)  ~ & *(Apricot) & *(Apricot) & *(Apricot)
\end{ytableau}}
\\[10pt]
Maximal
\\ $\quad$
\end{minipage}
        \caption{Some particular cases}
        \label{fig:particularYoung}
    \end{subfigure}
    \caption{Young tableaux encoding puncture data. The orange blocks correspond to bolt contributions.} 
    \label{fig:Young}
\end{figure}

We may now solve \eqref{eq:fluxN2} iteratively and obtain for $a=1,\dots,d$
\begin{equation}\label{haty}
    \hat{y}_a= l_a n_a +\sum_{b=1}^{a-1} k_{b} n_{b}\, .
\end{equation}
Notice that if the  fluxes $n_a>0$ are all 
positive integers,  the $\hat{y}_a$ are then a sequence of increasing positive integers, and hence the $y_a$ are a sequence of increasing coordinate values. 
This 
data was already shown in 
figure~\ref{fig:toric}, where one can plot the sequence of $n_a$ horizontally. 
Recalling that above we  also argued that 
$y_d=y_N=9\pi \lp^3 b_1b_2N$, we then have
\begin{equation}\label{sumtoN}
    N=\hat{y}_d=\sum_{a=1}^{d}k_a n_a\,, 
\end{equation}
where the second equality is just \eqref{haty} for $a=d$.
It follows that the fluxes $n_a$ and the orbifold ranks $k_a$ define a partition of $N$. Observe that there is a simple dictionary between the partition and a Young diagram which encodes the puncture data, see figure~\ref{fig:Young}. This is the well-known statement that an $\mathcal{N}=2$ puncture is specified by a Young diagram of SU$(N)$ \cite{Gaiotto:2009we}. The orbifold ranks $k_a$ are the widths of the blocks whilst the $n_a$ are the heights (and $l_a$ are the lengths of the first  $a$ blocks).

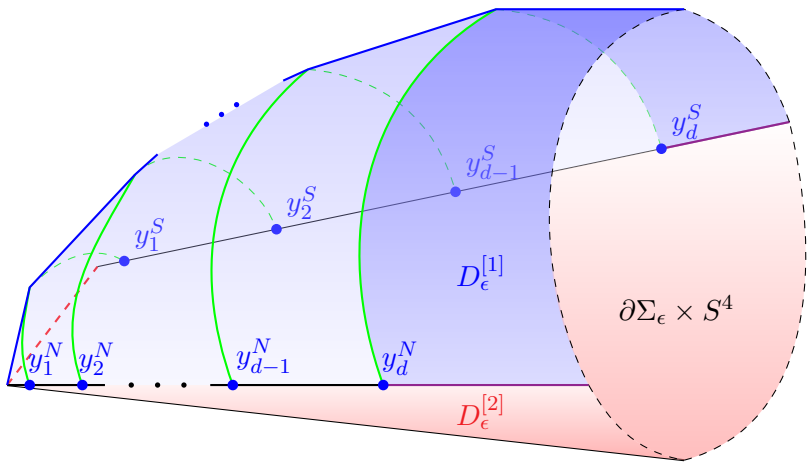
\begin{figure}[ht]
    \centering
\begin{tikzpicture}

\coordinate (y0) at (0,0);
\coordinate (y1) at (0.3,0);
\coordinate (y2) at (1,0);
\coordinate (yd-1) at (3,0);
\coordinate (yd) at (5,0);
\coordinate (yfinal) at (7.73,0);
\coordinate (my0) at (1.2,1.575);
\coordinate (my1) at (1.55705, 1.64971);
\coordinate (my2) at (3.58034, 2.07306);
\coordinate (myd-1) at (5.96067, 2.57112);
\coordinate (myd) at (8.69806, 3.14389);
\coordinate (myfinal) at (10.4,3.5);
\coordinate (kink1) at (0.3,1.3);
\coordinate (kink2) at (1.7,2.8);
\coordinate (kinkd-1) at (4,4.2);
\coordinate (kinkd) at (6.5,5);
\coordinate (kinkfinal) at (9,5);
\coordinate (bot) at (9,-1);
\coordinate (intUL) at (7.24248, 2.83932);

\draw[thick, black](y0)--(bot);

\draw[dashed,green] (kink1) to[out=45,in=150] (my1) ;
\draw[dashed,green] (kink2) to[out=45,in=110] (my2) ;
\draw[dashed,green] (kinkd-1) to[out=-5,in=105] (myd-1) ;
\draw[dashed,green] (kinkd) to[out=-5,in=105] (myd);

\draw[thick, dashed,red] (y0) -- (my0);
\draw[thick, Plum] (myd) -- (myfinal);
\draw[ black] (my0) -- (myd);

\fill[blue] (my1) circle(2pt) node[above right] {$ y_1^S$};
\fill[blue] (my2) circle(2pt) node[above right] {$y_2^S$};
\fill[blue] (myd-1) circle(2pt) node[above right] {$ y_{d-1}^{S}$};

\shade[blue!10,top color=blue!60,  bottom color=blue!10,fill opacity=0.3] (y0) -- (kink1) -- (kink2) -- (2.,3.07) --(3.67,4.05) -- (kinkd-1) -- (kinkd) -- (kinkfinal)to[out=190,in=117] (yfinal) -- (yfinal) -- cycle;

\shade[blue!30,top color=blue!70,  bottom color=blue!30,fill opacity=0.4]  (kinkd) -- (kinkfinal)to[out=190,in=117] (yfinal) -- (yfinal) -- (yd)to[out=110,in=210] (kinkd);

\fill[blue!30,top color=blue!60,  bottom color=blue!30,fill opacity=0.3](intUL)--(myfinal)to[out=110,in=-20] (kinkfinal)-- (kinkd)--(kinkd) to[out=-5,in=105] (myd)--(myfinal);

\fill[blue!10,top color=blue!40,  bottom color=blue!10,fill opacity=0.2](intUL)--(myfinal)to[out=110,in=-20] (kinkfinal)-- (kinkfinal)to[out=190,in=81] (intUL);

\shade[red!30, bottom color=red!30, top color=red!5,fill opacity=1] (y0)--(yfinal) to[out=300,in=170](bot)--(bot)--cycle;

\shade[red!50, bottom color=red!50, top color=red!0,fill opacity=0.5]  (bot) to[out=15,in=283](myfinal)--(intUL) to[out=266, in=117](yfinal)--(yfinal) to[out=300,in=170](bot);
\draw[thick, Plum] (myd) -- (myfinal);
\fill[blue!85] (myd) circle(2pt) node[above right] {$ y_d^{S}$};

\draw[thick,green] (yd) to[out=110,in=210] (kinkd) ;

\draw[thick,green] (yd-1) to[out=110,in=220] (kinkd-1) ;
\draw[thick,green] (y2) to[out=110,in=240] (kink2) ;
\draw[thick,green] (y1) to[out=110,in=260] (kink1) ;

\draw[dashed,black](bot) to[out=15,in=283](myfinal);
\draw[dashed,black](myfinal) to[out=110,in=-20] (kinkfinal);
\draw[dashed,black] (kinkfinal) to[out=190,in=117](yfinal);
\draw[dashed,black] (yfinal) to[out=300,in=170](bot);

\fill[blue] (2.65,3.47) circle(1pt);
\fill[blue] (2.85,3.6) circle(1pt);
\fill[blue] (3.05,3.73) circle(1pt);

\draw[thick,black] (y0)--(1.3,0);
\fill[black] (1.65,0) circle(1pt);
\fill[black] (2,0) circle(1pt);
\fill[black] (2.35,0) circle(1pt);
\draw[thick,black] (2.7,0)--(yd);
\draw[thick,Plum] (yd)--(yfinal);

\draw[thick, blue]  (y0) -- (kink1)  -- (kink2)--(2.,3.07);
\draw[thick, blue] (3.67,4.05) -- (kinkd-1) -- (kinkd)--(kinkfinal)
  node[midway, anchor=north] {};

\fill[blue] (y1) circle(2pt) node[above] {$\quad\, y_1^N$};
\fill[blue] (y2) circle(2pt) node[above] {$\quad y_2^N$};
\fill[blue] (yd-1) circle(2pt) node[above] {$\quad\quad\, y_{d-1}^{N}$};
\fill[blue] (yd) circle(2pt) node[above] {$\quad\, y_d^{N}$};

\node at (8.9, 1.0) {$\partial\Sigma_\epsilon\times S^4$};
\node at (6.3,1.5) {\textcolor{blue}{$D^{[1]}_{\epsilon}$}};
\node at (6.3,-0.35) {\textcolor{Red}{$D^{[2]}_\epsilon$}};
  
\end{tikzpicture}
 \caption{The 3d version of the toric diagram for the $\mathcal{N}=2$ puncture. The dark blue lines and the red line are the same as in figure~\ref{fig:toric}, with the former obtained by cutting the diagram vertically.
 The red face denotes the locus where $\partial_{\varphi_2}\rightarrow 0$.
    The four-cycle $D_{\epsilon}^{[1]}$ consists of the blue face between the black dashed line and the first green line from the right. The four-cycle $D_{\epsilon}^{[2]}$ consists of all of the red face. We can also easily identify compact two-cycles in the diagram, they are simply lines in the diagram between two nodes. The green lines are the $S^2_{R,a}$ used in equation \eqref{eq:intYN2} and \eqref{DeltaS2}.
    }
    \label{fig:toric3D} 
\end{figure}

There are two remaining four-cycles that we need to consider, which can be seen most easily by including the $S^2_R$ in figure~\ref{fig:toric}. The result is figure~\ref{fig:toric3D}.\footnote{This is not a toric diagram in the usual sense, since $S^4$ is not toric. Instead one builds the diagram by noting that we may draw a ``toric" diagram for $S^4$ as an ellipse. This is nothing but two copies of $\mathbb{C}^2$ glued together at an equatorial $S^3$. Our 3d ``toric diagram" is then obtained by gluing two copies of the product of $\mathbb{C}$ with the partial resolution of the $\mathbb{C}^2/\mathbb{Z}_K$ singularity.   } 
These four-cycles are different to those previously considered in that they are non-compact (hence not really ``cycles'', although we will use that terminology). The two four-cycles will fix the K\"ahler class $\omega$ through $\Sigma_{\epsilon}$ which will appear in the central charge. 
The first four-cycle is
$D^{[1]}_{\epsilon}
\equiv \D_{d+1}\times S^2_R$, where recall that $\D_{d+1}\cong \C/\Z_{k_{d}}=\Sigma_\epsilon$ is 
the Riemann surface direction.  
This flux is the analogue of \eqref{GfluxDirac}, which
recall is delicate since it is not invariant under large gauge transformations \eqref{banishtothebulk}.
Applying localization we find
\begin{align}\label{eq:fluxintomegaN2}
   N^{[1]}_{\epsilon}\equiv
   \frac{1}{(2\pi\lp)^3}\int_{D^{[1]}_{\epsilon}}G
   &=\frac{2}{(2\pi\lp)^3}\frac{2\pi}{b_2}\int_{\Sigma_\epsilon^N}\left[\frac{\omega}{6y_d}-\frac{2\pi c_1(\mathcal{L}_2)}{b_2}\frac{y_d}{9}\right] \nn\\ &=\frac{1}{108\pi^3\lp^6b_1 b_2^2 N } \int_{\Sigma_{\epsilon}^N}\omega\, ,
\end{align}
where we have used the $\mathcal{N}=2$ condition for the discrete Chern number fluxes \eqref{N2twists},
so that $\int_{\Sigma_\epsilon}c_1(\mathcal{L}_2)=0$. 
Further note there are two equal contributions from $\Sigma_\epsilon^N$ and $\Sigma_\epsilon^S$, giving the prefactor of 2 in the middle expression.
From this we deduce
\begin{equation}\label{intomegauseless}
    \int_{\Sigma_\epsilon^N}\omega=108\pi^3\lp^6b_1 b_2^2 N N^{[1]}_\epsilon\,,
\end{equation}
in terms of the currently undetermined flux $N^{[1]}_\epsilon$. 
This expression on its own is not particularly useful as {\it a priori} we do not know how $N^{[1]}_\epsilon$ is related to the other fluxes. There are two ways to proceed. On the one hand, the integral of $G$ through $D^{[1]}_\epsilon$ can be computed explicitly by writing down an ansatz for $G$ which fixes $N^{[1]}_{\epsilon}$ in terms of the other fluxes. We perform this analysis in appendix~\ref{app:explicitG}.  
On the other hand, and somewhat unique to the $\mathcal{N}=2$ case, we may fix this using the remaining non-compact four-cycle. Observe that we may define a four-cycle by taking the union of the two copies of the 4d toric diagram at the two poles of the $S^2_R$. These are glued together along the collapsed two-sphere (red line) in figure~\ref{fig:toric} (which is a copy of $S^1_{\varphi_1}\times \Sigma_\epsilon$) leaving a compact cycle in the non-disc directions. This cycle is the generalization of the the four-cycle $D_{[2]}$ considered in section \ref{sec:M5review} to the resolved geometry and we denote it by $D^{[2]}_\epsilon$ in the following, see figure~\ref{fig:toric3D}. The integral of the flux over this cycle receives contributions from the 2 bolts $\Sigma_\epsilon^N$, $\Sigma_\epsilon^S$ and the $2(d-1)$ nuts and we find
\begin{align}\label{eq:trivialfluxintN2}
   N^{[2]}_\epsilon\equiv\frac{1}{(2\pi\lp)^3}\int_{D^{[2]}_{\epsilon}}G
   &=\frac{2}{(2\pi\lp)^3}\left\{\frac{2\pi}{b_1}\int_{\Sigma_\epsilon^N}\left[\frac{\omega}{6y_d}-\frac{2\pi c_1(\mathcal{L}_1)}{b_1}\frac{y_d}{9}\right]+\sum_{a=1}^{d-1} \frac{1}{k_a}\frac{(2\pi)^2}{\epsilon_1^a\epsilon_2^a}\frac{y_a}{9}\right\}
   \nn\\ 
    &=\frac{1}{108\pi^3\lp^6 {b_1^2 b_2} N } \int_{\Sigma_{\epsilon}^N}\omega-\frac{b_2}{b_1}(N-n_d)\, ,
\end{align}
where we used $\int_{\Sigma_\epsilon} c_1(\mathcal{L}_1)=1-1/k_d$ as per \eqref{N2twists} after (partial) resolution. To obtain the final expression one uses \eqref{eq:fluxN2} to eliminate $y_a$, and in particular one finds the identity
\begin{align}
    \frac{2}{(2\pi\lp)^3}\sum_{a=1}^{d-1} \frac{1}{k_a}\frac{(2\pi)^2}{\epsilon_1^a\epsilon_2^a}\frac{y_a}{9}& =-\frac{b_2}{b_1}\left(\frac{N}{k_d}-n_d\right)\,.
\end{align}
In the $\mathcal{N}=2$ case this flux is necessarily trivial, $N^{[2]}_\epsilon=0$. As explained in appendix~\ref{app:explicitG}, see equation \eqref{Gansatz}, the flux $G$ has no support on this cycle since this would break the $\mathcal{N}=2$ symmetry to $\mathcal{N}=1$. This further allows us to solve for the K\"ahler class $\omega$ over $\Sigma_{\epsilon}$ giving
\begin{equation}\label{intomegaN2}
    \int_{\Sigma_{\epsilon}^N}\omega=\sG 108 \pi^3 \lp^6 b_1 b_2^2  N (N-n_d)\, .
\end{equation}
Comparing \eqref{intomegauseless} and \eqref{intomegaN2} we deduce $N^{[1]}_\epsilon=N-n_d$, which is confirmed by the explicit computation in appendix \ref{app:explicitG}, see \eqref{fluxboltexplicitN2}. We now have all the necessary ingredients to compute the central charge.

\subsection{Observables}

We may now finally turn to the computation of the central charge arising from the glued-in puncture geometry.
We have two bolt contributions along $\Sigma_\epsilon^{N,S}$ (where $y=\pm y_d$), along with $2(d-1)$ nut contributions to take into account. 
Putting all of these together we find that the total contribution to the $a$-central charge from a puncture is\footnote{In gravity, at two-derivative level, we are actually computing the quantity $4a-3c=\tfrac{1}{12}(4n_v-n_h)$, which in the large $N$ limit is precisely $a$. When comparing with the field theory results one should take this into account. This gives a more refined check of the duality than just taking a large $N$ limit and studying the $O(N^3)$ term. \label{foot:4a-3c}} 
\begin{align}\label{N2punct}
        \delta a
        &=\frac{2}{2(2\pi)^6\lp^9} \left[\frac{(2\pi)^2}{b_1b_2}\int_{\Sigma_\epsilon}\left[\Phi_2-2\pi\Big(\frac{c_1(\mathcal{L}_1)}{b_1}+\frac{c_1(\mathcal{L}_2)}{b_2}\Big)\Phi_0\right]+\sum_{a=1}^{d-1} \frac{1}{k_a}\frac{(2\pi)^3}{\epsilon_1^a\epsilon_2^a\epsilon_3^a}\Phi_0\Big|_{y_a}\right] \nn \\ 
        &=-\frac{9 b_1 b_2^2}{16}\bigg[\sG\Big(2-{3}\frac{n_d}{N}+\frac{1}{k_d}\Big)N^3- \sum_{a=1}^{d-1} \frac{k_a\hat{y}_a^3}{l_a l_{a+1}} \bigg]\,,
    \end{align}
where $\Phi_{2}, \Phi_0$ are written in \eqref{fixedstuff}, the weights $\epsilon_i^a$ are in \eqref{weightsN2}, $\hat y_a$ in \eqref{haty}, and we used \eqref{intomegaN2} and \eqref{N2twists} to evaluate the integrals. 

This is the local off-shell contribution to the central charge from the puncture. Since the local analysis here has the opposite orientation to the bulk contribution\footnote{This is the same minus sign appearing in \eqref{splitSigma}. } the final result for the central charge of the Riemann surface with $n$ $\mathcal{N}=2$ punctures is
\begin{equation}\label{eq:afullN=2}
   a=\abulk - \sum_{I=1}^n \delta a^{I}\, ,
\end{equation}
where the bulk contribution $\abulk$ is given in \eqref{eq:abulk}.

In the case where the bulk solution preserves $\mathcal{N}=2$ supersymmetry one finds that this exactly reproduces the field theory contribution in \cite{Chacaltana:2010ks}.\footnote{To show this it is simplest to use the form given in \cite{Couzens:2023kyf}, which can be written more simply in the form above. Also note that the field theory result is the on-shell one, whereas we have presented the off-shell result. For the $\mathcal{N}=2$ solution to go on-shell one sets $b_1=\tfrac{1}{3}$, $b_2=\tfrac{2}{3}$ and $p_1^{\mathrm{bulk}}=2-2g$. } In fact, our results are more general and give the contributions for the $\mathcal{N}=2$ puncture independently of the supersymmetry preserved in the bulk. Furthermore, recall that we have made a choice of the local orbifold action in \eqref{Ak}. Taking instead the quotient to act on $\mathbb{C}_2$ rather than $\mathbb{C}_1$ we obtain another local $\mathcal{N}=2$ preserving puncture giving the same central charge contribution as in \eqref{N2punct} after interchanging $b_1\leftrightarrow b_2$. In the literature these are sometimes referred to as $(1,0)$ and $(0,1)$ punctures respectively \cite{Bah:2018gwc}.
Despite the individual punctures preserving $\mathcal{N}=2$ supersymmetry, two punctures of different type placed at different (arbitrary) points on the Riemann surface imply that the global preserved supersymmetry is $\mathcal{N}=1$, irrespective of the bulk supersymmetry. 

It is important to emphasize that the result \eqref{eq:afullN=2} is an off-shell result. To go on-shell one needs to extremize over the R-symmetry parameters $b_1,b_2$ subject to the constraint $b_1+b_2=1$. Including only $\mathcal{N}=2$ punctures allows us to explicitly perform the extremization for an arbitrary configuration of punctures. Observe that the off-shell central charge for a Riemann surface with an arbitrary configuration of (partially-Higgsed) $\mathcal{N}=2$ punctures takes the form
\begin{equation}
    a= -\frac{9}{8}b_1b_2(b_1\mathfrak{p}_2+b_2\mathfrak{p}_1)N^3\, ,
\end{equation}
where the $\mathfrak{p}_i$ are independent of the R-symmetry mixing parameters, $b_i$. One finds\footnote{One should eliminate $\hat{y}$ in terms of the fluxes using \eqref{haty} which makes the result manifestly independent of the $b$'s; however, to keep the formulae more succinct we keep $\hat{y}$. }
\begin{equation}
    \mathfrak{p}_1=p_1^{\text{bulk}}-\frac{1}{2}\sum_{I\in(1,0) \text{ punct.}}\left (2-\frac{3 n^{I}_d}{N}+\frac{1}{k_d^I}-\sum_{a=1}^{d^I-1}\frac{k_a^I }{l_a^I l_{a+1}^I }\frac{\hat{y}_{a,I}^{3}}{N^3}\right)\, .
\end{equation}
and similarly for $\mathfrak{p}_2$ with $(1,0)\rightarrow (0,1)$. This is the extension of $p^{\text{tot}}$, (c.f. \eqref{bobthebuilder}) to the Higgsed punctures.
In particular, the off-shell central charge takes the same functional form as the smooth Riemann surface case with the replacement $p_i\rightarrow \mathfrak{p}_i$; as we will see shortly, this is not a generic feature for $\mathcal{N}=1$ punctures. 
Maximizing the central charge one finds 
\begin{equation}\label{extrbi}
    b_1=\frac{2 \mathfrak{p}_1-\mathfrak{p}_2+\sqrt{\mathfrak{p}_1^2- \mathfrak{p}_1\mathfrak{p}_2+\mathfrak{p}_2^2}}{3(\mathfrak{p}_1-\mathfrak{p}_2)}\, ,\quad  b_2=\frac{2 \mathfrak{p}_2-\mathfrak{p}_1+\sqrt{\mathfrak{p}_1^2- \mathfrak{p}_1\mathfrak{p}_2+\mathfrak{p}_2^2}}{3(\mathfrak{p}_2-\mathfrak{p}_1)}\, ,
\end{equation}
and the on-shell central charge is
\begin{equation}
    a=-\frac{3b_1 b_2(\mathfrak{p}_1+\mathfrak{p}_2-\sqrt{\mathfrak{p}_1^2- \mathfrak{p}_1\mathfrak{p}_2+\mathfrak{p}_2^2})}{8} N^3\, .
\end{equation}
Note that in general this includes terms beyond $\mathcal{O}(N^3)$, as one can see by performing a series expansion for large $N$, and it would be interesting to match this to field theory results in light of footnote \ref{foot:4a-3c}.

As a final observable to consider in the $\mathcal{N}=2$ puncture case let us compute the conformal dimensions of BPS operators in the CFT corresponding to M2-branes wrapping calibrated two-cycles in the geometry  \cite{Gauntlett:2006ai,Bah:2013wda}. The M2-branes sit at the centre of AdS$_5$ and wrap a two-cycle in the internal space which is calibrated with respect to the two-form $Y'$ defined in equation \eqref{bildef}. There are two classes of cycle depending on whether the M2-brane wraps the R-symmetry direction or not. For the cycle which does not wrap the R-symmetry direction the two-cycle is necessarily a fixed point set. Clearly, for the present set-up the only two-cycles satisfying this condition are the two copies of the bolt which must include all the bulk and puncture contributions. We therefore find that the conformal dimension for the BPS particle obtained by wrapping an M2-brane on the bolt is
\begin{align}
     \Delta\big[\Sigma_{g,n}\big]=-\frac{3}{2}(b_1\mathfrak{p}^{\mathrm{bolt}}_2+b_2 \mathfrak{p}^{\mathrm{bolt}}_1)N\, ,
\end{align}
where we have defined
\begin{align}
\mathfrak{p}^{\mathrm{bolt}}_1 \equiv 
p_1^{\mathrm{bulk}} + \sum_{I\in(1,0) \text{ punct.}} \left(1-\frac{n_d^I}{N}\right)\, ,
\end{align}
and similarly for $\mathfrak{p}^{\mathrm{bolt}}_2$ with $(1,0)\rightarrow (0,1)$.

Finally let us consider the two-cycles obtained by wrapping the R-symmetry direction. These are given by the various compact lines, between two nuts in figure~\ref{fig:toric3D}. There are two classes of two-cycle, those drawn in green which are the $S^2_R$ for fixed $y_a$ (i.e. the green line going between $y_a^N$ and $y_a^S$), and the black lines between fixed points which are spindles $\mathbb{WCP}^{1}_{[k_{a-1},k_a]}$ and are the resolution two-cycles from the blow-up. In each case the calibration condition requires us to take the cycle which gives a positive result. In the following we will focus on a $(1,0)$ puncture, with the $(0,1)$ puncture obtained by $b_1\leftrightarrow b_2$.
For the first class of two-cycle we have
\begin{equation}\label{DeltaS2}
    \Delta\big[S^2_{R,a}\big]=3 b_1 \hat{y}_a\, ,
\end{equation}
with $\hat{y}_a$ in \eqref{haty}.
The value of $b_1$ is determined by the extremization of the central charge, which depends on the particular punctures inserted  \eqref{extrbi}. For the full $\mathcal{N}=2$ bulk solutions, $b_1=1/3$, such that $\Delta_a=\hat y_a$, and agrees with the explicit line-charge analysis in \cite{Gaiotto:2009gz} once one notices that the $\hat{y}_a$ are precisely the values of the line charge at the kinks. For the spindle cycles one finds
\begin{equation}\label{Deltaspindle}
    \Delta\big[\mathbb{WCP}^1_{[k_{a-1},k_a]}\big]=\frac{3 b_2}{2}(n_a-n_{a-1})\, .
\end{equation}
Observe that for a pure $\mathcal{N}=2$ theory $b_2=2/3$ and the result for the conformal dimension is simply that it is the difference between the neighbouring fluxes. 

\subsection{\texorpdfstring{Full $\mathcal{N}=2$}{N=2} solutions}\label{sec:N=2sol}

To make the above formulae more explicit we will look at some particular choices of fluxes and give the dictionary to the field theory. While the $\mathcal{N}=2$ puncture geometries we just considered can in principle be inserted into both $\mathcal{N}=1$ and $\mathcal{N}=2$ bulk solutions, in the following we want to focus on the latter. This allows us to give simple examples and also recover results from the literature. Multiple such punctures, associated to possibly different partitions of $N$, can be inserted\footnote{Note that no other type (like $\mathcal{N}=2$ preserving the other circle direction, or $\mathcal{N}=1$) is allowed if one wants to preserve global $\mathcal{N}=2$.} and the central charge picks up the same contribution \eqref{N2punct}, with their respective puncture data inserted, for each of them. 
For concreteness let us take $p_1^{\text{bulk}}=2(1-g)$, $p_2^{\text{bulk}}=0$, then going on-shell fixes the R-symmetry parameters to be $b_1=1/3$, $b_2=2/3$ and the central charge becomes
\begin{equation}\label{apunct}
    a=\frac{1}{3}(g-1)N^3-\sum_{I=1}^n \delta a^{I}\,, \quad 
    \delta a=-\sG\frac{1}{12}\left[\Big(2-{3}\frac{n_d}{N}+\frac{1}{k_d}\Big)N^3-\sum_{a=1}^{d-1}\frac{k_a}{l_al_{a+1}}\hat y_a^3\right]\,.
\end{equation}

As a first example consider the somewhat trivial case where the we take only unresolved singularities $\mathbb{C}^2/\mathbb{Z}_k$. This of course is exactly what was discussed in section \ref{sec:orbifoldsection}; however, as a consistency check we must be able to recover those results from \eqref{apunct}. These punctures correspond to rectangular partitions of the Young tableau, i.e. a single block ($d=1$) of length $k$ and height $n=N/k$, see figure~\ref{fig:particularYoung}. Such punctures lead to an SU$(k)$ flavour symmetry in the dual field theory. The rectangular punctures are particularly simple since they only receive bolt contributions and the contribution from a rectangular puncture is
\begin{equation}
    \delta a^{\mathrm{rectangular}}=-\frac{1}{6}\Big(1-\frac{1}{k}\Big)N^3\,,
\end{equation}
which is precisely \eqref{eq:aorb} once the $\mathcal{N}=2$ condition is imposed. As a further consistency check a trivial puncture must return a trivial result. In this case we have $k=1$, $n=N$ and indeed $\delta a$ vanishes identically. Punctures were not all created equally and our final rectangular puncture, the maximal puncture, has a somewhat special standing. For the maximal puncture we have $k=N$ and $n=1$ and the contribution to the anomaly polynomial is
\begin{equation}
    \delta a^\text{max}=-\frac{1}{6}\left(N^3-N^2\right)\,.
\end{equation}
Taking the Riemann surface to be a two-sphere with three maximal punctures the dual field theory is the $T_N$ theory, and it is simple to check that the large $N$ central charge matches.

The final explicit puncture we consider is the minimal puncture which has
\begin{equation}
    d=2\,,\,k=(1,1)\,,\,n=(1,N-1)\,,\,l=(2,1)\,,\,\hat y=(2,N)\implies \delta a^\text{min}=\frac{1}{4} N^2-\frac{1}{3}\,,
\end{equation}
and in particular the central charge receives nut contributions. Here again we match the field theory results even at $\mathcal{O}(1)$. In particular, we reproduce the analysis of \cite{Bah:2019jts} for these various cases.

As a final aside, note that the puncture contribution in \eqref{apunct} can be written in the form
\begin{equation}
    \delta a=-\sG\frac{1}{12}\left[\Big(2-3\frac{n_d}{N}\Big)N^3+\sum_{a=1}^{d}\frac{\hat y_a^3-\hat y^3_{a-1}}{l_a}\right]\,.
\end{equation}
Using that
\begin{equation}
    \hat y_a=l_an_a+M_{a-1}\,, \quad M_a\equiv\sum_{b=1}^ak_bn_b\,,
\end{equation}
(note that $M_d=N$),
the sum in terms of flux numbers $n_a$ reads
\begin{align}
    \delta a=-\sG\frac{1}{4}\Bigg[&\Big(\frac{2}{3}-\frac{n_d}{N}\Big)N^3\nn \\ &+\sum_{a=1}^{d}\frac{1}{3}l_a^2(n_a^3-n_{a-1}^3)+l_aM_{a-1}(n^2_a-n_{a-1}^2)+M_{a-1}^2(n_a-n_{a-1})\Bigg]\,.
\end{align}
This reproduces the results of \cite{Bah:2019jts, Bah:2022yjf} which studied $\mathcal{N}=2$ punctures using anomaly inflow.\footnote{We use the same variables but different notation. The dictionary between our results and those in  \cite{Bah:2019jts, Bah:2022yjf}
is $\hat y_a\to N_a,\,  n_a\to\omega_a,\, M_a\to y_a.$
} 
This may also be obtained from a line charge description using the explicit supergravity solutions in \cite{Gaiotto:2009gz}. 
Our results present a clean derivation (and extension) of these results which we will now further generalize to new punctures which only preserve $\mathcal{N}=1$ supersymmetry.


\section{\texorpdfstring{$\mathcal{N}=1$}{N=1} punctures}\label{sec:N=1}

We return to the study of $\mathbb{C}^3/\mathbb{Z}_K$ singularities introduced in section \ref{sec:orbifoldsection}, where the quotient is a finite subgroup of SU$(3)$ acting as
\begin{align}\label{eq:SU(3)action}
(w,z_1,z_2)\mapsto(\ex^{2\pi \ii /K} w, \ex^{2\pi \ii{\alpha_1} /K} z_1,\ex^{2\pi \ii{\alpha_2} /K} z_2)\, ,
\end{align}
with 
\begin{equation}\label{alphaconstr}
  \alpha_1+\alpha_2+1=K\,.  
\end{equation}
Note that if one of the $\alpha_i$ is trivial then the quotient is a finite subgroup of SU$(2)$ and there is an enhancement of supersymmetry to $\mathcal{N}=2$. Given that this was considered previously we will ignore this case here. 

\subsection{\texorpdfstring{Geometry of $\mathbb{C}^3/\mathbb{Z}_K$ singularities}{Geometry of C3/ZK singularities}}

\begin{figure}[ht!]

 \begin{minipage}{.32\textwidth}

\begin{center}
\tdplotsetmaincoords{78}{-60}

\begin{tikzpicture}
		[tdplot_main_coords,
			cube/.style={very thick,black},
			axisb/.style={->,blue,thick},
            axisr/.style={->,red,thick},
            axisg/.style={->,Green,thick},
			inf/.style={dashed,black},scale=.75,every node/.style={scale=1.}]

\shadedraw[white,top color=white, bottom color=red!50,fill opacity=0.5] (0,0,0)--(3,0,0)--(4,3,0)--(1,3,0)--(0,0,0);
\shadedraw[white,right color=blue!40, left color=white,fill opacity=0.8] (0,0,0)--(1,3,0)--(2/3,4,-2)--(-4/3,-2,-2)--(0,0,0);
\shadedraw[white,left color=Green!70, right color=Green!0,fill opacity=0.5] (0,0,0)--(3,0,0)--(3,-2,-2)--(-4/3,-2,-2)--(0,0,0);


\draw[black, thick](-4/3,-2,-2)--(0,0,0);
\draw[black, thick](0,0,0)--(1,3,0);
\draw[Plum,thick](0,0,0)--(3,0,0);

\draw[axisg] (3,0,-1.5)--(3,-.8,-0.7);
\draw[axisb] (-0.5,1.5,-1)--(-2,2.,-0.5);
\draw[axisr] (2,1.5,0)--(2,1.5,1);

\fill[black] (0,0,0) circle(2pt) node[below] {}; 
\node[Plum] at (1.4,0,.4) {$\Sigma_\epsilon$};

\node at (1.5,1.5,1.35) {\color{red}$v_1$};
\node at (3.,-1.,-0.4) {\color{Green}$v_2$};
\node at (-1.5,2.2,-0.2) {\color{blue}$v_3$};

\end{tikzpicture}
\end{center}
\end{minipage}~
 \begin{minipage}{.32\textwidth}

\begin{center}
\tdplotsetmaincoords{78}{-60}

\begin{tikzpicture}
		[tdplot_main_coords,
			cube/.style={very thick,black},
			axisb/.style={->,blue,thick},
            axisr/.style={->,red,thick},
            axisg/.style={->,Green,thick},
			inf/.style={dashed,black},scale=.75,every node/.style={scale=1.}]

\shadedraw[white,top color=white, bottom color=red!50,fill opacity=0.5] (.5,0,0)--(3,0,0)--(4,3,0)--(1,3,0)--(.5,1.5,0)--(.5,0,0);
\shadedraw[white,right color=blue!40, left color=white,fill opacity=0.8] (.5,1.5,0)--(1,3,0)--(2/3,4,-2)--(-4/3,-2,-2)--(-1,-1.5,-1.5)--(.5,1.5,0);
\shadedraw[white,left color=Green!70, right color=Green!0,fill opacity=0.5] (.5,0,0)--(3,0,0)--(3,-2,-2)--(-4/3,-2,-2)--(-1,-1.5,-1.5)--(.5,0,0);
\fill[orange!70,opacity=.5] (.5,0,0)--(.5,1.5,0)--(-1,-1.5,-1.5)--(.5,0,0);

\draw[black, thick](-4/3,-2,-2)--(-1,-1.5,-1.5);
\draw[black, thick](.5,1.5,0)--(1,3,0);
\draw[Plum,thick](.5,0,0)--(3,0,0);
\draw[black,thick](.5,0,0)--(.5,1.5,0)--(-1,-1.5,-1.5)--(.5,0,0);

\draw[axisg] (3,0,-1.5)--(3,-.8,-0.7);
\draw[axisb] (-0.5,1.5,-1)--(-2,2.,-0.5);
\draw[axisr] (2.,1.5,0)--(2.,1.5,1);
\draw[->,RedOrange,thick] (0,.2,-.2)--(-1,.2,.8);

\fill[black] (.5,0,0) circle(2pt) node[below] {}; 
\fill[black] (.5,1.5,0) circle(2pt) node[below] {}; 
\fill[black] (-1,-1.5,-1.5) circle(2pt) node[below] {}; 
\node[Plum] at (1.4,0,.4) {$\Sigma_\epsilon$};

\node at (1.5,1.5,1.35) {\color{red}$v_1$};
\node at (3.,-1.,-0.4) {\color{Green}$v_2$};
\node at (-1.5,2.2,-0.2) {\color{blue}$v_3$};
\node at (-1.1,.5,1) {\color{RedOrange}$v_4$};

\end{tikzpicture}
\end{center}
\end{minipage}~
 \begin{minipage}{.32\textwidth}

\begin{center}
\tdplotsetmaincoords{78}{-60}

\begin{tikzpicture}
		[tdplot_main_coords,
			cube/.style={very thick,black},
			axisb/.style={->,blue,thick},
            axisr/.style={->,red,thick},
            axisg/.style={->,Green,thick},
			inf/.style={dashed,black},scale=.75,every node/.style={scale=1.}]

\shadedraw[white,top color=white, bottom color=red!50,fill opacity=0.5] (.5,0,0)--(3,0,0)--(4,3,0)--(1,3,0)--(.5,1.5,0)--(.5,0,0);
\shadedraw[white,right color=blue!40, left color=white,fill opacity=0.8] (.5,1.5,0)--(1,3,0)--(2/3,4,-2)--(-2/3,0,-2)--(.5,1.5,0);
\shadedraw[white,left color=Green!70, right color=Green!0,fill opacity=0.5] (.5,0,0)--(3,0,0)--(3,-2,-2)--(0,-2,-2)--(.5,0,0);
\shadedraw[white,top color=orange!70, bottom color=orange!10,fill opacity=0.5] (.5,0,0)--(.5,1.5,0)--(-2/3,0,-2)--(0,-2,-2)--(.5,0,0);

\draw[black, thick](.5,1.5,0)--(1,3,0);
\draw[Plum,thick](.5,0,0)--(3,0,0);
\draw[black,thick](.5,0,0)--(.5,1.5,0)--(-2/3,0,-2);
\draw[black,thick] (0,-2,-2)--(.5,0,0);

\draw[axisg] (3,0,-1.5)--(3,-.8,-0.7);
\draw[axisb] (-0.5,1.5,-1)--(-2,2.,-0.5);
\draw[axisr] (2.,1.5,0)--(2.,1.5,1);
\draw[->,RedOrange,thick] (0.5,.2,-1.2)--(-.2,.2,-.5);

\fill[black] (.5,0,0) circle(2pt) node[below] {}; 
\fill[black] (.5,1.5,0) circle(2pt) node[below] {}; 
\node[Plum] at (1.4,0,.4) {$\Sigma_\epsilon$};

\node at (1.5,1.5,1.35) {\color{red}$v_1$};
\node at (3.,-1.,-0.4) {\color{Green}$v_2$};
\node at (-1.5,2.2,-0.2) {\color{blue}$v_3$};
\node at (0,0.5,-0.3) {\color{RedOrange}$v_4$};

\end{tikzpicture}
\end{center}
\end{minipage}

\caption{Toric diagram for $\mathbb{C}^3/\mathbb{Z}_K$ (left) and two types of (partial) resolution (centre, right). }
\label{fig:Zkres}

\end{figure}
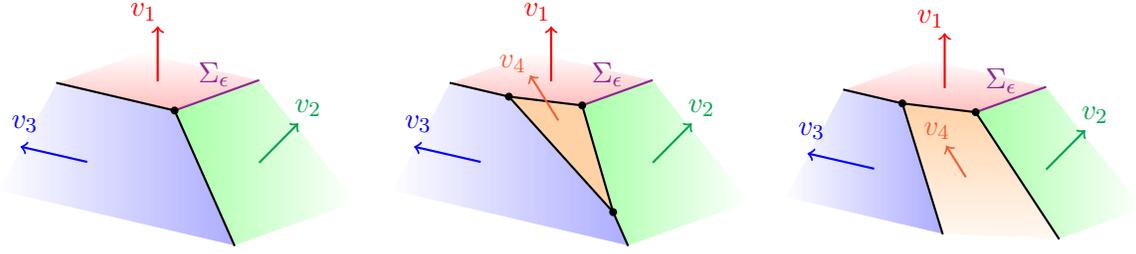

Introducing coordinates $(w,z_1,z_2)\in \C^3$ and writing $w=|w|\me^{\ii \psi}$, $z_i=|z_i|\me^{\ii \varphi_i}$, we may take the following generating vector fields for the $T^3=$U$(1)^3$ action:
\begin{equation}\label{eq:basisZk}
     e_1=\partial_{\varphi_1}\, ,\quad e_2=\partial_{\varphi_2}-\partial_{\varphi_1}\, ,\quad e_3= \frac{1}{K}(\partial_{\psi}+(\alpha_1-K) \partial_{\varphi_1}+\alpha_2\partial_{\varphi_2})\, .
\end{equation}
Note that this basis is normalized on $\mathbb{C}^3/\mathbb{Z}_K$ so that exponentiating all three vectors by $2\pi$ generates an effective torus action and the $(3,0)$-form $\Omega=\dd w\wedge \dd z_1\wedge \dd z_2$ has charge $1$ under $e_1$ and is uncharged under both $e_2$ and $e_3$.
There are three toric divisors, the first and second at $\{z_i=0\}$ with normal vector $v_i=\partial_{\varphi_i}$ respectively and the third at $\{w=0\}$ with normal vector $v_3=\partial_{\psi}$. Inverting the relations \eqref{eq:basisZk} for the basis vectors, one finds the toric vectors
\begin{equation}\label{eq:toricZk}
    v_1=\del_{\varphi_1}=(1,0,0)\, ,\quad v_2=\del_{\varphi_2}=(1,1,0)\, ,\quad v_3=\del_{\psi}=(1,-\alpha_2,K)\, .
\end{equation}
This is a convenient choice of basis which defines the $\mathbb{C}^3/\mathbb{Z}_K$ singularity. 
The degree of the orbifold singularity can be read off from the toric data by taking the determinant of the three vectors, $|\det(v_1,v_2,v_3)|=K$.
The associated toric diagram is the left diagram in figure~\ref{fig:Zkres}. 

We can now ask about resolutions of the above singularity. This is given by a fine regular star triangulation of the 3d polytope, as described in \cite{ROAN1996489} (see also \cite{Eckhard:2020jyr} for a review). Pictorially it amounts to adding planes in figure~\ref{fig:Zkres} (left). One choice is to ``cut the corner" of the singularity, as shown in figure~\ref{fig:Zkres} (centre), however planes could also be added which cut off a line rather than the corner (figure~\ref{fig:Zkres} right) in the spirit of the $\mathcal{N}=2$ case. 
The addition of planes introduces additional nuts, which we label by $a=1,\dots ,d$, which may or may not be smooth points. Such points arise at the intersection of three planes, defined by the normal vectors $v_{\{1,2,3\}}^a$, with the triplet defining a maximal cone in the fan. Points arising from the intersection of four or more planes are not smooth, even in the orbifold sense, and thus we ignore this possibility. 
Between neighbouring points there exist compact edges, defined by the intersection of two planes, and thus two vectors, which give rise to compact two-cycles in the geometry.
We restrict to crepant resolutions in the following which implies that all vectors have first component $1$, 
but are otherwise arbitrary. These then define an orbifold singularity for each maximal cone of degree
\begin{equation}
   d_0^a= |\det(v_1^a,v_2^a,v_3^a)|\equiv k_a\,.
\end{equation}
By adding enough planes the singularity is resolved if all the points are smooth ($k_a=1$ $\forall a$), but this need not be the case in the following. 

\subsection{Local \texorpdfstring{$\mathcal{N}=1$}{N=1} punctures}\label{sec:N=1glue}

The $\mathbb{C}^3/\mathbb{Z}_K$ geometry is the local geometry at one of the poles of the $S^4$ and an arbitrary point on the Riemann surface $\Sigma_g$, where we can take the latter point to be at the centre of a disc with local coordinate $w\in\C$.
We first explain how the partially resolved $\mathbb{C}^3/\Z_K$ geometry is glued in at a pole (say, the north pole), and then how we glue north and south poles together and into the bulk geometry.

We fix the north pole of $S^4$, with local complex coordinates $z_1,z_2$ on the tangent directions and coordinate $w$ on the Riemann surface, and wish to glue in the partially resolved $\C^3/\Z_K$ geometry in figure~\ref{fig:Zkres}. 
This requires identifying the ``bolt'' direction, namely the $\Sigma_\epsilon$ direction. This is an interval closed at one end and open at the other in the toric diagram, where it glues onto the bulk Riemann surface and is shown in figure~\ref{fig:Zkres}. 
With our labelling this edge is specified by the vectors $v_1$, $v_2$. 
We furthermore label the vertices in the partially resolved geometry so that the origin of $\Sigma_\epsilon$ is the vertex $d$, which we take to be 
an orbifold singularity 
of degree  $k_d$. Note that this will locally be described as $\C^3/\Z_{k_d}$, where the quotient in \eqref{eq:SU(3)action} changes after resolution. We denote the new parameters by $\alpha_i^d$, which obey the constraint
\begin{equation}
    \alpha_1^d+\alpha_2^d+1=k_d\,.
\end{equation}
In particular if the singularity is fully resolved, $k_d=1$ such that both $\alpha_i=0$.
Moreover, the first Chern classes of the line bundles over the bolt after resolution are then
\begin{equation}\label{porbN1}
    \int_{\Sigma_\epsilon}c_1(\mathcal{L}_i)=\frac{\alpha_i^d}{k_d}\,.
\end{equation}
The toric vectors at the $d$'th vertex are
\begin{equation}\label{eq:vbolt}
    v_1^d=v_1=(1,0,0)\, ,\quad v_2^d=v_2=(1,1,0)\, ,\quad v_3^d=(1,-\alpha_2^d,k_d)\, .
\end{equation}
The value of $y$ at this point will be denoted $y_d$, and as for the $\mathcal{N}=2$ punctures will play a distinguished role.

To construct the full puncture we need to glue together two copies of this geometry, one at the north pole and one at the south pole of $S^4$. 
In principle we may use different partial resolutions at each pole, and thus all quantities in the above paragraphs will have north and south labels, e.g. $\Sigma_\epsilon^N$, $\Sigma_\epsilon^S$, values of $y$ at the vertices $y_a^N$, $y_a^S$, with the index $a$ running from 1 to $\dN$, $\dS$, etc.\footnote{For fully resolved geometries, with all $k_a=1$, the Calabi--Yau three-folds will necessarily have the same Euler numbers, even if their topologies are different, and hence $\dN=\dS$.} The boundary around this geometry is $S^4\times S^1$, where it glues into the bulk. 
Describing this is quite subtle, since $S^4$ is not toric, so we will carefully discuss this now. 

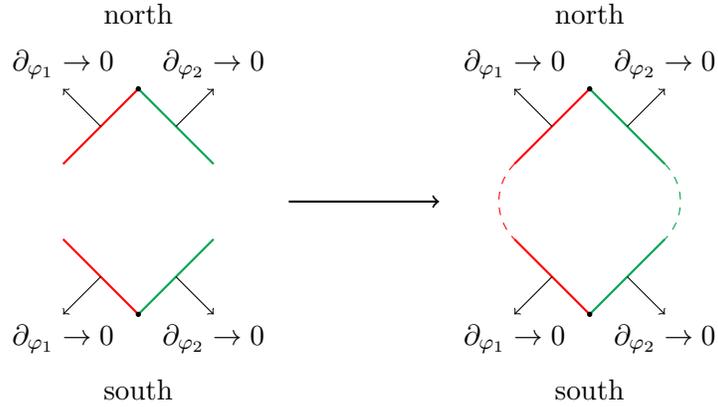
\begin{figure}[ht]
    \centering
\begin{tikzpicture}

\coordinate (origin) at (0,0);
\coordinate (xaxis) at (1,1);
\coordinate (midxaxis) at (1/2,1/2);
\coordinate (yaxis) at (-1,1);
\coordinate (midyaxis) at (-1/2,1/2);

\coordinate (yaxis2) at (-1,2);
\coordinate (xaxis2) at (1,2);
\coordinate (origin2) at (0,3);
\coordinate (midxaxis2) at (1/2,2.5);
\coordinate (midyaxis2) at (-1/2,2.5);


\draw[red,thick] (yaxis)--(origin);
\draw[Green,thick] (origin)--(xaxis);
\draw[->] (midxaxis)--(1,0);
\draw[->] (midyaxis)--(-1,0);
\node[below] at (-1,0) {$\partial_{\varphi_1}\rightarrow 0$};
\node[below ] at (1,0) {$\partial_{\varphi_2}\rightarrow 0$};

\draw[red,thick] (yaxis2)--(origin2);
\draw[Green,thick] (origin2)--(xaxis2);
\draw[->] (midxaxis2)--(1,3);
\draw[->] (midyaxis2)--(-1,3);
\node[above] at (-1,3) {$\partial_{\varphi_1}\rightarrow 0$};
\node[above ] at (1,3) {$\partial_{\varphi_2}\rightarrow 0$};

\node[] at (0,-1) {$\text{south}$};
\node[] at (0,4) {$\text{north}$};

\fill[black] (origin2) circle(1pt) node[] {};
\fill[black] (origin) circle(1pt) node[] {};

\draw[->,thick] (2,1.5)--(4,1.5);

\coordinate (origin3) at (6,0);
\coordinate (xaxis3) at (7,1);
\coordinate (midxaxis3) at (6.5,1/2);
\coordinate (yaxis3) at (5,1);
\coordinate (midyaxis3) at (5.5,1/2);

\coordinate (yaxis4) at (5,2);
\coordinate (xaxis4) at (7,2);
\coordinate (origin4) at (6,3);
\coordinate (midxaxis4) at (6.5,2.5);
\coordinate (midyaxis4) at (5.5,2.5);


\draw[Green,thick] (origin3)--(xaxis3);
\draw[red,thick] (yaxis3)--(origin3);
\draw[->] (midxaxis3)--(7,0);
\draw[->] (midyaxis3)--(5,0);
\node[below] at (5,0) {$\partial_{\varphi_1}\rightarrow 0$};
\node[below ] at (7,0) {$\partial_{\varphi_2}\rightarrow 0$};

\draw[red,thick] (yaxis4)--(origin4);
\draw[Green,thick] (origin4)--(xaxis4);
\draw[->] (midxaxis4)--(7,3);
\draw[->] (midyaxis4)--(5,3);
\node[above] at (5,3) {$\partial_{\varphi_1}\rightarrow 0$};
\node[above ] at (7,3) {$\partial_{\varphi_2}\rightarrow 0$};

\node[] at (6,-1) {$\text{south}$};
\node[] at (6,4) {$\text{north}$};

\fill[black] (origin3) circle(1pt) node[] {};
\fill[black] (origin4) circle(1pt) node[] {};

\draw[dashed,red] (yaxis4) to[out=225,in=135] (yaxis3);
\draw[dashed,Green] (xaxis4) to[out=315,in=45] (xaxis3);
        
\end{tikzpicture}
\caption{We show how we glue the two patches of $\mathbb{C}^2$ together to form a ``toric diagram'' for $S^4$.  Observe that we take a copy of $\mathbb{C}^2$, truncated in extent for artistic reasons, and take a mirror image. We then join the lines corresponding to the \emph{same} degenerating vector field together. In this way no new fixed points are generated. The result is an ellipse, where the red and green edges correspond to copies of $S^2_i\subset \C_i\oplus \R$, for $i=1,2$, respectively, thinking of $S^4\subset \C_1\oplus\C_2\oplus \R$. These two-spheres meet at the two poles of $S^4$. When computing the weights we must take into account that 
when gluing we need to reverse the orientation in the southern patch, giving a mirror image, in order to obtain a consistent overall orientation.}
\label{fig:S4}
\end{figure}

We begin by describing an $S^4$ in terms of a ``toric diagram" before explaining how to construct the puncture geometries. We think of an $S^4$ topologically as two copies of $\mathbb{C}^2$ glued together (non-holomorphically), which we denote by north and south. We take the coordinates on $\mathbb{C}^2$ to be $z_i=|z_i|\me^{\ii \varphi_i}$. The toric diagram of $\mathbb{C}^2$ is simply a 2d ``corner'' (the positive quadrant in $\R^2$), see figure~\ref{fig:S4}, with the faces being the loci $z_i=0$. To construct the $S^4$ we take a second, mirror copy of $\mathbb{C}^2$, and identify the non-compact edges in the two diagrams, see figure~\ref{fig:S4}. This compactifies the geometry, and since we are identifying the edges does not introduce any additional fixed points.\footnote{The more canonical way of gluing toric diagrams together introduces new fixed points because one does not identify the same faces in the way we do here.}
Note that it is important that we identify the \emph{same} faces, i.e. the normal vectors. However, 
we will see below that once we partially resolve the copies of $\C^3/\Z_\K$ at the poles of $S^4$ this does not require us to take a mirror copy of the toric diagram. 

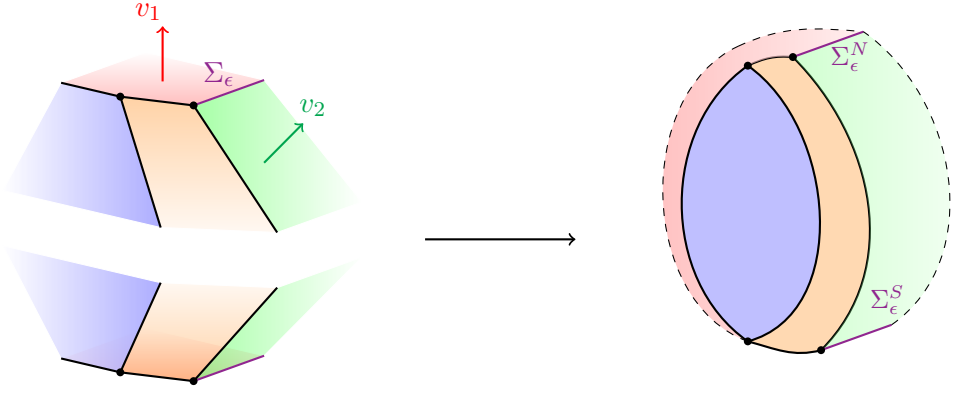
\begin{figure}[ht]
    \centering

 \begin{minipage}{.32\textwidth}
\tdplotsetmaincoords{78}{-60}

\begin{center}
\begin{tikzpicture}
		[tdplot_main_coords,
			cube/.style={very thick,black},
			axisb/.style={->,blue,thick},
            axisr/.style={->,red,thick},
            axisg/.style={->,Green,thick},
			inf/.style={dashed,black},scale=.75,every node/.style={scale=1.}]

\coordinate (Norigin) at (0.5,0,0);

\shadedraw[white,top color=white, bottom color=red!50,fill opacity=0.5] (0.5,0,0)--(3,0,0)--(4,3,0)--(1,3,0)--(.5,1.5,0)--(.5,0,0);
\shadedraw[white,right color=blue!40, left color=white,fill opacity=0.8] (.5,1.5,0)--(1,3,0)--(2/3,4,-2)--(-2/3,0,-2)--(.5,1.5,0);
\shadedraw[white,left color=Green!70, right color=Green!0,fill opacity=0.5] (.5,0,0)--(3,0,0)--(3,-2,-2)--(0,-2,-2)--(.5,0,0);
\shadedraw[white,top color=orange!70, bottom color=orange!10,fill opacity=0.5] (.5,0,0)--(.5,1.5,0)--(-2/3,0,-2)--(0,-2,-2)--(.5,0,0);

\draw[black, thick](.5,1.5,0)--(1,3,0);
\draw[Plum,thick](.5,0,0)--(3,0,0);
\draw[black,thick](.5,0,0)--(.5,1.5,0)--(-2/3,0,-2);
\draw[black,thick] (0,-2,-2)--(.5,0,0);

\shadedraw[white,top color=white, bottom color=red!50,fill opacity=0.5] (0.5,0,-5)--(3,0,-5)--(4,3,-5)--(1,3,-5)--(.5,1.5,-5)--(.5,0,-5);
\shadedraw[white,right color=blue!40, left color=white,fill opacity=0.8] (.5,1.5,-5)--(1,3,-5)--(2/3,4,-3)--(-2/3,0,-3)--(.5,1.5,-5);
\shadedraw[white,left color=Green!70, right color=Green!0,fill opacity=0.5] (.5,0,-5)--(3,0,-5)--(3,-2,-3)--(0,-2,-3)--(.5,0,-5);
\shadedraw[white,bottom color=orange!70, top color=orange!10,fill opacity=0.5] (.5,0,-5)--(.5,1.5,-5)--(-2/3,0,-3)--(0,-2,-3)--(.5,0,-5);

\draw[black, thick](.5,1.5,-5)--(1,3,-5);
\draw[Plum,thick](.5,0,-5)--(3,0,-5);
\draw[black,thick](.5,0,-5)--(.5,1.5,-5)--(-2/3,0,-3);
\draw[black,thick] (0,-2,-3)--(.5,0,-5);

\draw[axisg] (3,0,-1.5)-- (3,-.8,-0.7);
\draw[axisr] (2.,1.5,0)--(2.,1.5,1);

\fill[black] (.5,0,0) circle(2pt) node[below] {}; 
\fill[black] (.5,1.5,0) circle(2pt) node[below] {}; 
\fill[black] (.5,0,-5) circle(2pt) node[below] {}; 
\fill[black] (.5,1.5,-5) circle(2pt) node[below] {}; 

\node[Plum] at (1.4,0,.4) {$\Sigma_\epsilon$};

\node at (1.5,1.5,1.35) {\color{red}$v_1$};
\node at (3.,-1.,-0.4) {\color{Green}$v_2$};

\end{tikzpicture}

\end{center}
\end{minipage}
\begin{minipage}{.2\textwidth}
\begin{center}

\begin{tikzpicture}
\node at (0,0) {};
\draw[->,thick] (0,-1)--(2,-1);
    
\end{tikzpicture}

\end{center}
\end{minipage}
\begin{minipage}{.32\textwidth}
\begin{center}
\tdplotsetmaincoords{78}{-60}
\begin{tikzpicture}
	[tdplot_main_coords,
			cube/.style={very thick,black},
			axisb/.style={->,blue,thick},
            axisr/.style={->,red,thick},
            axisg/.style={->,Green,thick},
			inf/.style={dashed,black},scale=.75,every node/.style={scale=1.}]

\fill[blue!50,opacity=0.5] (0.5,1.5,0) to[out=215,in=145] (0.5,1.5,-5)--(0.5,1.5,-5) to[out=15,in=325] (0.5,1.5,0);


\fill[orange!60,opacity=0.5] (0.5,1.5,0) to[out=25,in=175]   (-0.5,0,0.5)--(-0.5,0,0.5)  to[out=-45,in=45] (0.5,0,-5)--(0.5,0,-5)  to[out=195,in=-15]  (0.5,1.5,-5)--  (0.5,1.5,-5) to[out=15,in=325] (0.5,1.5,0) ;

\draw[-,thick,black] (0.5,1.5,0) to[out=25,in=175]   (-0.5,0,.5);
\draw[-,thick,black] (0.5,0,-5)  to[out=195,in=-15]  (0.5,1.5,-5);
\draw[-,thick,black] (-0.5,0,0.5)  to[out=-45,in=45] (0.5,0,-5) ;
\draw[-,thick,black] (0.5,1.5,-5) to[out=15,in=325] (0.5,1.5,0) ;

\shade[green!10,left color=green!60,  right color=green!10,fill opacity=0.3] (3,0,-5) to [out=35,in=325] (2,0,.5)-- (-0.5,0,0.5) to[out=-45,in=45] (0.5,0,-5) --(3,0,-5)to [out=35,in=325] (2,0,.5);


\fill[red!10,left color=red!60,  right color=red!10,fill opacity=0.4] (2,0,0.5)--    (-0.5,0,0.5) to[out=175,in=25] (0.5,1.5,0) to[out=215,in=145] (0.5,1.5,-5) to[out=160,in=205]  (0.5,1.7,0.33) to[out=25,in=175] (2.,0,0.5);

\draw[-,thick,black] (0.5,1.5,0) to[out=215,in=145] (0.5,1.5,-5);

\draw[dashed,black] (0.5,1.5,-5) to[out=160,in=205] (0.5,1.7,0.33) ;
\draw[dashed, black]  (0.5,1.7,0.33) to[out=25,in=175] (2.,0,0.5);

\draw[-,thick,Plum](-0.5,0,0.5)--(2.,0,0.5) node[below] {\small $\Sigma_\epsilon^N\quad$};
\draw[-,thick,Plum](0.5,0,-5)--(3,0,-5) node[above] {\small $\Sigma_\epsilon^S\,\,$};

\draw[dashed, black] (3,0,-5) to [out=35,in=325] (2,0,0.5);

\fill[black] (-.5,0,0.5) circle(2pt) node[below] {}; 
\fill[black] (.5,1.5,0) circle(2pt) node[below] {}; 
\fill[black] (.5,0,-5) circle(2pt) node[below] {}; 
\fill[black] (.5,1.5,-5) circle(2pt) node[below] {}; 

\end{tikzpicture}

\end{center}
\end{minipage}

\caption{We depict the gluing of two copies of toric diagrams. Observe that the red and green faces are the loci where the two U$(1)$'s of the $S^4$ shrink, and are determined by the choice of R-symmetry vector. 
These are glued together as shown. 
The other non-compact cycles are required to match in such a way that they glue together into compact four-cycles. }

\label{fig:S4fromC3}
\end{figure}

Having explained how to construct an $S^4$ we now proceed to construct the puncture geometries. At each of the poles of the $S^4$ and centre of the disc in the Riemann surface,
 we have a copy of $\mathbb{C}^3/\mathbb{Z}_K$. We want to glue these together to obtain the puncture geometry. We require that the boundary is $S^4\times S^1$ which uniquely fixes how we glue the two pieces together. Consider gluing two copies of $\mathbb{C}^3/\mathbb{Z}_K$ or their resolutions. We take $v_1$ and $v_2$ to be the normals to the faces where $\partial_{\varphi_1}$ and $\partial_{\varphi_2}$ shrink respectively -- these are determined unambiguously since we fix the R-symmetry using these vector fields. Following the prescription for constructing an $S^4$ we need to identify these faces. We must also identify any other non-compact faces so that the only remaining non-compact faces are those normal to $v_1$ and $v_2$, which define the $S^4$ coordinates, see figure \ref{fig:S4fromC3}. Observe that the requirement for the gluing is that the non-compact faces agree: this does not impose any constraints on the compact faces, and these may differ between the north and south geometries, cf.\ figure~\ref{fig:toric3D}.

The R-symmetry vector is given by
\begin{equation}
    \xi=b_1\del_{\varphi_1}+b_2\del_{\varphi_2}=(b_1+ b_2, b_2,0)\,,
\end{equation}
where we used \eqref{eq:toricZk} in the second equality.
The action of $\xi$ 
generically has fixed points at  the intersection of three planes defined by a maximal cone. Computing the weights at the fixed points in the north patch is simple using the formula 
\begin{equation}\label{weightsnutsN1}
    (\epsilon_1^a,\epsilon_2^a,\epsilon_3^a) =\xi\cdot(v_1^a,v_2^a,v_3^a)^{-1}=\frac{1}{k_a}(\det(\xi,v_2^a,v_3^a),\det(v_1^a,\xi,v_3^a),\det(v_1^a,v_2^a,\xi))\,.
\end{equation}
One needs to be careful when computing the weights in the south patch. Following the bulk computation in section~\ref{sec:M5review}, the product of the weights at an isolated fixed point in the south patch are taken to be the negative of the product of the weights given by the above formula \eqref{weightsnutsN1}, i.e.\ minus the product of weights in the north patch, due to the orientation reversal. This extends to fixed points of cycles too. For the southern bolt the weights are slightly more subtle. The product of the two non-trivial weights picks up a sign as before, and in addition one of the two first Chern classes of the normal bundles, which are used in the BVAB theorem, also picks up an additional sign correlated with the weights (as in section~\ref{sec:M5review}). The weights at the bolt fixed point sets $\Sigma_\epsilon^N$, 
$\Sigma_\epsilon^S$
may hence be taken to be
\begin{equation}
       (\epsilon_1^d,\epsilon_2^d,\epsilon_3^d) =\xi\cdot(v_1^d,v_2^d,v_3^d)^{-1}=(b_1,\pm b_2,0)\,,
\end{equation}
taking the $+$ sign for the northern bolt and $-$ for the southern bolt. In order to glue the puncture geometry into the bulk we must align the bolts with the poles of the $S^4$ fixing $\hat{y}_{\dN}^N=\hat{y}_N=N$ and $\hat{y}_{\dS}^S=\hat{y}_S=-\hat{y}_N=-N$,  where we have again defined the rescaled variables
\begin{equation}\label{eq:yhatdef}
    \hat{y}_a\equiv \frac{y_a}{9 \pi \lp^3 b_1 b_2}\, .
\end{equation}
This follows since the flux of $G$ through the $S^4$ fibre over a point on the Riemann surface \eqref{eq:S4flux} is independent of the latter choice of point. Therefore, in order to glue the 
$S^4$ fibres over $\Sigma_\epsilon$ with the $S^4$ fibres over the bulk Riemann surface we must impose this consistency condition where they meet.

As a warm-up let us analyse the unresolved geometry. This will recover the $\mathcal{N}=1$ orbifold analysis in section \ref{sec:orbi2},  but in doing so this sets up the analysis of the more general punctures. First notice that since the boundary is $S^4\times S^1$ there are two linearly embedded $S^2$'s which are trivial in homology. Integrating the two-form $Y$ over these two cycles implies that $\hat{y}_S^2=\hat{y}_N^2$, and further requiring that the conformal dimension of the operator dual to this two cycle is non-zero, implies $\hat{y}_N=-\hat{y}_S$. This will be a generic feature of all the puncture geometries that we will consider; that is, the ``origin'' of the bolt in the north and south patch satisfies $N=\hat{y}_N=-\hat{y}_S= \hat{y}_1^N=-y_1^S\equiv y_1$, where note that with no partial resolution $\dN=\dS=1$. 

There are four U$(1)^2$-invariant four-cycles in the puncture geometry (three of which are immediately visible in figure \ref{fig:N1res} for $K=3$). The first four-cycle is the smooth $S^4$ on the boundary through which the flux is $\hat y_1=N$ as fixed by the bulk analysis. There is also a quotiented copy of the $S^4$ at the centre of the disc, the blue face of figure \ref{fig:N1res}, with flux
\begin{equation}
    \frac{1}{(2\pi\lp)^3}\int_{S^4/\mathbb{Z}_K}G=\frac{N}{K}\,.
\end{equation}
Finally, there are two non-compact four-cycles which are $S^2$ orbibundles over the bolt $\Sigma_\epsilon$. These non-compact four-cycles are the faces with normals $v_1$ and $v_2$ which we denote by $D^{[1]}_\epsilon$ and $D^{[2]}_\epsilon$ respectively. From \eqref{GfluxDirac} we have 
\begin{equation}
    \frac{1}{(2\pi\lp)^3}\int_{D^{[1]}_\epsilon} G=\frac{\alpha_1}{K}N\,,
\end{equation}
whilst localization gives the formula
\begin{equation}
    \frac{1}{(2\pi\lp)^3}\int_{D^{[1]}_{\epsilon}} G=\frac{2}{(2\pi\lp)^3}\frac{2\pi}{b_2}\bigg[\int_{\Sigma_{\epsilon}}\frac{\omega}{6y_1}-\frac{2\pi}{b_2}\frac{ \alpha_2}{K}\frac{y_1}{9}\bigg]\,,
\end{equation}
where we used \eqref{intc1Li} and the same identities hold for $1\leftrightarrow 2$. 
Combining these gives
\begin{equation}\label{eq:omegaorbiN1}
    \int_{\Sigma_\epsilon} \omega =108\pi^3\lp^6 b_1 b_2   \left( b_1\frac{\alpha_2}{ K}+ b_2\frac{\alpha_1}{ K}\right)N^2\,,
\end{equation}
and the central charge is
\begin{align}\label{N1orby}
        \delta a^\mathrm{orb}
        &=\frac{2}{2(2\pi)^6\lp^9} \frac{(2\pi)^2}{b_1b_2}\int_{\Sigma_\epsilon}\left[-\frac{y_1}{36}\omega+2\pi\Big(\frac{c_1(\mathcal{L}_1)}{b_1}+\frac{c_1(\mathcal{L}_2)}{b_2}\Big)\frac{y_1^3}{162}\right] \nn \\ 
        &=-\frac{9}{8}b_1b_2\Big(b_1\frac{\alpha_2}{K}+b_2\frac{\alpha_1}{K}\Big)N^3\,,
    \end{align}
reproducing \eqref{eq:aorb}. We now turn to general $\mathcal{N}=1$ punctures.

\subsection{Flux quantization}

There are two types of compact cycles in the above puncture geometries: those contained in a single copy of the toric geometry, and those extending between the north and south geometries. The former include an arbitrary number of $y_a^N$ (or $y_a^S$), while the latter have contributions from both north and south vertices. 
A generic cycle (of either type) $D_A$ with normal vector $v_A$ is enclosed by a set of vertices $y_a$, such that $v^a_i=v_A$ for one of the three $i$, i.e.\ $v_A$ is in the cone corresponding to the fixed point $y_a$. We denote the set of such vertices by $a\in\mathcal{I}_A$, where by an abuse of notation $a$ also labels whether $y$ is in the $N$ or $S$ copy. The quantization of the flux through $D_A$ then reads 
\begin{equation}\label{generalfluxN1}
    N_A=\frac{1}{(2\pi\lp)^3}\int_{D_A}G=\frac{1}{(2\pi\lp)^3}\sum_{a\in\mathcal{I}_A} \frac{1}{k_a}\frac{(2\pi)^2}{\epsilon_{i_a}^a\epsilon_{j_a}^a}\frac{y_a}{9}\, .
\end{equation}
Here the two weights $\epsilon_{i_a}^a, \epsilon_{j_a}^a$ entering at each nut are those involving $v_A$ in their determinant (see \eqref{weightsnutsN1}).
Unlike the $\mathcal{N}=2$ punctures, we cannot solve these linear constraints on the $y_a$ iteratively and obtain a general expression for $\hat y_a$ in terms of only $N_A$ -- as we will see, there are not enough equations. 
We will comment on this further in the next subsection.

Finally one can also quantize the flux through $D^{[i]}_\epsilon$, the four-cycles normal to $v_i$. These pick up contributions from the bolts at $\Sigma_\epsilon^N$ and $\Sigma_\epsilon^S$. Note that  
$y_{\dN}^N=-y_{\dS}^S\equiv y_d$ and $b_2^N=-b_2^S\equiv b_2$. On the other hand the line bundles may be different. There are also nut contributions from the vertices surrounding the face $D^{[i]}_\epsilon$, i.e.\ the isolated fixed points $y_a$ which have $v_i$ in their cone. We denote this set by $\mathcal{I}^i$. Then localization gives
\begin{align}\label{fluxD1N1}
    N^{[1]}_\epsilon\equiv\frac{1}{(2\pi\lp)^3}\int_{D^{[1]}_\epsilon} G&=\frac{1}{(2\pi\lp)^3}\left\{\frac{4\pi}{b_2}\bigg[\frac{1}{2}\int_{\Sigma_{\epsilon}^N\cup\Sigma_\epsilon^S}\frac{\omega}{6y_d}-\frac{2\pi }{b_2}\frac{y_d}{9}\porb_2\bigg]+\sum_{a\in \mathcal{I}^1} \frac{1}{k_a}\frac{(2\pi)^2}{\epsilon_1^a\epsilon_2^a}\frac{y_a}{9}\right\}\nn \\  &=\frac{1}{108\pi^3\lp^6b_1 b_2^2 N } \frac{1}{2}\int_{\Sigma_{\epsilon}^N\cup \Sigma_\epsilon^S}\omega-\frac{b_1}{b_2}\Big(\porb_2 N+\pnut_1\Big)\,,
\end{align}
where we define the quantities
\begin{equation}
    \porb_i\equiv\frac{1}{2}\int_{\Sigma_\epsilon^N\cup\Sigma_\epsilon^S}c_1(\mathcal{L}_i)=\frac{1}{2}\left(\int_{\Sigma_\epsilon^N}c_1(\mathcal{L}_i)+\int_{\Sigma_\epsilon^S}c_1(\mathcal{L}_i)\right)=\frac{1}{2}\left(\frac{\alpha_i^{\dN}}{k_{\dN}}+\frac{\alpha_i^{\dS}}{k_{\dS}}\right)\,,
\end{equation}
(the factor $1/2$ is included in order to have $\piorb=\alpha_i^d/k_d$ in case of identical copies), 
and also
\begin{equation}\label{defn1}
    \pnut_1\equiv -\frac{b_2}{b_1}\frac{1}{(2\pi\lp)^3}\sum_{a\in \mathcal{I}^1} \frac{1}{k_a}\frac{(2\pi)^2}{\epsilon_1^a\epsilon_2^a}\frac{y_a}{9}=-\sum_{a\in \mathcal{I}^1} \frac{b_2^2}{2\epsilon_1^a\epsilon_2^a}\frac{\hat{y}_a}{k_a}\,.
\end{equation}
For $D^{[2]}_\epsilon$ we again get the flux integral \eqref{fluxD1N1} with $1\leftrightarrow 2$.
These relations for $N^{[1]}_\epsilon$  and $N^{[2]}_\epsilon$ then respectively give 
\begin{equation}\label{eq:JbulkpunctN11}
    \frac{1}{2}\int_{\Sigma_\epsilon^N\cup\Sigma_\epsilon^S}\omega=108\pi^3\lp^6b_1b_2\left[b_1 \big(\porb_2 N+\pnut_1\big)+b_2N^{[1]}_\epsilon\right]N\,,
\end{equation}
and
\begin{equation}\label{eq:JbulkpunctN12}
    \frac{1}{2}\int_{\Sigma_\epsilon^N\cup\Sigma_\epsilon^S}\omega=108\pi^3\lp^6b_1b_2\left[b_1 N^{[2]}_\epsilon+b_2\big(\porb_1N+\pnut_2\big)\right]N\,.
\end{equation}
The two expressions for $\int\omega$ should be the same, such that the two non-compact fluxes are not independent 
\begin{equation}\label{homologyN1}
     b_2N^{[1]}_\epsilon+b_1 \Big(\porb_2 N+\pnut_1\Big)=b_1 N^{[2]}_\epsilon+b_2\Big(\porb_1 N+\pnut_2\Big)\,.
\end{equation}

Note that the $\mathcal{N}=2$ discussion 
perfectly fits in this picture. First the north and south copies are identical such that $\piorb=\alpha_i^d/k_d$. Then recall that supersymmetry imposes 
$\porb_1=1-\frac{1}{k_d}$, $\porb_2=0$, and $N_\epsilon^{[2]}=0$. Moreover the sums over nut contributions give 
\begin{equation}\label{bob}
   \pnut_1=0\,,\quad \pnut_2 = \sum_{a=1}^{d-1}\frac{k_a\hat y_a}{l_a l_{a+1}}=\frac{N}{k_d}-n_d \,,  
\end{equation}
such that,
\begin{equation}\label{harriet}
    N_\epsilon^{[1]}=\porb_1 N+\pnut_2=N-n_d\,.
\end{equation}
Here $\mathcal{N}=2$ supersymmetry implies (by the SO$(3)_R$ symmetry) 
that 
the north and south copies have to be identical, and in the sum over nuts in $\pnut_2$ one obtains the same contribution twice. The key element here is to be able to evaluate the sum in $\pnut_2$, which requires a dictionary between $y_a$ and $n_a$ which is not available (in general) for $\mathcal{N}=1$ punctures. Notice
in particular that 
the $\mathcal{N}=1$ result depends on one of the non-compact fluxes
(say $N_\epsilon^{[1]}$). 

\subsection{Observables}

Finally, we can compute the contribution of an $\mathcal{N}=1$ puncture to the central charge. The same comments as for \eqref{fluxD1N1} about the bolt contributions apply here also, and we obtain 
\begin{align}\label{N1punct}
        \delta a
        &=\frac{1}{2(2\pi)^6\lp^9} \left[\frac{2(2\pi)^2}{b_1b_2}\left[-\frac{1}{2}\int_{\Sigma_\epsilon^N\cup\Sigma_\epsilon^S}\frac{y_d}{36}\omega+2\pi\Big(\frac{\porb_1}{b_1}+\frac{\porb_2}{b_2}\Big)\frac{y_d^3}{162}\right]-\sum_a \frac{1}{k_a}\frac{(2\pi)^3}{\epsilon_1^a\epsilon_2^a\epsilon_3^a}\frac{y_a^3}{162}\right] \nn \\ 
        &=-\frac{9}{16} b_1 b_2\left[3\Big(b_1\pnut_1+b_2N^{[1]}_\epsilon\Big)N^2+\Big(2b_1\porb_2-b_2\porb_1\Big)N^3+\frac{1}{2}\sum_a \frac{1}{k_a} \frac{b_1^2b_2^2}{\epsilon_1^a\epsilon_2^a\epsilon_3^a}\hat y_a^3\right]\,.
    \end{align}
Here the sum runs over all isolated fixed points, the weights $\epsilon^a_i$ are given in \eqref{weightsnutsN1}, and $\mathfrak{n}_1$ in \eqref{defn1}. 
The integral of $\omega$ in \eqref{N1punct} has been evaluated using \eqref{eq:JbulkpunctN11}. One could have equivalently used \eqref{eq:JbulkpunctN12}. It is this choice that creates the asymmetry between $b_1$ and $b_2$ in the result on the second line of \eqref{N1punct}. 
Recall from the discussion in section~\ref{sec:orbi2} that the fluxes $N_\epsilon^{[i]}$
are not quantized or gauge-invariant, only becoming quantized and
gauge-invariant when added to the rest of the bulk flux along these cycles. 
They should be regarded as part of the boundary data for a particular puncture, which we then glue to the bulk.

The $\hat y_a$ are functions of the fluxes through \eqref{generalfluxN1}.
Focusing on the resolved geometry at (say) the north pole, for a fully resolved geometry $\mathcal{X}$ we have $\dN=\chi(\mathcal{X})=1+b_2(\mathcal{X})+b_4(\mathcal{X})$, where 
$b_i(\mathcal{X})$ denote Betti numbers. 
There are $b_4(\mathcal{X})$ compact four-cycles, and hence \eqref{generalfluxN1} allows one to eliminate $b_4(\mathcal{X})$ of the $\hat{y}_a$ variables 
associated to that pole. Given that also $\hat{y}^N_{\dN}=N$, this generically leaves $b_2(\mathcal{X})$ of the $\hat{y}_a$ undetermined. 
As we will see in the examples, and as summarized in the introduction, the remaining 
$b_2(\mathcal{X})$ variables should be extremized over, and this is equivalent to imposing that $\Phi^{*G}$ (or equivalently $\Phi^Y$) is equivariantly closed at the level of the $\hat{y}_a$ variables. 
More precisely, we find that the quadratic equations one obtains by extremizing $\delta a$ over the $b_2(\mathcal{X})$ unconstrainted $\hat{y}_a$ variables are equivalent to 
equations obtained by integrating $\Phi^Y$ 
over all toric two-cycles using localization, and imposing that the latter are related by their homology relations!

Finally the total central charge with $n$ punctures is 
\begin{equation}\label{eq:afull}
   a=\abulk - \sum_{I=1}^n \delta a^{I}\, ,
\end{equation}
with $\abulk$ the central charge of a smooth Riemann surface given in \eqref{eq:abulk}, 
and $\delta a^I$ given by \eqref{N1punct}. 
Only after adding all contributions should one then 
extremize the central charge over the $b_i$, subject to the constraint $b_1+b_2=1$, to find 
the on-shell central charge. 

From the general $\mathcal{N}=1$ puncture expression \eqref{N1punct} it is straightforward to recover the previous cases we have analysed, including the  orbifold and $\mathcal{N}=2$ results as special cases.
Indeed, the unresolved orbifold case simply corresponds to turning off the nut contributions, consequently also setting  $\mathfrak{n}_i=0$, and taking 
$N_\epsilon^{[i]}=\porb_i N = \frac{\alpha_i}{K} N$
-- then \eqref{N1punct} immediately reduces to \eqref{N1orby}. 
Alternatively, for (1,0) $\mathcal{N}=2$ punctures we may substitute using 
\eqref{bob}, \eqref{harriet}, to reproduce \eqref{N2punct}. 

The final observables we wish to compute are the conformal dimensions of BPS operators dual to M2-branes wrapped on two-cycles. On every toric two-cycle $\Sigma$ (i.e.\ every edge/line of the doubled toric diagram) one can wrap an M2-brane. For calibrated submanifolds $\Sigma$ this corresponds to a BPS operator in the field theory, whose conformal dimension 
$\Delta_\Sigma$ is given in equation \eqref{fixedstuff}.  
One possibility is to wrap an M2-brane on a copy of the punctured bolt, at the poles of the $S^4$. Using 
equation \eqref{eq:JbulkpunctN11} we may compute
\begin{align}
\frac{1}{2}\left(\Delta\big[\Sigma_{g, \mathrm{punct}}^N\big] +\Delta\big[\Sigma_{g,\mathrm{punct}}^S\big]\right)=-\frac{3}{2}(b_1\mathfrak{p}^{\mathrm{bolt}}_2+b_2 \mathfrak{p}^{\mathrm{bolt}}_1)N\, ,
\end{align}
where we have defined
\begin{align}
\mathfrak{p}^{\mathrm{bolt}}_1 \equiv 
p_1^{\mathrm{bulk}} + \sum_{I=1}^n \frac{N_\epsilon^{[1],I}}{N}\, , \qquad  \mathfrak{p}^{\mathrm{bolt}}_2 \equiv 
p_2^{\mathrm{bulk}} + \sum_{I=1}^n \left(p_2^{\mathrm{orb},I} + \frac{\mathfrak{n}_1^I}{N}\right) \, ,
\end{align}
which is a straightforward generalization of the $\mathcal{N}=2$ result.

There are also two-cycles which are acted on non-trivially by the R-symmetry, where only the last term in the expression for $\Delta[\Sigma]$ in \eqref{fixedstuff} enters. 
A compact cycle will necessarily have the topology of a weighted projective space 
$\mathbb{WCP}^1_{[k_a,k_b]}$, 
where the the poles/fixed points are associated to $\hat{y}_a$ and $\hat{y}_b$, 
which are orbifold singularities of degrees $k_a, k_b\in\mathbb{N}$, respectively. The R-symmetry rotates the weighted projective space, fixing its poles.
Notice these latter fixed points can either be both in the north or both in the south patches, or one in each. We then have
the general formula
\begin{align}\label{Deltagen}
  \Delta[\Sigma] = 
  -\frac{3}{2}b_1b_2\left(\frac{\hat{y}_a}{k_a\epsilon_a}+\frac{\hat{y}_b}{k_b\epsilon_b}\right)\, .
\end{align}
Here $\epsilon_a$, $\epsilon_b$ are the weights of the R-symmetry vector on 
the tangent space to this two-cycle, where note that necessarily $k_a\epsilon_a=-k_b\epsilon_b$. More precisely, the weight $\epsilon_a$ that enters out of $\epsilon^a_{1,2,3}$ defined in \eqref{weightsnutsN1} is precisely that which does not involve the pair of normal vectors that define the edge connecting $\hat{y}_a$ to $\hat{y}_b$. The $\mathcal{N}=2$ results we presented are special cases of the more general formula \eqref{Deltagen}.

\subsection{Examples}\label{sec:Z3ex}

In this subsection we will apply the general procedure outlined in the previous subsections to 
two examples.

\subsubsection{\texorpdfstring{$\mathbb{Z}_3$}{Z3} quotient}

\begin{figure}[ht!]

 \begin{minipage}{.48\textwidth}

\begin{center}
\tdplotsetmaincoords{78}{-60}

\begin{tikzpicture}
		[tdplot_main_coords,
			cube/.style={very thick,black},
			axisb/.style={->,blue,thick},
            axisr/.style={->,red,thick},
            axisg/.style={->,Green,thick},
			inf/.style={dashed,black}]

\shadedraw[white,top color=white, bottom color=red!50,fill opacity=0.5] (0,0,0)--(3,0,0)--(4,3,0)--(1,3,0)--(0,0,0);
\shadedraw[white,right color=blue!40, left color=white,fill opacity=0.8] (0,0,0)--(1,3,0)--(2/3,4,-2)--(-4/3,-2,-2)--(0,0,0);
\shadedraw[white,left color=Green!70, right color=Green!0,fill opacity=0.5] (0,0,0)--(3,0,0)--(3,-2,-2)--(-4/3,-2,-2)--(0,0,0);


\draw[black, thick](-4/3,-2,-2)--(0,0,0);
\draw[black, thick](0,0,0)--(1,3,0);
\draw[Plum,thick](0,0,0)--(3,0,0);

\draw[axisg] (3,0,-1.5)--(3,-.8,-0.7);
\draw[axisb] (-0.5,1.5,-1)--(-2,2.,-0.5);
\draw[axisr] (2,1.5,0)--(2,1.5,1);

\fill[black] (0,0,0) circle(2pt) node[below] {}; 
\node[Plum] at (1.4,0,.3) {$\Sigma_\epsilon$};

\node at (1.5,1.5,1.5) {\color{red}$v_1=(1,0,0)$};
\node at (3.,-1.,-0.4) {\color{Green}$v_2=(1,1,0)$\textcolor{white}{push}};
\node at (-1.5,2.2,-0.2) {\color{blue}$v_3=(1,-1,3)$};

\end{tikzpicture}
\end{center}
\end{minipage}~
 \begin{minipage}{.48\textwidth}

\begin{center}
\tdplotsetmaincoords{78}{-60}

\begin{tikzpicture}
		[tdplot_main_coords,
			cube/.style={very thick,black},
			axisb/.style={->,blue,thick},
            axisr/.style={->,red,thick},
            axisg/.style={->,Green,thick},
			inf/.style={dashed,black}]

\shadedraw[white,top color=white, bottom color=red!50,fill opacity=0.5] (.5,0,0)--(3,0,0)--(4,3,0)--(1,3,0)--(.5,1.5,0)--(.5,0,0);
\shadedraw[white,right color=blue!40, left color=white,fill opacity=0.8] (.5,1.5,0)--(1,3,0)--(2/3,4,-2)--(-4/3,-2,-2)--(-1,-1.5,-1.5)--(.5,1.5,0);
\shadedraw[white,left color=Green!70, right color=Green!0,fill opacity=0.5] (.5,0,0)--(3,0,0)--(3,-2,-2)--(-4/3,-2,-2)--(-1,-1.5,-1.5)--(.5,0,0);
\fill[orange!70,opacity=.5] (.5,0,0)--(.5,1.5,0)--(-1,-1.5,-1.5)--(.5,0,0);

\draw[black, thick](-4/3,-2,-2)--(-1,-1.5,-1.5);
\draw[black, thick](.5,1.5,0)--(1,3,0);
\draw[Plum,thick](.5,0,0)--(3,0,0);
\draw[black,thick](.5,0,0)--(.5,1.5,0)--(-1,-1.5,-1.5)--(.5,0,0);

\draw[axisg] (3,0,-1.5)--(3,-.8,-0.7);
\draw[axisb] (-0.5,1.5,-1)--(-2,2.,-0.5);
\draw[axisr] (2.,1.5,0)--(2.,1.5,1);
\draw[->,RedOrange,thick] (0,.2,-.2)--(-1,.2,.8);

\fill[black] (.5,0,0) circle(2pt) node[below] {$\qquad y_3$}; 
\fill[black] (.5,1.5,0) circle(2pt) node[below] {$y_2\quad$}; 
\fill[black] (-1,-1.5,-1.5) circle(2pt) node[left] {$y_1$}; 
\node[Plum] at (1.4,0,.3) {$\Sigma_\epsilon$};

\node at (1.5,1.5,1.5) {\color{red}$v_1=(1,0,0)$};
\node at (3.,-1.,-0.4) {\color{Green}$v_2=(1,1,0)$};
\node at (-1.5,2.2,-0.2) {\color{blue}$v_3=(1,-1,3)$};
\node at (-1,1.5,1) {\color{RedOrange}$v_4=(1,0,1)$};

\end{tikzpicture}
\end{center}
\end{minipage}

\caption{Toric diagram for $\mathbb{C}^3/\mathbb{Z}_3$ (left) and its $\mathcal{O}(-3)\rightarrow \mathbb{CP}^2$ resolution (right).}
\label{fig:toricC3}

\end{figure}
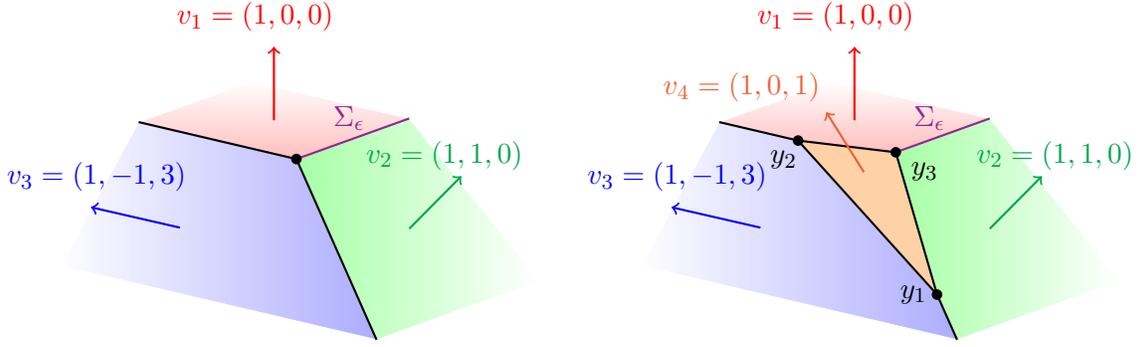

\begin{figure}[ht!]

   \begin{minipage}{0.49\textwidth}
\begin{center}
\pgfdeclarelayer{pre main}
\begin{tikzpicture}
 \pgfsetlayers{pre main,main}
 
    \shadedraw[blue!10,top color=blue!40,  bottom color=blue!10,fill opacity=.8] (-.5,.5)--(0.5,2)--(1,.5)--(-.5,.5);
   \shadedraw[shading=axis,left color=green!0,right color=green!70,
  shading angle=116.565051+90,
  fill opacity=0.5] (1,.5) -- (0.5,2) -- (4,2) ;
    \draw[Plum,thick] (0.5,2)--(4,2);
    \draw[thick] (1,.5)--(0.5,2)--(-.5,.5);
    \fill[black] (0.5,2) circle(2pt); 

    \shadedraw[blue!10,bottom color=blue!40,  top color=blue!10,fill opacity=0.8] (-.5,-.5)--(0.5,-2)--(1,-.5)--(-.8,-.5);
     \shadedraw[shading=axis,left color=green!0,right color=green!70,
  shading angle=-116.565051+90,
  fill opacity=0.5] (1,-.5) -- (0.5,-2) -- (4,-2) ;
    \draw[Plum,thick] (0.5,-2)--(4,-2);
    \draw[thick] (1,-.5)--(0.5,-2)--(-.5,-.5);
    \fill[black] (0.5,-2) circle(2pt); 

    \draw[dashed, name path=r1] plot [smooth] coordinates {(1,.5) (1.1,0) (1,-.5)};
    \draw[dashed, name path=l1] plot [smooth] coordinates {(-.5,-.5) (-.65,0) (-.5,.5)};
    \tikzfillbetween[of= l1 and r1]{blue!9};

    \definecolor{lightgreen}{RGB}{180,253,180}
    \draw[dashed, name path=r2] plot [smooth] coordinates {(4,2) (4.5,.5) (4.6,0) (4.5,-.5) (4,-2)};
    \tikzfillbetween [of= r1 and r2]{color=lightgreen,fill opacity=.5};

     \draw[dashed] plot [smooth] coordinates {(4,2) (3.3,.6) (3.2,0) (3.3,-.6) (4,-2)};

    \node[blue] at (0.2,0) {$S^4/\mathbb{Z}_3$};
    \node[Green] at (2.25,0) {$\Sigma_\epsilon\ltimes S^2_1$};
    \node[red] at (1.5,2.7) {$\Sigma_\epsilon\ltimes S^2_2$};
    \draw[->,red] (1.5,2.5) to[out=190,in=90] (1.3,2.1);
    \node[Plum] at (2.75,2.25) {$\Sigma_\epsilon$};
    \node at (5.2,1.2) {$\partial\Sigma_\epsilon\times S^4$};

\end{tikzpicture}
\label{fig:N1unres}
\end{center}

\end{minipage}~
\begin{minipage}{0.49\textwidth}

\begin{center}
\pgfdeclarelayer{pre main}
\begin{tikzpicture}
 \pgfsetlayers{pre main,main}

    \draw[dotted] (-.5,.5)--(0.5,2)--(4,2);
    \draw[dotted] (1,.5)--(0.5,2);

    \fill[orange!70,opacity=.5] (0.1,1.4)--(.9,.8)--(2,2)--cycle;
    \draw[thick]  (0.1,1.4)--(.9,.8)--(2,2)--cycle;

    \draw[dotted] (-.5,-.5)--(0.5,-2)--(4,-2);
    \draw[dotted] (1,-.5)--(0.5,-2);

    \fill[orange!70,opacity=.5] (0.1,-1.4)--(.9,-.8)--(2,-2)--cycle;
    \draw[thick]  (0.1,-1.4)--(.9,-.8)--(2,-2)--cycle;
 
    \shadedraw[blue!10,top color=blue!40,  bottom color=blue!10,fill opacity=.8] (-.5,.5)--(0.1,1.4)--(.9,.8)--(1,.5)--(-.5,.5);
   \shadedraw[shading=axis,left color=green!0,right color=green!70,
  shading angle=116.565051+90,
  fill opacity=0.5] (1,.5) -- (.9,.8)--(2,2) -- (4,2) ;
    \draw[thick] (-.5,.5)--(0.1,1.4)--(.9,.8)--(1,.5);
    \draw[Plum,thick] (2,2) -- (4,2);
    
    \fill[black] (0.1,1.4) circle(2pt); 
    \fill[black] (.9,.8) circle(2pt); 
    \fill[black] (2,2) circle(2pt); 

   \shadedraw[blue!10,bottom color=blue!40,  top color=blue!10,fill opacity=.8] (-.5,-.5)--(0.1,-1.4)--(.9,-.8)--(1,-.5)--(-.5,-.5);
   \shadedraw[shading=axis,left color=green!0,right color=green!70,
  shading angle=-116.565051+90,
  fill opacity=0.5] (1,-.5) -- (.9,-.8)--(2,-2) -- (4,-2) ;
    \draw[thick] (-.5,-.5)--(0.1,-1.4)--(.9,-.8)--(1,-.5);
    \draw[Plum,thick] (2,-2) -- (4,-2);
    
    \fill[black] (0.1,-1.4) circle(2pt); 
    \fill[black] (.9,-.8) circle(2pt); 
    \fill[black] (2,-2) circle(2pt); 

    \draw[dashed, name path=r1] plot [smooth] coordinates {(1,.5) (1.1,0) (1,-.5)};
    \draw[dashed, name path=l1] plot [smooth] coordinates {(-.5,-.5) (-.65,0) (-.5,.5)};
    \tikzfillbetween[of= l1 and r1]{blue!9};

    \definecolor{lightgreen}{RGB}{180,253,180}
    \draw[dashed, name path=r2] plot [smooth] coordinates {(4,2) (4.5,.5) (4.6,0) (4.5,-.5) (4,-2)};
    \tikzfillbetween [of= r1 and r2]{color=lightgreen,fill opacity=.5};
     
    \node[RedOrange] at (.9,1.35) {$\mathbb{CP}^2$};
    \node[blue] at (0.2,0) {$D_0$};
    \node[Green] at (2.45,0) {$D^{[2]}_\epsilon$};
     \node[red] at (2.7,2.8) {$D^{[1]}_\epsilon$};

\draw[->,red] (2.5,2.5) to[out=190,in=90] (2.3,2.1);
    
    \node[Plum] at (3.25,2.25) {$\Sigma_\epsilon$};
    \node at (5.2,1.2) {$\partial\Sigma_\epsilon\times S^4$};

    \draw[dashed] plot [smooth] coordinates {(4,2) (3.3,.6) (3.2,0) (3.3,-.6) (4,-2)};
    
\end{tikzpicture}

\end{center}

\end{minipage}%

\caption{
Puncture geometry: two copies of the (un)resolved geometries in figure~\ref{fig:toricC3} are smoothly glued together on the (left) right. The orientation and tilts are modified for a better visualization, such that the red face there is now hidden behind.}
\label{fig:N1res}

\end{figure}
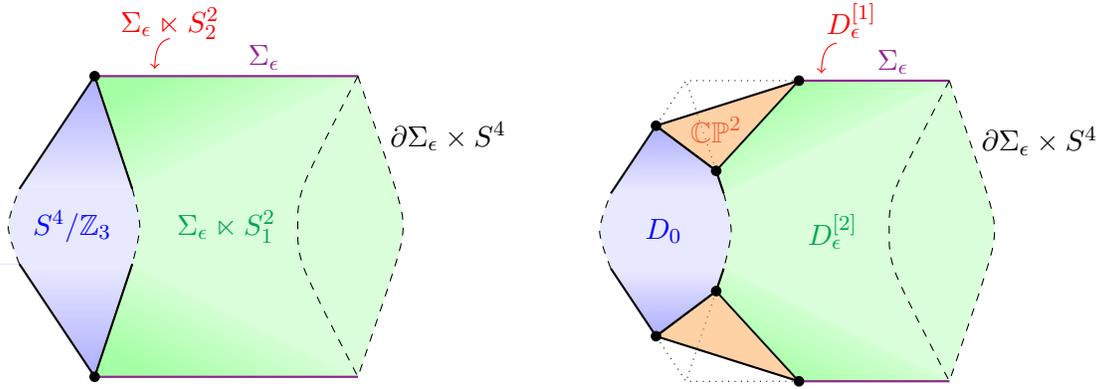

Let us first consider the simplest $\mathcal{N}=1$ preserving quotient, namely the diagonal action of $\mathbb{Z}_3$ on $\C^3$. Before resolving the singularity, in the notation of section \ref{sec:orbifoldsection} we have $d=1$, $k_1=K=3$. The weights of the action are uniquely fixed by \eqref{alphaconstr} to be $\alpha_1=\alpha_2=1$. Then the contribution to the central charge from the puncture is
\begin{equation}
    \delta a=-\frac{3}{8} b_1 b_2(b_1+b_2) N^3\,,
\end{equation}
where recall that $b_1+b_2=1$.

Consider now resolving the quotient. There is a unique crepant resolution of the singularity by blowing up a $\mathbb{CP}^2$ at the singular point. The total space is then the non-compact Calabi--Yau, $\mathcal{X}=\mathcal{O}(-3)\rightarrow\mathbb{CP}^2$. The resolution is implemented by introducing a plane which ``cuts the corner'' off the toric diagram of $\mathbb{C}^3/\mathbb{Z}_3$, see figure \ref{fig:toricC3}. The resultant toric vectors are taken to be
\begin{equation}
    v_1=(1,0,0)\, ,\quad v_2=(1,1,0)\, ,\quad v_3=(1,-1,3)\,,\quad v_4=(1,0,1)\, ,
\end{equation}
where $v_4$ is the normal direction to the added plane, $3v_4=v_1+v_2+v_3$. There are now three fixed points, arising from the intersection of the plane normal to $v_4$ with the planes normal to any two of $v_1,v_2,v_3$. To each fixed point there is an associated maximal cone consisting of the three vectors defining the normals to the intersecting planes, with each cone containing $v_4$. We take the maximal cones and the $y$ coordinate at the fixed point to be
\begin{equation}
    \langle v_1,v_2, v_4\rangle~ \leftrightarrow~ y_3\, ,\quad \langle v_2,v_3, v_4\rangle ~\leftrightarrow ~y_1\, ,\quad \langle v_3,v_1, v_4\rangle ~\leftrightarrow ~y_2\, ,
\end{equation}
see figure \ref{fig:toricC3}. It is simple to see that each of the fixed points is smooth, $k_a=1$ and the singularity has been fully resolved.
The weights at each vertex are given by
\begin{align}\label{weightsCP2}
    &(\epsilon_1^1,\epsilon_2^1,\epsilon_3^1)=\xi\cdot(v_4,v_2,v_3)^{-1} = (3b_1,b_2-b_1,-b_1)\nn \\
    &(\epsilon_1^2,\epsilon_2^2,\epsilon_3^2)=\xi\cdot(v_1,v_4,v_3)^{-1} = (b_1-b_2,3b_2,-b_2)\nn \\
    &(\epsilon_1^3,\epsilon_2^3,\epsilon_3^3)=\xi\cdot(v_1,v_2,v_4)^{-1} = (b_1,b_2,0)\,.
\end{align}
Note also that the
geometry in a neighbourhood of the 
bolt is simply the direct product of the disc and $\mathbb{R}^4$, and in particular the orbifold first Chern class of the normal bundle is zero ($\alpha_i^3=0$). 

We now need to glue together two of these orbifolds or their resolutions to make the puncture geometry. As we emphasized in section \ref{sec:N=1glue}, in order to perform the gluing we require that the non-compact faces are all identical, but in general compact faces may differ. This means that there are two distinct options: either we glue two copies of the resolved orbifold, or we glue together the orbifold and its resolution. In the following we will present the former option, see figure \ref{fig:N1res}, gluing together two of the resolved geometries, but we stress that this is a choice here. We will carefully denote which fixed points correspond to the north and south parts of the puncture geometry. The weights in the north copy are those in \eqref{weightsCP2}, while the product of the south pole weights has the opposite sign.
For each vector we also have an associated four-cycle, see figure \ref{fig:N1res}: $D_1\equiv D^{[1]}_\epsilon$, $D_2\equiv D^{[2]}_\epsilon$, $D_3\equiv D_0$, 
$D_4^N\equiv {\mathbb{CP}^2_N}$, $D_4^S\equiv {\mathbb{CP}^2_S}$.

First let us quantize the fluxes. Recall that the matching condition for the $S^4$ fibre fixes 
\begin{equation}
   \hat{y}_3^N=-\hat{y}_3^S=N \,,
\end{equation}
and we will use this consistently in the following. Note that there is no such condition for $y_1$ and $y_2$, and they can in general differ between the north and south. Computing the flux threading through the two resolution $\mathbb{CP}^2$'s we find
\begin{equation}\label{eq:fluxCP2}
    N_{\mathbb{CP}^2_N} =\frac{1}{(2\pi\lp)^3}\left(\frac{(2\pi)^2}{\epsilon_2^1\epsilon_3^1}\frac{y_1^N}{9}+\frac{(2\pi)^2}{\epsilon_1^2\epsilon_3^2}\frac{y_2^N}{9}+\frac{(2\pi)^2}{\epsilon_1^3\epsilon_2^3}\frac{y_3^N}{9}\right)=\frac{N}{2}+\frac{b_2 \hat{y}_1^N-b_1 \hat{y}_2^{N}}{2(b_1-b_2)}\, .
\end{equation}
Here the weights are given in \eqref{weightsCP2}, and the two weights that enter at a given fixed point are those in \eqref{weightsCP2} that do not include the normal vector $v_4$ to $\mathbb{CP}_N^2$. 
There is a similar expression for $N_{\mathbb{CP}^2_S}$ with the exchange $N\rightarrow S$, and the product of weights changing sign.
We use these fluxes to eliminate $\hat{y}_2^{N}$ and $\hat{y}_2^{S}$ giving
\begin{equation}\label{eq:y2rule}
    \hat y_2^N = \Big(1-\frac{b_2}{b_1}\Big)( N - 2N_{\mathbb{CP}^2_N})+\frac{b_2}{b_1}\hat y_1^N\,, \quad \hat y_2^S = -\Big(1-\frac{b_2}{b_1}\Big)( N - 2N_{\mathbb{CP}^2_S})+\frac{b_2}{b_1}\hat y_1^S\,.
\end{equation}
Further, we can consider the flux threading through $D_0$ which is the face normal to $v_3$. This is topologically an $S^2$ bundle over $S^2$ (in fact naturally the third Hirzebruch surface $\mathbb{F}_3$, which is diffeomorphic to the non-trivial $S^2$ bundle over $S^2$, and like $\mathbb{CP}^2$ is also not a spin manifold).
We find
\begin{equation}
    N_{D_0}=\frac{b_1(\hat{y}_{2}^N-\hat{y}_{2}^S)-b_2(\hat{y}_{1}^N-\hat{y}_1^S)}{6(b_1-b_2)}\, .
\end{equation}
This is not an independent flux but rather satisfies
\begin{equation}
    3 N_{D_0}+N_{\mathbb{CP}^2_N}+N_{\mathbb{CP}^2_S}=N\, .
\end{equation}
Finally, the fluxes threading through the two non-compact cycles are 
\begin{align}
    N_{\epsilon}^{[1]}&=\frac{1}{108 \pi^3 \lp^6 b_1 b_2^2 N}\frac{1}{2}\int_{\Sigma_{\epsilon}^N\cup \Sigma_\epsilon^S}\omega -\frac{b_1}{ b_2}\frac{\hat{y}_2^N-\hat{y}_2^S}{6}\, ,\\ \label{eq:N2epsilonZ3}
      N_{\epsilon}^{[2]}&=\frac{1}{108 \pi^3 \lp^6 b_1^2 b_2 N}\frac{1}{2}\int_{\Sigma_{\epsilon}^N\cup \Sigma_\epsilon^S}\omega -\frac{b_2}{ b_1}\frac{\hat{y}_1^N-\hat{y}_1^S}{6}\, ,
\end{align}
where we have used that $\int_{\Sigma_{\epsilon}}c_{1}(L_i)=0$. 
Recall that these fluxes are not gauge-invariant since the cycle has a boundary. Eliminating the integral of $\omega$ implies a consistency condition between the two fluxes, in particular one has 
\begin{equation}\label{eq:N1-2eps}
b_1N_{\epsilon}^{[2]}+b_2\frac{\hat{y}_1^N-\hat{y}_1^S}{6}= b_2 N_{\epsilon}^{[1]}+b_1\frac{\hat{y}_2^N-\hat{y}_2^S}{6}\, .
\end{equation}
We may pick one of the fluxes, say \eqref{eq:N2epsilonZ3}, and use this to give an expression for the integral over the K\"ahler class in terms of $N_\epsilon^{[2]}$ and the free parameters 
$\hat{y}_{1}^N,~\hat{y}_{1}^S$.

The contribution of the puncture to the central charge, $\delta a$, though complicated looking, is simply obtained giving
\begin{align}\label{deltaaCP2}
  \delta a=  \frac{3}{32} b_2 \Big[2 b_1^2 \Big\{&N^3-3 N^2 \big(3 N_{\epsilon}^{[2]}+N_{\mathbb{CP}^2_N}+N_{\mathbb{CP}^2_S}\big)+6 N
\Big(N_{\mathbb{CP}^2_N}^2+N_{\mathbb{CP}^2_S}^2\Big)-4 \Big(N_{\mathbb{CP}^2_N}^3+N_{\mathbb{CP}^2_S}^3\Big)\!\Big\}\nonumber\\
+\frac{b_2^2}{2} \big(&2 N-2 N_{\mathbb{CP}^2_N}-2N_{\mathbb{CP}^2_S}-\hat{y}_{1}^{N}+\hat{y}_{1}^{S}\big) \Big\{(2N_{\mathbb{CP}^2_N}+\hat{y}_1^N-2N_{\mathbb{CP}^2_S}+\hat{y}_1^S)^2 \nonumber\\
&+(N-2N_{\mathbb{CP}^2_N}-\hat{y}_1^N)^2+(N-2N_{\mathbb{CP}^2_S}+\hat{y}_1^S)^2\Big\}\nonumber\\
-4 b_1 b_2 \Big\{&N^3-3 N^2\Big(N_{\mathbb{CP}^2_N}+N_{\mathbb{CP}^2_S}\Big)- 4 N_{\mathbb{CP}^2_N}^3 -4 N_{\mathbb{CP}^2_S}^3-3 N_{\mathbb{CP}^2_N}^2 \hat{y}_{1}^{N}+3 N_{\mathbb{CP}^2_S}^2 \hat{y}_{1}^{S}\nonumber\\
&+3 N \left(N_{\mathbb{CP}^2_N} \big(2 N_{\mathbb{CP}^2_N}+\hat{y}_{1}^{N}\big)+N_{\mathbb{CP}^2_S} \big(2 N_{\mathbb{CP}^2_S}-\hat{y}_{1}^{S}\big)\right)\!\Big\}\Big]\,.
\end{align}
Notice that this is cubic in the $b_i$'s, as one expects for the central charge and its dependence on the R-symmetry. Moreover, note that there are now terms of the form $b_2^3$ which were absent in the $\mathcal{N}=2$ case. Further, observe that the central charge still depends on the two parameters $\hat{y}_1^N$ and $\hat{y}_1^S$. 
Note that one could have equally used $N^{[1]}_\epsilon$ rather than $N^{[2]}_\epsilon$ to evaluate the integral of $\omega$ as these are related through \eqref{eq:N1-2eps}. Similarly  one could have eliminated $\hat{y}_1$ in \eqref{eq:fluxCP2} rather than $\hat{y}_2$, such that the final result would be in terms of $N^{[1]}_\epsilon$, and the extremization over $\hat y_2^{N,S}$. In particular this would give terms in $b_1^3$ instead.

An interesting observation is that extremizing \eqref{deltaaCP2} over the 
parameters $\hat y_1^{N,S}$ effectively imposes homology relations on the integrals of $\Phi^Y$ (or equivalently $\Phi^{*G}$) in the puncture geometry. Indeed 
since the second Betti
number $b_2(\mathbb{CP}^2)=1$, the three toric two-cycles in the blown-up $\mathbb{CP}^2$ are necessarily all homologous. As such, if we integrate the closed two-form $Y$ over these homologous cycles they must be equal; that is,
\begin{equation}
    \int_{D_1\cap D_4^N}\Phi^Y= \int_{D_2\cap D_4^N}\Phi^Y= \int_{D_3\cap D_4^N}\Phi^Y\, ,
\end{equation}
and similarly for the southern cycle.
Evaluating the integrals using localization gives quadratic expressions in $y_a$,  and
the two equalities give rise to a single independent constraint, which reads
\begin{equation}\label{CP2relZ3}
    (b_1-b_2)(y_3^N)^2=b_1 ({y}_2^N)^2-b_2 ({y}_1^N)^2\, ,
\end{equation}
and similarly for the southern cycle.  Going to the hatted variables and further eliminating $\hat{y}_2$ using \eqref{eq:y2rule}, these constraints are then exactly the conditions obtained by extremizing the central charge over $\hat{y}_1^N$ and $\hat{y}_1^S$!
Finally, one should still extremize the full central charge over $b_1,b_2$ subject to $b_1+b_2=1$. 

The general expression \eqref{deltaaCP2} is a little unwiedly, and for illustration
we consider a straightforward simplification to spell out the extremization. We take the fluxes through the north and south copies of $\mathbb{CP}^2$'s to be equal $N_{\mathbb{CP}^2}^{N}=N_{\mathbb{CP}^2}^{S}\equiv N_{\mathbb{CP}^2}$. This actually sets $y_i^N=-y_i^S\equiv y_i$. For readability we will also directly consider this symmetric case in the example in the next subsection. The shift of the central charge then simplifies to 
\begin{align}\label{deltaasym}
  \delta a=  \frac{3}{16} b_2 \Big[ b_1^2 \Big\{&N^3-3 N^2 \big(3 N_{\epsilon}^{[2]}+2N_{\mathbb{CP}^2}\big)+12 N
N_{\mathbb{CP}^2}^2-8 N_{\mathbb{CP}^2}^3\!\Big\}+{b_2^2} \big( N-2 N_{\mathbb{CP}^2}-\hat{y}_{1}\big)^3\nonumber\\
-2 b_1 b_2 \Big\{&N^3-6 N^2N_{\mathbb{CP}^2}- 8 N_{\mathbb{CP}^2}^3-6 N_{\mathbb{CP}^2}^2 \hat{y}_{1}+6 N N_{\mathbb{CP}^2} \big(2 N_{\mathbb{CP}^2}+\hat{y}_{1}\big)\Big\}\Big]\,,
\end{align}
where one should still extremize over $\hat y_1$. The extremization gives
\begin{equation}
    \hat y_1^*=N-2N_{\mathbb{CP}^2}\pm2\sqrt{\frac{b_1}{b_2}N_{\mathbb{CP}^2}(N_{\mathbb{CP}^2}-N)}\,.
\end{equation}
Substituting this back into \eqref{eq:y2rule} gives\footnote{Recall that one needs to pick the solution which maximizes the central charge. Since we are looking only at the puncture contribution, which we take away from the bulk central charge, maximizing the bulk central charge is equivalent to minimizing the puncture contribution. We are therefore lead to pick the root with the minus sign.}
\begin{equation}
    \hat y_2^*=N-2N_{\mathbb{CP}^2}\pm2\sqrt{\frac{b_2}{b_1}N_{\mathbb{CP}^2}(N_{\mathbb{CP}^2}-N)}\,,
\end{equation}
such that 
\begin{equation}
    b_1 y_2^{*2}-b_2 y_1^{*2}=(b_1-b_2)N^2\,,
\end{equation}
which corresponds to the homology relation \eqref{CP2relZ3} as claimed.
Finally, the extremized central charge is given by inserting $y_1^*$ in \eqref{deltaasym} giving
\begin{align}
   \delta a^* =&\frac{3}{16} b_1 b_2 \bigg[b_1 \big(N^3-3 N^2 (3 N^{[2]}_\epsilon+2 N_{\mathbb{CP}^2})+12 N N_{\mathbb{CP}^2}^2-8 N_{\mathbb{CP}^2}^3\big)  \\ \nonumber
   &-2 b_2 \big(N^3-6 N N_{\mathbb{CP}^2}^2+4 N_{\mathbb{CP}^2}^3\big)-16 N_{\mathbb{CP}^2}(N_{\mathbb{CP}^2}-N) \sqrt{b_1 b_2 N_{\mathbb{CP}^2}(N_{\mathbb{CP}^2}-N)}\bigg]
\end{align}
Note that the \textit{on-shell} result still requires an extremization over $b_i$ which should be conducted for the total central charge $a=a^\mathrm{bulk}-\delta a^*$. Note that the asymmetry between $b_1$ and $b_2$ comes from using $N^{[2]}_\epsilon$.

A final simplification is to turn off the fluxes through the $\mathbb{CP}^2$ cycles: $N_{\mathbb{CP}^2}^{N}=N_{\mathbb{CP}^2}^{S}=0$. This further sets $\hat y_1=\hat y_2=\hat y_3=N$ (such that $N^{[1]}_\epsilon=N^{[2]}_\epsilon=N/3$) and the contribution to the central charge reduces to 
\begin{equation}
    \delta a=-\frac{3}{8} b_1 b_2(b_1+b_2) N^3\,,
\end{equation}
which is the result for the unresolved $\mathbb{Z}_3$ singularity. This is a non-trivial check: 
 we recover the orbifold result as a limit of the resolved result. In this orbifold case there is only the extremization over $b_i$ left to perform.


\subsubsection{\texorpdfstring{$\mathbb{Z}_5$}{Z5} quotient}\label{sec:Z5ex}

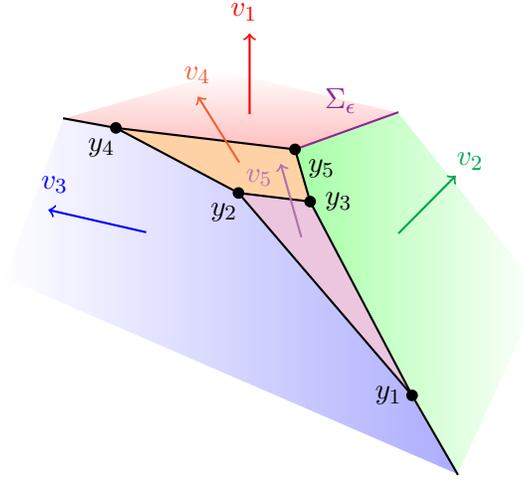
\begin{figure}[ht!]

\begin{center}
\tdplotsetmaincoords{78}{-60}

\begin{tikzpicture}
		[tdplot_main_coords,
			cube/.style={very thick,black},
			axisb/.style={->,blue,thick},
            axisr/.style={->,red,thick},
            axisg/.style={->,Green,thick},
			inf/.style={dashed,black},scale=2.2]

\shadedraw[white,top color=white, bottom color=red!50,fill opacity=0.5] (.25,0,0)--(3/2,0,0)--(2,1.5,0)--(1/3,5/3,0)--(.25,1.25,0)--(.25,0,0);
\shadedraw[white,right color=blue!40, left color=white,fill opacity=0.8] (.25,1.25,0)--(1/3,5/3,0)--(1/5,2,-1)--(-2/3,-5/3,-5/3)--(-.5,-1.25,-1.25)--(0,.25,-.25)--(.25,1.25,0);
\shadedraw[white,left color=Green!70, right color=Green!0,fill opacity=0.5] (.25,0,0)--(1.5,0,0)--(1.5,-1,-1)--(-2/3,-5/3,-5/3)--(-.5,-1.25,-1.25)--(0,-.25,-.25)--(.25,0,0);
\fill[orange!70,opacity=.5] (.25,0,0)--(.25,1.25,0)--(0,.25,-.25)--(0,-.25,-.25)--(.25,0,0);
\fill[Thistle!90,opacity=.5] (0,.25,-.25)--(0,-.25,-.25)--(-.5,-1.25,-1.25)--(0,.25,-.25);

\draw[black, thick] (1/3,5/3,0)--(.25,1.25,0)--(0,.25,-.25)--(-.5,-1.25,-1.25)--(-2/3,-5/3,-5/3);
\draw[black, thick](-.5,-1.25,-1.25)--(0,-.25,-.25)--(.25,0,0)--(.25,1.25,0);
\draw[black,thick](0,.25,-.25)--(0,-.25,-.25);
\draw[Plum,thick](.25,0,0)--(1.5,0,0);

\draw[axisg] (1.5,0,-.75)--(1.5,-.4,-0.35);
\draw[axisb] (-0.25,.75,-.5)--(-1,1.,-0.25);
\draw[axisr] (1.,.75,0)--(1.,.75,.5);
\draw[->,RedOrange,thick] (-.25,.1,0)--(-.75,.1,.5);
\draw[->,Orchid,thick] (0.5,0.1,-.6)--(0.25,0.1,-.1);

\fill[black] (.25,0,0) circle(1pt) node[below] {$\,\,\,\,\,\quad y_5$}; 
\fill[black] (.25,1.25,0) circle(1pt) node[below] {$y_4\quad$};
\fill[black] (0,-.25,-.25) circle(1pt) node[right] {$\, y_3$}; 
\fill[black] (0,.25,-.25) circle(1pt) node[below] {$y_2\quad$};
\fill[black] (-.5,-1.25,-1.25) circle(1pt) node[left] {$y_1$}; 
\node[Plum] at (.8,0,.2) {$\Sigma_\epsilon$};

\node at (.5,.5,.75) {\color{red}$v_1
$};
\node at (1.5,-.5,-0.25) {\color{Green}$v_2
$};
\node at (-.75,1.1,-0.15) {\color{blue}$v_3
$};
\node at (-.5,.25,.57) {\color{RedOrange}$v_4
$};
\node at (0.25,0.25,-.2) {\color{Orchid}$v_5
$};

\end{tikzpicture}

\end{center}

\caption{Toric diagram for the resolution of the $\mathbb{C}^3/\mathbb{Z}_5$ singularity with $\alpha_2=1$. The orange cycle has topology $\mathbb{F}_3$  and the lilac one $\mathbb{CP}^2$.}
\label{fig:toricC3Z5}

\end{figure}

As a second example, we consider the resolution of $\mathbb{C}^3/\mathbb{Z}_5$. We take toric data 
\begin{equation}
    v_1=(1,0,0)\,,\quad  v_2=(1,1,0)\,, \quad v_3=(1,-1,5)\,,
\end{equation}
so that 
before resolution $\alpha_2=1$, $K=5$ (implying $\alpha_1=3$) and the shift to the central charge is given by the orbifold formula
\begin{equation}
    \delta a^\mathrm{orb}=-\frac{9}{40} b_1 b_2(b_1+3b_2) N^3\,.
\end{equation}
We now resolve by adding to this the following vectors
\begin{equation}
    v_4=(1,0,1)\,,\quad v_5=(1,0,2)\,.
\end{equation}
See figure \ref{fig:toricC3Z5} for the polytope.\footnote{In general there will be different polytopes for different ranges of K\"ahler parameters, corresponding to different triangulations of the toric diagram. However, here there is only one triangulation.  } 
This fully resolves the singularity. Indeed one can check that the determinants of the vectors $k_a=1$ at each vertex. Consequently, $\alpha_i^d=0$ or equivalently the local geometry in a neighbourhood of the bolt is a direct product. 
The weights at the vertices are given by
\begin{align}\label{weightsZ5}
    &(\epsilon_1^1,\epsilon_2^1,\epsilon_3^1) =\xi\cdot(v_5,v_2,v_3)^{-1}= (5b_1,b_2-2b_1,-2b_1) \nn \\
    &(\epsilon_1^2,\epsilon_2^2,\epsilon_3^2)=\xi\cdot(v_3,v_4,v_5)^{-1} = (-b_2,2b_1-b_2,-b_1+3b_2)
    \nn \\
    &(\epsilon_1^3,\epsilon_2^3,\epsilon_3^3)=\xi\cdot(v_4,v_2,v_5)^{-1} = (2b_1,b_2,-b_1)\nn \\
    &(\epsilon_1^4,\epsilon_2^4,\epsilon_3^4)=\xi\cdot(v_1,v_4,v_3)^{-1} = (b_1-3b_2,5b_2,-b_2)\nn \\
    &(\epsilon_1^5,\epsilon_2^5,\epsilon_3^5)=\xi\cdot(v_1,v_2,v_4)^{-1} = (b_1,b_2,0)\,.
\end{align}
As a quick check of the Calabi--Yau property, note that the weights all sum to $b_1+b_2$. The fifth vertex is where the toric geometry meets the bolt Riemann surface, and has $\hat y_5=N$.

As explained earlier, we could consider  gluing different geometries in the north and south. Here we make the choice of gluing two copies of the fully resolved geometry we have just described. 
The weights at the north copy are those in \eqref{weightsZ5}, while their product pick up a minus sign in the south patch.
To each toric vector is associated a four-cycle: 
$D_1\equiv D^{[1]}_\epsilon$, $D_2\equiv D^{[2]}_\epsilon$, $D_3\equiv D_0$, 
$D_4^N\equiv {\mathbb{F}^3_N}$, $D_4^S\equiv {\mathbb{F}^3_S}$, $D_5^N\equiv {\mathbb{CP}^2_N}$, $D_5^S\equiv {\mathbb{CP}^2_S}$.
Moreover, we take the fluxes through the north and south compact cycles to be equal: $N_{\mathbb{F}^3}^{N}=N_{\mathbb{F}^3}^{S}\equiv N_{\mathbb{F}^3}$, $N_{\mathbb{CP}^2}^{N}=N_{\mathbb{CP}^2}^{S}\equiv N_{\mathbb{CP}^2}$. Again this is not necessary but will simplify the expressions. Indeed it sets $y_a^N=-y_a^S\equiv y_a$.

First we quantize the flux through the compact cycles. Localization gives 
\begin{align}\label{fluxCP2Z5}
 N_{\mathbb{CP}^2} 
 & =\frac{\hat{y}_3}{4}-\frac{2b_1\hat{y}_2-b_2\hat{y}_1}{4(2b_1-b_2)}\,, \\
\label{fluxF3Z5}
 N_{\mathbb{F}_3} 
   & =\frac{N-\hat y_3}{2}+\frac{b_1(\hat{y}_2-\hat{y}_4)}{2(b_1-3b_2)}\,.
 \end{align}
 The final compact cycle $D_0$ picks up six nut contributions, which combine to give
\begin{align}
    N_{D_0}= \frac{2 b_1^2 \hat y_4+3 b_2^2 \hat y_1-b_1 b_2 (\hat y_1+5 \hat y_2+\hat y_4)}{5 (b_1-3 b_2) (2 b_1-b_2)}\,.
    \end{align}
This does not provide an additional constraint as, similarly to the previous example, the fluxes are related: 
\begin{equation}
    5 N_{D_0}  +4N_{\mathbb{CP}^2} +2 N_{\mathbb{F}_3}  =N\,.
\end{equation}
Next we compute the two non-compact fluxes (taking the 
integrals of $\omega$ over $\Sigma_\epsilon^N$ and $\Sigma_\epsilon^S$ to be equal by symmetry)
\begin{align}
    N_{\epsilon}^{[1]}&=\frac{1}{108 \pi^3 \lp^6 b_1 b_2^2 N}\int_{\Sigma_{\epsilon}}\omega -\frac{b_1}{ b_2}\frac{\hat{y}_4}{5}\, ,\\ \label{eq:N2epsilonZ5}
      N_{\epsilon}^{[2]}&=\frac{1}{108 \pi^3 \lp^6 b_1^2 b_2 N}\int_{\Sigma_{\epsilon}}\omega -\frac{b_2}{b_1}\Big(\frac{\hat{y}_1}{10}+\frac{\hat{y}_3}{2}\Big)\, ,
\end{align}
which satisfy
\begin{align}
    b_2N_{\epsilon}^{[1]}+{b_1}\frac{\hat{y}_4}{5}=
      b_1N_{\epsilon}^{[2]}+{b_2}\Big(\frac{\hat{y}_1}{10}+\frac{\hat{y}_3}{2}\Big)\, .
\end{align}
Having this relation between the two fluxes means that it does not matter which one of the two we use to replace the integral of $\omega$ in the following computation of the central charge.

We now compute the central charge. This consists of the bolt contribution and five nut contributions (doubled to account for the north and south copies). This gives
\begin{align}\label{deltaaZ5}
  \delta a=  \frac{9}{80} b_2\Big[ b_1^2 \Big\{& N^3-3 N^2 (5 N^{[2]}_\epsilon+4 N_{\mathbb{CP}^2}+2 N_{\mathbb{F}^3})+12 N (2 N_{\mathbb{CP}^2}+N_{\mathbb{F}^3})^2-8 (2 N_{\mathbb{CP}^2}+N_{\mathbb{F}^3})^3 \!\Big\}\nonumber\\[0.5ex]
+\frac{b_2^2}{4} \Big\{(N&-4 N_{\mathbb{CP}^2}-2 N_{\mathbb{F}^3}-\hat y_1)^3-5 (N-2 N_{\mathbb{F}^3}-\hat y_3)^2 (7N-12 N_{\mathbb{CP}^2}-14 N_{\mathbb{F}^3}-3 \hat y_1-4 \hat y_3)\!\Big\}\nonumber\\[0.5ex]
-6b_1b_2  \Big\{& N^3+ N (2 N_{\mathbb{CP}^2}+N_{\mathbb{F}^3}) (4 N_{\mathbb{CP}^2}+12 N_{\mathbb{F}^3}+\hat y_1+5 \hat y_3)\nonumber\\[0.5ex] 
&  -N_{\mathbb{F}^3} \left(6N^2+16 N_{\mathbb{CP}^2}^2+8 N_{\mathbb{F}^3}^2+ (\hat y_1+5 \hat y_3 ) (4 N_{\mathbb{CP}^2}+ N_{\mathbb{F}^3})\right) \nonumber \\[0.5ex]
 &-N_{\mathbb{CP}^2}(7N^2 + \frac{32}{3} N_{\mathbb{CP}^2}^2+28  N_{\mathbb{F}^3}^2+4 \hat y_1 N_{\mathbb{CP}^2}+5 \hat y_3^2) \!\Big\}\Big]\,.
\end{align}
We have used \eqref{fluxCP2Z5} and \eqref{fluxF3Z5} to replace $\hat y_2$ and $\hat{y}_4$ and \eqref{eq:N2epsilonZ5} for the integral of $\omega$ over the bolt. One could equivalently pick another set of variables, but we find that this choice 
makes the expressions more easy on the eye.  The free variables left are $\hat y_1$ and $\hat{y}_3$, and they should be extremized over.

The extremization leads to a pair of quadratic constraints, which are not easily solved in closed form. However, as already mentioned, these constraints actually correspond to homology relations for $\Phi^Y$ integrals in the resolution. Let us see that explicitly.
First, the divisor $D_5$ is a triangle, and so must be  a copy $\mathbb{CP}^2$, with normal bundle $\mathcal{O}(-3)\rightarrow\mathbb{CP}^2$. 
The edges of the triangle are all copies of $\mathbb{CP}^1$ whose homology classes are equal, so
\begin{align}
\int_{D_2\cap D_5}\Phi^Y =    \int_{D_3\cap D_5}\Phi^Y =\int_{D_4\cap D_5}\Phi^Y \, .
\end{align}
Using localization to evaluate the integrals, 
this imposes a single independent constraint, which reads
\begin{align}\label{CP2relation}
(2b_1-b_2)y_3^2=2b_1 y_2^2-b_2 y_1^2\, .
\end{align}
Second the divisor $D_4$ is a quadrilateral. We can determine its topology by writing the projected vectors
\begin{align}
u_1=v_1-v_4\, , \quad u_2 = v_2-v_4\, , \quad u_3 = v_3-v_4\, , \quad u_4=v_5-v_4\, .
\end{align}
The kernel of the linear map specified by the $u_a$ 
is generated by $\{(0,1,1,-3),(1,0,0,1)\}$, which defines the third Hirzebruch surface $\mathbb{F}_3$: taking the K\"ahler quotient of $\C^4$ by 
$(0,1,1,-3)$ gives $\C\times (\mathcal{O}(-3)\rightarrow\mathbb{CP}^1)$, and then quotienting by $(1,0,0,1)$ precisely projectivizes the fibres $\C^2$ to copies of $\mathbb{CP}^1$. 
Then the homology relations for the two two-cycles within $D_4\cong \mathbb{F}_3$  follow from the kernel for the $u_a$ vectors. 
Namely we can label
\begin{align}
F_1 = D_3\cap D_4\, ,\quad 
F_2 = D_2\cap D_4\, , \quad S_1 = D_1\cap D_4\, , \quad S_2 = D_5\cap D_4\, . 
\end{align}
Here $F_1$, $F_2$ are the fibres of the $\mathbb{CP}^1$ bundle over the two poles of the base $\mathbb{CP}^1$, respectively. They have self-intersection number zero.  
Then $S_1$, $S_2$ are the two sections of this bundle 
at the poles of the fibre and have self-intersection number $+3$ and $-3$ respectively. 
 It follows\footnote{A more complete discussion may be found in section 5.1.3 of \cite{Gauntlett:2019pqg}.} that the homology relations are $[F_1]=[F_2]$ and $[S_1]-[S_2]=3[F_1]$, which give
\begin{align}\label{fibrerelation}
\frac{1}{b_1-3b_2}(y_4^2-y_2^2) & = \frac{1}{b_1}(y_5^2-y_3^2)\,,\\
\label{F3relation}
\frac{1}{b_2}(y_5^2-y_4^2)-\frac{1}{b_2}(y_3^2-y_2^2)   & = 3\cdot \frac{1}{b_1-3b_2}(y_4^2-y_2^2)\, .
\end{align}
Note that the last relation is redundant -- it is already implied by \eqref{CP2relation} and \eqref{fibrerelation}.
We stress that these constraints do not need to be imposed by hand, but rather directly come out of extremizing the central charge.
Indeed, one can easily check that the extremization with respect to $\hat{y}_1$
imposes a linear combination of \eqref{CP2relation} and \eqref{fibrerelation}, while extremization with respect to $\hat{y}_3$ gives rise to \eqref{F3relation}.

\section{Conclusion}\label{sec:conclusion}

In this paper we have shown how equivariant localization can be used to compute BPS observables in gravity for M5-brane SCFTs associated to Riemann surfaces with punctures. For $\mathcal{N}=1$ punctures this leads to entirely new results, and our work opens up a number of interesting directions for future research.

Firstly, it would clearly be of great interest to reproduce our gravity results from a dual field theory computation of the central charge. 
 One prediction of our work is that $a$-maximization in this set-up can be formulated to include 
$b_2(\mathcal{X})$ variables $\hat{y}_a$, that are also extremized over. Although notice that while our 
 $a$ is cubic in both the
 $b_i$ and $\hat{y}_a$ variables, it is not a cubic function of $(b_i,\hat{y}_a)$ as a single set of variables. 
As mentioned in the introduction, relating the 
holographic anomaly inflow method \cite{Bah:2018gwc} more directly to equivariant localization would surely help understand better the anomaly polynomial calculation for the configuration of M5-branes with punctures. 

In addition to the $\mathbb{C}^3/\mathbb{Z}_{K}$ quotients considered here, one can also consider quotients of the form $\mathbb{C}^3/(\mathbb{Z}_{K_1}\times \mathbb{Z}_{K_2})$. This naturally leads one to consider taking $S^4/\mathbb{Z}_{K}$ bundles over a possibly punctured Riemann surface and then resolving the singularities there too.\footnote{Some work using anomaly inflow along this direction for smooth Riemann surfaces has been conducted in \cite{Bah:2021brs}.} It is interesting to realize this using our equivariant localization technology, where we will be able to generalize this picture, and this is a direction we hope to report on soon. In addition to giving rise to new interesting classes to study this will also help explain classes of explicit solutions constructed in \cite{Gauntlett:2004zh}, whose physical origin is not understood. Furthermore this will shed light on the so-called $(p,q)$ punctures considered in \cite{Bah:2018gwc}. These can be understood as a specific choice of action for the $\mathcal{N}=1$ quotients discussed here. Using our results we are able to recover the central charge computed there using anomaly inflow. 

We have focused on regular punctures, but gravity duals to  M5-branes probing an $\mathcal{N}=2$ irregular puncture have been constructed in 
\cite{Bah:2021mzw, Bah:2021hei,Couzens:2022yjl,Bah:2022yjf}.\footnote{See also \cite{Kim:2025ziz} for an $\mathcal{N}=1$ prototype of an irregular type puncture. Equivariant localization can be used to recover the central charge presented there.} 
These solutions involve a fibration over a  disc with explicit M5-brane sources over the disc boundary and are the holographic duals of Class S Argyres--Douglas theories.
Equivariant localization also extends to this setting \cite{BCK}, where it is shown how to recover the exact (to order $1$) central charge for a subset of these theories and other $\mathcal{N}=2$ preserving solutions. For the Argyres--Douglas theories there are two classes of irregular puncture that can be considered, known as type I and type III\footnote{In the literature one also sees type II punctures; however, as explained in \cite{Beem:2023ofp} these are a special class of type I punctures.}, see for example \cite{Xie:2017vaf}. The type I puncture is well understood field theoretically, however the type III puncture is far less well understood with general results for the central charge of the theories within this class unknown. The explicit solutions \cite{Bah:2021mzw, Bah:2021hei,Couzens:2022yjl,Bah:2022yjf} and work in \cite{BCK} only realize the type I puncture and it would be interesting to use equivariant localization to make predictions for the central charges for the type III class of theories. 

More broadly, equivariant localization makes clear that certain BPS observables in supergravity can be computed by cutting and gluing manifolds together, with fixed point contributions in each piece. In simple settings the 
building blocks have been called 
``gravitational blocks'' 
(as introduced in \cite{Hosseini:2019iad}). Ultimately we believe that everything should be expressible in terms of equivariant cohomology, with the same then being true in field theory. We leave these very interesting questions for future work.

\section*{Acknowledgements}
We would like to thank Chris Beem, Pieter Bomans, Adam Kmec, Matteo Sacchi, Zhenghao Zhong for helpful discussions. 
JFS would like to thank the Centro Internacional de F\'isica, Universidade de Bras\'ilia, and the Department of Mathematics, University of Turin, for hospitality. JFS is supported in part by STFC grant ST/X000761/1. AL is supported by a Palmer Scholarship.


\appendix

\section{Explicit flux quantization for \texorpdfstring{$\mathcal{N}=2$}{N=2} punctures}\label{app:explicitG}

In this appendix we show how to introduce an explicit ansatz that represents the cohomology class of $G$ for $\mathcal{N}=2$ punctures. One can recover various formulae in the main text by direct integration of this four-form.

Recall that the local $\mathcal{N}=2$ puncture geometry involves a 
partial resolution of the $\C^2/\Z_\K$ singularity, shown in figure~\ref{figpartial}. 
We begin by introducing
Poincar\'e duals 
$\Psi_a$ for each of the compact 
toric divisors $\D_a$, $a=2,\ldots,d$, in this 4d geometry. 
By definition the $\Psi_a$ are closed two-forms that are supported (non-zero)
only in a neighbourhood of each 
toric divisor, and integrate 
to 1 along the normal directions. For illustration, the support of the Poincar\'e dual $\Psi_2$ to $\D_2$ is shown as the shaded region in figure \ref{figpartial}.
These duals then satisfy\footnote{Notice that this intersection matrix appeared in equivariant localization in supergravity, but in a different setting, 
in \cite{BenettiGenolini:2024hyd}.}
\begin{align}\label{Psi}
\int_{\D_b} \Psi_{a} = \begin{cases}\  \frac{1}{k_{a-1}} & \ \  b= a-1\, ,\\ 
\ \frac{1}{k_a} & \ \  b= a+1\, ,\\ 
\ -(\frac{1}{k_{a-1}}+\frac{1}{k_a}) & \ \ b=a\, ,\end{cases}
\end{align}
for $a=2,\ldots, d$. 
The first two lines follow immediately from the definition, 
noting that 
the normal directions to $\D_a$  over its poles (blue dots in figure~\ref{figpartial}) are the same as the tangent directions to $\D_{a-1}$, $\D_{a+1}$ at those same poles 
(which are orbifold singularities of order $k_{a-1}$, $k_{a}$, respectively).
For the last line of \eqref{Psi} we note $\chi(\D_a)=  
\frac{1}{k_{a-1}}+\frac{1}{k_a}$ is the Euler number of the weighted projective space $\D_a=\mathbb{WCP}^1_{[k_a,k_{a-1}]}$, 
where recall that the normal bundle of $\D_a$ is (necessarily 
for a Calabi--Yau two-fold) its cotangent bundle. 

We can then write the following ansatz for the cohomology class of $G$:
\begin{align}\label{Gansatz}
G = G_{\mathrm{bulk}}+(2\pi\lp)^3\left[\sum_{a=2}^{d} \left(\sum_{b=1}^{a-1} k_bn_b\right)\Psi_{a} \wedge \vol_{S^2_R}  \right]\, ,
\end{align}
where similarly to \eqref{Gformula} 
we define 
\begin{align}\label{Gbulk}
G_{\mathrm{bulk}} \equiv 
\frac{(2\pi\lp)^3}{2\pi}N\left[\diff\rho\wedge (\diff\varphi_1-A_1)-\rho F_1\right]\wedge \vol_{S^2_R}\,.
\end{align}
Here $\vol_{S^2_R}$ is any volume form on $S^2_R$ that integrates to 1, while we take $\rho$ to be a function defined in the 4d geometry that is $-1$ near to $\D_{d+1}=\Sigma_\epsilon\cong \C/\Z_{k_d}$, and rapidly decays to 0 as one moves away from it. 
The first term 
in \eqref{Gbulk} then 
integrates to $N$ over the copy of $S^4$ on the dotted black line in figure~\ref{fig:toric}. 
This expression for 
$G_{\mathrm{bulk}}$ 
is closed and well-defined in the whole puncture geometry region; in particular, notice that although $\vol_{S^2_R}$ 
is only well-defined where $S^2_R$ is not collapsed, 
both $\rho$ and $\diff\rho$ are only supported near to $\D_{d+1}=\Sigma_\epsilon$ (corresponding to the shaded region in figure~\ref{fig:toric}).  
Since $\D_{d+1}\cong \C/\Z_{k_d}$, as in \eqref{GfluxDirac} we have 
\begin{align}\label{Diracbit}
\frac{1}{(2\pi\lp)^3}\int_{D^{[1]}_\epsilon}G_{\mathrm{bulk}} =\sG N\int_{\D_{d+1}} \frac{F_1}{2\pi} 
=\sG\left(1-\frac{1}{k_d}\right)N\, ,
\end{align}
where recall that $D^{[1]}_\epsilon \equiv \D_{d+1}\times S^2_R$, and
where we have used \eqref{N2twists} to evaluate the integral.

The second term in \eqref{Gansatz} precisely produces the fluxes 
in \eqref{n1def}, \eqref{eq:fluxN2}. 
To see this, first recall that 
the Poincar\'e duals $\Psi_a$ are supported 
only in a neighbourhood of each compact toric divisor, 
and satisfy \eqref{Psi}. For example, the 
only form supported along $\D_1\cong\C/\Z_{k_1}$ 
is $\Psi_2$, which integrates to $1/k_1$. Thus
\begin{align}
\frac{1}{(2\pi\lp)^3}\int_{D_1} G = 
\int_{\D_1} (k_1n_1)\Psi_2 = n_1\, ,
\end{align}
where recall $D_1\cong S^4/\Z_{k_1}$. 
Then for $D_a=\D_a\times S^2_R$, $a=2,\ldots,d$, we have
\begin{align}\label{computeflux}
\frac{1}{(2\pi\lp)^3}\int_{D_a} G & = 
\int_{\D_a} \left[\left(\sum_{b=1}^{a-2}k_bn_b\right)\Psi_{a-1} +
\left(\sum_{b=1}^{a-1}k_bn_b\right)\Psi_{a}+\left(\sum_{b=1}^{a}k_bn_b\right)\Psi_{a+1}\right]\nonumber\\
& = \frac{1}{k_{a-1}}\left(\sum_{b=1}^{a-2}k_bn_b\right) -\left(\frac{1}{k_{a-1}}+\frac{1}{k_a}\right)\left(\sum_{b=1}^{a-1}k_bn_b\right) + \frac{1}{k_a}\left(\sum_{b=1}^{a}k_bn_b\right)\nonumber\\
& = n_a-n_{a-1}\, .
\end{align}
Notice here that $D_2$ and $D_d$ are special: for $D_2$ the first sum on the right hand side of \eqref{computeflux} is empty (and indeed we did not define $\Psi_1$). Instead for $D_d$ 
the last term $\Psi_{d+1}$ 
in \eqref{computeflux} is identified with the first term in $G_{\mathrm{bulk}}$ in \eqref{Gbulk}. 
Finally the full $G$-flux along 
$D_{d+1}=D^{[1]}_\epsilon$ is then, using 
\eqref{Diracbit} for the first bulk contribution, 
\begin{align}\label{fluxboltexplicitN2}
N^{[1]}_\epsilon\equiv\frac{1}{(2\pi\lp)^3}\int_{D_{d+1}}G & =\sG\left(1-\frac{1}{k_d}\right)N + 
\int_{\D_{d+1}}\left(\sum_{b=1}^{d-1}k_bn_b\right)\Psi_d\nonumber\\
& = \sG\left(1-\frac{1}{k_d}\right)N+\sG
\frac{1}{k_d}\left(\sum_{b=1}^{d-1}k_bn_b\right)=N-n_d\, ,
\end{align}
where we used \eqref{sumtoN} in the last step.

\bibliographystyle{JHEP}

\bibliography{22Loc-2}

\end{document}